\documentclass[12pt]{article} 
\usepackage{amsmath}
\usepackage{latexsym}
\usepackage{slashed}
\usepackage{graphicx}
\begin{document} 
\bibliographystyle{unsrt} 
 
\begin{center} 
{\LARGE \bf Effective Hadronic Lagrangians Based on QCD: Potential Models and Skyrmions}\\[10mm] 
 
Sultan Catto$^{\dag}$\\{\it Physics Department\\ The Graduate School and University Center\\365 Fifth Avenue\\
New York, NY 10016-4309\\ and \\ Center for Theoretical Physics \\The Rockefeller University \\
1230 York Avenue\\ New York NY 10021-6399}\\[6mm] 
\end{center} 
\vbox{\vspace{5mm}}

\begin{abstract} An approximate hadronic symmetry based on spin and flavor independence and broken by spin and mass dependent terms is shown to follow from QCD. This symmetry justifies the $SU(6)$ classification scheme, but is more general in allowing its supersymmetric extension based on a diquark-antiquark symmetry. It will be shown that the same supersymmetry is also implied in the skyrmion type effective Lagrangian which could be extracted from QCD. Predictions of the Skyrme model is improved by using different realizations of the chiral group.

\end{abstract}
\vbox{\vspace{5mm}}


\vbox{\vspace{10mm}}

$^\dag$ Work supported in part by DOE contracts No. DE-AC-0276 ER 03074 and 03075, and PSC-CUNY Research Awards.
\newpage

\section*{Introduction}

Inside rotationally excited baryons, QCD leads to the formation of diquarks well separated from the remaining quark. At this separation the scalar, spin independent, confining part of the effective QCD potential is dominant. Since QCD forces are also flavor independent, the force between the quark ($q$) and the diquark ($qq$) inside an excited baryon is essentially the same as the one between $q$ and the antiquark $\bar{q}$ inside an excited meson. Thus, the approximate spin-flavor independence of hadronic physics expressed by $SU(6)$ symmetry gets extended to the $SU(6/21)$ supersymmetry through a symmetry between $\bar{q}$ and $(qq)$, resulting into the parallelism of mesonic and baryonic Regge trajectories. Various aspects and implications of this approximate effective supersymmetry and its breaking are discussed in the first section. 

In the next section we discuss the Skyrmion approach to effective hadronic Lagrangians.
In the early $1960$'s Skyrme$^{\cite{skyrme}}$ proposed a model in which baryons can be understood as solitons in an effective Lagrangian for bosons. He added a fourth term to the non-linear sigma model$^{\cite{gursey1}}$ and proposed that the topological soliton solution of the meson field could be identified as a nucleon, and that the winding number is the baryon number. This pioneering work of Skyrme was largely ignored until two decades later when Pak and Tze$^{\cite{pak}}$, and Gipson and Tze$^{\cite{gipson}}$ re-expressed Skyrme's ideas in the modern context of QCD and studied their implications for weak interactions.  In an attempt to verify one of the Skyrme's conjectures that the winding numbers are the baryon numbers, Balachandran and his collaborators$^{\cite{bala}}$ considered a system of interacting quarks and solitons. Using some results of Goldstone and Wilczek$^{\cite{gold}}$, they showed that, in the presence of the classical soliton field, the Dirac sea of quarks carries a baryonic current which is same as the topological current. The possible connections of QCD to effective meson theories are due to 't Hooft$^{\cite{thooft}}$ and Witten$^{\cite{witten}}$. They showed that in the limit of large number of colors $N_c$, QCD can be reduced to a theory of weakly interacting mesons and glue balls. Using $1/N_c$ as an expansion parameter, Witten argued that to leading order, baryons emerge as solitons in the weakly coupled phase of mesons if some topological properties and stability conditions are assumed. Later Witten$^{\cite{witten2}}$ showed that it is necessary to introduce a Wess-Zumino term in the effective chiral theory in order to predict some observed physical processes allowed by the non-abelian anomaly in QCD. He also demonstrated that the gauged W-Z action can lead to the identification of the topological current with the baryonic current. To verify another Skyrme conjecture about topological solitons being fermions, he rotated a skyrmion adiabatically through $2\pi$-radians and found that the skyrmion had picked up a phase of $N_c\pi$. Hence a skyrmion is a fermion if $N_c$ is odd.

Afterwards, Adkins, Nappi and Witten$^{\cite{adkins}}$ systematically studied some static properties of nucleons in an explicit model in which the Skyrme Lagrangian is invariant under $SU(2)\times SU(2)$. These results agreed with experiment to 30 \%. Since then, there has been a tremendous interest in the Skyrme model. A good number of modifications to the model have been proposed, such as including the effect of the pion mass$^{\cite{adkins2}}$, adding higher order terms$^{\cite{adkins3}}$, extending the symmetry to $SU(3)\times SU(3)$$^{\cite{pras}}$, including vector mesons$^{\cite{adkins4}}$, hidden symmetry approach$^{\cite{bando}}$, bound state approach to strangeness$^{\cite{callan}}$, and non linear supersymmetric models and relation to skyrmions$^{\cite{marchildon} \cite{bergshoeff}}$. A recent review of supersymmetric chiral models was given by  Perelomov$^{\cite{perelomov}}$. 

Although the Skyrme model provides a reasonably good description of baryons, there are many problems which remain to be solved. In the Skyrme model, if the masses of the low lying baryons $N$ and $\Delta$ are taken as  input parameters, the predicted value of the pion decay constant $F_\pi$ is found to be too high; or if the experimental value of $F_\pi$ is taken as an input, the predicted masses of the nucleons turn out to be too high. Because the value $F_\pi$ is too large, many other predictions which depend on $F_\pi$ disagree with experiment. The model also predicts a tower of baryon states with equal spin and isospin ($s=i=5/2, 7/2, \ldots$) which do not exist in nature. They are considered artifacts of the model. 

Quark model, on the other hand, is regarded as an effective theory when the gluon degrees of freedom are integrated out. If both the Skyrme model and the quark model are effective models of QCD, they should have common symmetry of hadrons, such as chiral symmetry, The hadronic approximate $SU(6)$ symmetry$^{\cite{85}}$ and $SU(6/21)$ symmetry$^{\cite{88}}$ have been established for the quark model but ways to restore these symmetries for the Skyrme model have yet to be found. 

In this paper we address some of these problems, give partial solutions to some of the difficulties by improving the Skyrmion model for hadrons and suggest a strategy toward building a full fledged quantitatively viable skyrmionic description of hadronic physics.

\section{Effective Hadronic Supersymmetry}
Supersymmetry is a symmetry between fermions of half odd integer spin and bosons of integer spin. In supergravity, for example, we have an action that is invariant under operations transforming fields of spin $0$, $\frac{1}{2}$, $1$, $\frac{3}{2}$, and $2$ among themselves$^{\cite{freed}}$. Here we recall that low lying hadrons have the same range of spin values with $s=\frac{1}{2}$ and $\frac{3}{2}$ baryons interlaced with $s=0, 1$ and $2$ mesons. Groups that transform $s=0$ and $\frac{1}{2}$ mesons into each other and simultaneously mix $s=\frac{1}{2}$ and $\frac{3}{2}$ baryons have been proposed long ago$^{\cite{gr}}$. This kind of symmetry had to be broken symmetry since no degeneracy is observed between particles of different spin. However, a cursory glance at the Chew-Frautschi plot of hadronic trajectories$^{\cite{hendry}}$ reveals the following features:

\begin{itemize}

\item (a) All trajectories with $j$ (hadronic spin) versus $m^2$ ($m=$ hadronic mass) are approximately linear.

\item (b) Leading mesonic trajectories associated with lowest spin $0, 1$ and $2$ are parallel among themselves.

\item (c) Leading baryonic trajectories associated with $s=\frac{1}{2}$, $\frac{3}{2}$ are also parallel among themselves.

\item (d) Mesonic and baryonic trajectories are approximately parallel to each other with a universal slope of $\alpha \approx 0.9$ (GeV)$^{-2}$. 

\item (e) The separation between mesonic trajectories is nearly the same as the one between baryonic trajectories. 

\end{itemize}

Now, properties (b) and (c) suggest the existence of a phenomenological symmetry between mesons of different spin, which also operates on baryons with different spin.  For hadrons composed of light quarks $u$, $d$ and $s$, this symmetry is expressed by the group $SU(6)\times O(3)$ where $O(3)$ describes the rotational excitations on the leading trajectories, and the spin-flavor group $SU(6)$ classifies the lowest elements of the trajectories into particle multiplets. The property (d) on the other hand, tells us that there is a new kind of symmetry (supersymmetry) between the bosonic mesons and fermionic baryons. The universal Regge slope for hadrons is a supersymmetric observable.

The meaning of property (e) is that, the physical mechanism that breaks the $SU(6)$ symmetry must also be responsible for breaking of its supersymmetric extension. Finally, from property (a) we infer that the potential binding the quarks is approximately linear and we have to apply relativistic quantum mechanics appropriate to light quarks while the Schroedinger nonrelativistic theory is sufficient for the description of quarkonium systems for heavy quarks.

According to the quark model of Gell-Mann and Zweig, mesons and baryons are described respectively by bound ($\bar{q} q$) and $(qqq)$ systems. Any symmetry between mesons and baryons must correspond at the quark level to a supersymmetry between $\bar{q}$ (antiquarks) and bound $(qq)$ states (diquarks). Now $\bar{q}$, with $s=\frac{1}{2}$ and unitary spin associated with the triplet ${\bf (3)}$ representation of the flavor $SU(3)$ belongs to the ${\bf (6)}$ representation of $SU(6)$. The low lying baryons are in its ${\bf (56)}$ representation. Since ${\bf (56)}$ is contained in ${\bf 6 \times 21=56 + 70}$, the diquark with $s=0$ or $1$ must be in the symmetric ${\bf (21)}$ representation of $SU(6)$. Hence the hadronic supersymmetry we are seeking must transform the ${\bf (\bar{6})}$ and ${\bf (21)}$ $SU(6)$ multiplets, both color antitriplets into each other and therefore must be $27$ dimensional with $6$ fermionic and $21$ bosonic states. This supergroup is $SU(6/21)$. It was first introduced by Miyazawa$^{\cite{miyazawa}}$ as a generalization of the hadronic $SU(6)$ symmetry, following earlier attempts by Hwa and Nuyts$^{\cite{hwa}}$. 

The $(\bar{q})-(qq)$ symmetry that also implies $(q)-(\bar{q} \bar{q})$ symmetry will transform the meson ($q\bar{q})$ in general not only to baryons $(qqq)$ and antibaryons $(\bar{q} \bar{q} \bar{q})$ but also to exotic mesons $D-\bar{D}$ ($D=qq$, and $\bar{D}=\bar{q} \bar{q}$) that belong to the $SU(6)$ representations ${\bf 1, 35}$ and ${\bf 405}$. The ${\bf 1}$ and $({\bf 35})$ are $0^{+}$ and $1^{+}$ mesons while the (${\bf 405}$) also includes mesons with spin $2^{+}$ and isospin $2$. All the low lying hadrons will now be in the adjoint representation of $SU(6/21)$ with both spin and isospin taking values $0, \frac{1}{2}, 1, \frac{3}{2}$ and $2$.

The next introduction of supersymmetry into physics was within the context of dual resonance models that evolved later into string models. These theories due to Ramond$^{\cite{ramond}}$ and Neveu and Schwarz$^{\cite{neveu}}$ lead naturally to linear baryon and meson trajectories that are parallel. The string models were not local and they were not relativistic in four dimensions. 

Examples of renormalizable relativistic local quantum field theories involving $s=0, \frac{1}{2}$ and $1$ fields with interaction were first constructed by Wess and Zumino$^{\cite{wess}}$ following initial attempts by other authors$^{\cite{akulov}}$. Wess and Zumino based their work on the super-Poincar\'e algebra that is a supersymmetric generalization of the infinitesimal Poincar\'e group and sits in the superconformal algebra. The final step of incorporating $s=2$ and 
$s=\frac{3}{2}$ fields in a local relativistic field theory was taken by the discoverers of supergravity$^{\cite{freed}}$ and extended supergravity who were able to supersymmetrize Einstein's general relativity and Kaluza-Klein theories. All these local supersymmetric quantum field theories have fascinating convergence and symmetry properties but they are far from describing properties of hadrons or even quarks and leptons. If these fundamental fields have any physical reality at all, they may be associated with preons (or haplons) that would be hypothetical constituents of quarks, leptons and fundamental gauge bosons. 

Returning to the more concrete world of hadrons, we may try to see if the phenomenological approximate group $SU(6)$ and its supersymmetric extension can be justified within the standard theory of colored quarks interacting through gluons that are associated with the color gauge group $SU(3)^c$. The justification of $SU(6)$ and the derivation of its breaking was given by Georgi, Glashow and de Rujula$^{\cite{georgi}}$. 

On the other hand, a phenomenological supersymmetry in nuclear physics between odd and even nuclei was discovered and formulated by means of supergroups$^{\cite{baha}}$ $U(m/n)$ by Balantekin, Bars and Iachello$^{\cite{baha2}}$. It was natural to see if such groups could also describe hadronic supersymmetry. In the first part of this paper we propose to extend the approach of Georgi et al. to see how far it can provide a basis for the existence of an approximate hadronic supersymmetry. It turns out that most of the ingredients for this approach are already in the literature. 

The key concepts are:

\begin{itemize}

\item (a) {\bf The approximate validity of the string theory} as an approximation to QCD. This was shown by the lattice gauge theory as a strong coupling approximation to QCD, following the pioneering work of Wilson$^{\cite{wilson}}$ and also by the elongated bag model of Johnson and Thorn$^{\cite{johnson}}$ following the bag model$^{\cite{chodos}}$ approximation to QCD.

\item (b) {\bf The emergence of the diquark structure} in QCD through the string approximation. This was done by Eguchi$^{\cite{eguchi}}$ and also by Johnson and Thorn$^{\cite{johnson}}$.

\item (c) The vector (spin dependent) nature of the Coulomb part together with the scalar (spin independent) nature of the confining part of the $q-\bar{q}$ potential. These properties were collectively worked out by many authors using both perturbation theory and lattice gauge theory methods$^{\cite{otto}}$.

\item (d) The necessity for introducing exotic $D-\bar{D}$ mesons in QCD through the bag approximation$^{\cite{jaffe}}$, the string picture$^{\cite{harari}, {\cite{rosner}}}$ or the confining potential model$^{\cite{mulders}}$.

\item (e) Deviations from linearity of Regge trajectories within the context of QCD$^{\cite{hendry}}$.
\end{itemize}

Once all the parts of this jigsaw puzzle is put together, a rather simple picture emerges. The diquark behaves very much like the antiquark in the strong coupling regime because QCD forces are flavor independent and the confining part of QCD potential is spin independent. This immediately leads to a $\bar{q}-(qq)$ effective supersymmetry at large separation. At short distances, spin is approximately conserved because of asymptotic freedom but spin independence gets broken through one-gluon exchange in the perturbation theory regime. The diquark structure also disappears at short distance, leading to the breaking of both $SU(6)$ and $SU(6/21)$ at low energies. There is another difficulty associated with a supersymmetric extension of $SU(6)$ noted by Salam and Sthrathdee$^{\cite{salam}}$. Because of the anticommutativity of the Grassmann numbers and also of supercharges, antisymmetrical representations of $SU(6)$ like ${\bf (15)}$ for diquarks and ${\bf (20)}$ for baryons will occur in hadronic supermultiplets. In the case of colored quarks, however, it is possible to introduce non associative Grassmann numbers $u_1, u_2$ and $u_3$ constructed out of octonion units$^{\cite{gunaydin}}$. They transform like a triplet under the $SU(3)$ subgroup of the automorphism group $G_2$ of octonions. Then the octonionic quarks 
$\mbox{\boldmath$u$} \cdot \mbox{\boldmath$q$}_\alpha$ and $\mbox{\boldmath$u$} \cdot \mbox{\boldmath$q$}_\beta$ will commute unlike $q^i_\alpha$ and $q^j_\beta$ that are anticommutative ($\alpha$ and $\beta$ are combined spin-flavor indices). This procedure which is the basis of color algebra not only converts antisymmetrical spin-flavor group representations into symmetrical ones, but suppresses the color sextet ${\bf (6)}$ representations for diquarks that would otherwise be allowed under the rule for the preservation of the Pauli principle for the colored quark states. With the introduction of octonionic quarks, diquarks, mesons and baryons can all be viewed as elements of an octonionic superalgebra$^{\cite{88}}$.

The simplest supersymmetric Hamiltonian is obtained starting from semirelativistic dynamical models of quarks and diquarks already used by Lichtenberg et al.,$^{\cite{lichtenberg}}$ for an approximate calculation of baryonic masses. In the following subsections we will first give a derivation of hadronic mass formulae using semi-relativistic and relativistic formulation. Later, in section 2 using the skyrmion model we will show that we obtain the same mass formulae.

\section*{The Diquark and the Semirelativistic Hamiltonians for ($qq$) and $q-(qq)$ Systems}. 

In QCD, both $q-\bar{q}$ and $q-q$ forces are attractive, the color factor of the latter being half of that of the former. Thus, the formation of diquarks inside a baryon is a definite possibility. Various strong coupling approximations to QCD, like lattice gauge theory$^{\cite{wilson}}$ $^{\cite{otto}}$, 't Hooft's $\frac{1}{N_c}$ approximation (when $N_c$, the number of colors is very large), or the elongated bag model$^{\cite{johnson}}$ all give a linear potential between widely separated quarks and an effective string that approximates the gluon flux tube. In such a theory, Eguchi$^{\cite{eguchi}}$ has shown that it is energetically favorable for three quarks in a baryon to form a linear structure with a quark in the middle and two at the ends or, for high rotational excitation, a bilocal linear structure with one quark at one end and a diquark at the other end. This was reconfirmed by Johnson and Thorn independently in the bag model when the bag is deformed and elongated in a rotationally excited baryon$^{\cite{johnson}}$. Thus if we move along a leading baryon Regge trajectory to a region of high $j$, we are likely to find a baryon as a bilocal object consisting of a quark and a diquark instead of a trilocal object that represents a ground state baryon more accurately. On the other hand a meson is a bilocal object consisting of a quark and an antiquark interacting via a linear potential at large separation. Consider then a bilocal object with constituents with respective masses $m_1$, $m_2$, spins ${\bf s}_1$, ${\bf s}_2$ and color representations ${\bf 3}$ and ${\bf \bar{3}}$. One of the constituents can be a quark $q$ or the anti-diquark ${\bar{D}}$ (both color triplets) while the other can be an antiquark $\bar{q}$ or a diquark $D$ which are both color antitriplets. The QCD force between the two constituents will be flavor independent and it will consist of two parts: a Coulomb like part $V_c$ which transforms like the time component of a $4$-vector (due to the exchange of single $s=1$ gluons at short separations) and a confining part $V_s$ which is largely a relativistic scalar, is spin independent and is due to the exchange of a great many gluons that form a flux tube or an elongated bag at large separation of the constituents.

Let $\mbox{\boldmath$p$}_1 =\mbox{\boldmath$p$}$ and $\mbox{\boldmath$p$}_2 = -\mbox{\boldmath$p$}$ be the center of mass momenta of the two constituents. The quantity which is canonically conjugate to the relative coordinate 
$ \mbox{\boldmath$r$} ={\mbox{\boldmath$r$}}_1-{\mbox{\boldmath$r$}}_2 $ is 

\begin{equation}
-i~\mbox{\boldmath$\nabla$} =-i~\frac{\partial}{\partial \mbox{\boldmath$r$}}=\frac{1}{2}~(\mbox{\boldmath$p$}_1-
\mbox{\boldmath$p$}_2)=  \mbox{\boldmath$p$}   
\end{equation}
Ignoring the center of mass motion, following Lichtenberg et al.$^{\cite{88}}$, we can write a semi-relativistic wave equation for the wave function $\psi_{12}(\mbox{\boldmath$r$})$ of the bilocal object with energy eigenvalues $\Omega_{12}$, namely

\begin{equation}
(\Omega_{12}-V_c)\psi_{12}=\{ [(m_1+\frac{1}{2}~V_s)^2 -\nabla^2]^{\frac{1}{2}} + [(m_2 +\frac{1}{2}~V_s)^2 -\nabla^2)]^{\frac{1}{2}} \} \psi_{12} \label{eq:bir}
\end{equation}

The scalar and vector potentials are given by

\begin{equation}
V_s =br,~~~~~~V_c= -\frac{4}{3}~\frac{\alpha_s}{r} + \kappa_{12}~\frac{\mbox{\boldmath$s$}_1 \cdot
\mbox{\boldmath$s$}_2}{m_1 m_2 }
\end{equation}
where $\frac{4}{3}$ is the color factor, $\alpha_s$ is the strong coupling constant at the energy $\Omega_{12}$, and the spin dependent part of the vector potential is the hyperfine structure correction due to gluon exchange with $\kappa_{12}=|\psi_{12}(0)|^2$. We see that at large $r$, neglecting the mass difference $(m_2-m_1)$, we find the same equation for both the $(q-\bar{q})$ and the $q-D$ system, except for the presence of the hyperfine term that breaks the symmetry between $\bar{q}$ and $D$. To this approximation, we can transform the second constituent $\bar{q}$ into $D$ and vice-versa without changing the energy eigenvalue $\Omega$. This means that the system admits the approximate $SU(6/21)$ supersymmetry transformation

\begin{equation}
\delta \bar{q}_\alpha^i = \bar{b}_{\alpha\beta\gamma} ~(D^i)^{\beta\gamma} \nonumber
\end{equation}
   
\begin{equation}
\delta(D^i)^{\beta\gamma} = b^{\alpha\beta\gamma}~\bar{q}_\alpha^i   \label{eq:iki}
\end{equation}
in addition to the $SU(6)$ transformation
   
\begin{equation}
\delta \bar{q}_\alpha^i = m_\alpha^\beta~\bar{q}_\beta^i  \nonumber
\end{equation}
   
\begin{equation}
\delta(D^i)^{\beta\gamma}= n_{\rho \sigma}^{\beta\gamma}~(D^i)^{\rho\sigma}   \label{eq:uc}
\end{equation}

The breaking of both $SU(6)$ and $SU(6/21)$ is due to the hyperfine term while the supersymmetry is further broken by the quark-diquark mass difference $m_1-m_2$.

We could also have brought into play the wave functions $\psi_D$ and $\psi_{\bar{q}}$ of the diquark and antidiquark at point $\mbox{\boldmath$r$}_2$ in the field of the quark at point $\mbox{\boldmath$r$}_2$. The masses $m_1$ and $m_2$ must then be replaced by the reduced masses

\begin{equation}
\mu_D= \frac{m_q m_D}{m_q + m_D},~~~~~~\mu_{\bar{q}}=\frac{m_q m_{\bar{q}}}{m_q+m_{\bar{q}}} =\frac{1}{2}~m_q
\end{equation}

In this case, the wave function belongs to the fundamental $27$ dimensional representation of $SU(6/21)$ and the hamiltonian commutes with the supersymmetry transformation Eq.(\ref{eq:iki}) of the wave function except for the difference in the reduced masses $\mu_D$ and $\mu_q$ and the hyperfine structure term in the Hamiltonian.

The breaking of supersymmetry results in the mass difference

\begin{equation}
\Delta m= m_2-m_1 + \kappa~\frac{\mbox{\boldmath$s$}_1 \cdot \mbox{\boldmath$s$}_2}{m_1 m_2}
\end{equation}
at high energies both for baryons and mesons. At low energies, the baryon mass becomes a trilocal object with three quarks and the mass splitting is given by

\begin{equation}
\Delta m_{123} = \frac{1}{2}~ \kappa~ (\frac{\mbox{\boldmath$s$}_1 \cdot \mbox{\boldmath$s$}_2}{m_1 m_2}+
\frac{\mbox{\boldmath$s$}_2 \cdot \mbox{\boldmath$s$}_3}{m_2 m_3} +
\frac{\mbox{\boldmath$s$}_3 \cdot \mbox{\boldmath$s$}_1}{m_3 m_1})
\end{equation}
where $m_1$, $m_2$ and $m_3$ are the masses of the three different quark constituents. 

Going back to the first formulation which is more symmetrical, we can write the Hamiltonian associated with Eq.(\ref{eq:bir}) in a $SU(6/21)$ covariant form: we have four cases for $\psi_{12}$, namely $\psi_{q \bar{q}}$, $\psi_{D\bar{D}}$, $\psi_{qD}$ and $\psi_{\bar{D} \bar{q}}$ that represents mesons, exotic mesons, baryons and antibaryons with respective dimensions $6\times 6$, $21\times 21$, $6\times 21$, and $21\times 6$ that all fit into the following $27\times 27$ adjoint representation of $SU(6/21)$:

\begin{equation}
\psi(r) = \left(
\begin{array}{cc}
\psi_{q\bar{q}}(r) & \psi_{qD} (r) \\
\psi_{D \bar{q}}(r) & \psi_{\bar{D} D} 
\end{array}  \right) 
\end{equation}

The wave equation for the hadronic wavefunction $\psi$ can now be written as
   
\begin{equation}
i~\frac{\partial \psi}{\partial t} = H \psi = [K,\psi]+\mbox{\boldmath$S$} \psi \mbox{\boldmath$S$}  \label{eq:sekiz}
\end{equation}
where $K$ is the diagonal matrix
   
\begin{equation}
K= \left(
\begin{array}{cc}
K_q~I^{(6)} & 0 \\
0 & K_D~I^{(21)}
\end{array} \right),~~~~~~I^{(n)}= n\times n ~~~{\rm unit matrix}
\end{equation}
with
   
\begin{equation}
K_q= -\frac{2}{3}~\frac{\alpha_s}{r} + [(m_q +\frac{1}{2}~V_s)^2 -\nabla^2]^{\frac{1}{2}}  \nonumber
\end{equation}
   
\begin{equation}
K_D= -\frac{2}{3}~\frac{\alpha_s}{r} + [(m_q +\frac{1}{2}~V_s)^2 -\nabla^2]^{\frac{1}{2}}
\end{equation}
and

\begin{equation}
\mbox{\boldmath$S$} = \kappa^{\frac{1}{2}} ~\left(
\begin{array}{cc}
m_q^{-1} {\mbox{\boldmath$S$}}_q~I^{(6)} & 0 \\
0 & m_D^{-1} {\mbox{\boldmath$S$}}_D~I^{(21)}
\end{array} \right)   
\end{equation}
 
The second, spin dependent term on the right hand side of Eq.(\ref{eq:sekiz}) is a symmetry breaking term for both $SU(6)$ and $SU(6/21)$. The first term is $SU(6)$ symmetrical. It also preserves the supersymmetry in the limit $m_q=m_D$. Hence if the quark-diquark mass difference and the spin dependent terms are neglected, Eq.(\ref{eq:sekiz}) is invariant under the $SU(6/21)$ infinitesimal transformation

\begin{equation}
\delta \psi= [N, \psi]  \label{eq:onbir}
\end{equation}   
where

\begin{equation}
N= \left(
\begin{array}{cc}
M & b \\
\bar{b} & N
\end{array} \right)
\end{equation}   
where $M$ and $N$ have elements $m_\alpha^\beta$ and $n_{\rho \sigma}^{\beta \gamma}$ respectively in Eq.(\ref{eq:uc}) while the rectangular matrices $b$ and $\bar{b}$ have elements $b^{\alpha\beta\gamma}$ and ${\bar{b}}_{\alpha\beta\gamma}$ occurring in Eq.(\ref{eq:iki}). In that limit, the Hamiltonian $H$ commutes with the generators of the transformation Eq.(\ref{eq:onbir}).

\section*{The Hadronic Regge Trajectories}

Consider the Hamiltonian of Eq.(\ref{eq:bir}). We can write

\begin{equation}
-\nabla^2=\mbox{\boldmath$p$}^2 = p_r^2 + \frac{\ell (\ell+1)}{r^2}
\end{equation}
where $\ell$ is associated with the orbital excitation of the system. For high rotational excitations, the expectation value of $r$ is large, corresponding to a stretched string. The angular momentum $\ell$ is also large. The value of the centrifugal energy which is proportional to $\frac{\ell(\ell+1)}{r^2}$ has a similarly large value. Since $V_s$ that is proportional to $r$ will also have a high absolute value, the constituent masses become negligible in the high relativistic limit. On the other hand, the radial excitation term $p_r^2$ can be neglected on the leading trajectory associated with the lowest radial energy.

The ground state energy eigenvalue $E$ of the Hamiltonian                              
can be estimated by using the Heisenberg uncertainty principle. This leads                              
to the replacement of $r$ by $\Delta r$ and $p_{r}$ by                              
\begin{equation}                              
\Delta p_{r} = \frac{1}{2} (\Delta r)^{-1}, ~~~~~~(h = 1).                              
\end{equation}                              
                              
Then $E$ as a function of $\Delta r$ is minimized for the value of                              
$r_{0}$ of $\Delta r$. The $r_{0}$ corresponds to the Bohr radius for                              
the bound state. The confining energy associated with this Bohr radius                               
is obtained from the linear confining potential $S(r) = br$, so that the effective masses of the constituents become                              
\begin{equation}                              
M_{1} = m_{1} + \frac{1}{2} S_{0}, ~~~~                              
M_{2} = m_{2} + \frac{1}{2} S_{0}, ~~~~(S_{0} = b r_{0})
\end{equation}                                                            
For a meson $m_{1}$ and $m_{2}$ are the current quark masses while                              
$M_{1}$ and $M_{2}$ can be interpreted as the constituent quark                                
masses. Note that even in the case of vanishing quark masses associated                              
with perfect chiral symmetry, confinement results in non zero                              
constituent masses that spontaneously break the $SU(2) \times SU(2)$                               
symmetry of the $u$, $d$ quarks.                              
                              
Let us illustrate this method on the simplified spin free Hamiltonian                              
involving only the scalar potential. In the center of mass system,                                
$\mbox{\boldmath $p$}^{(1)}+\mbox{\boldmath $p$}^{(2)}=0$, or                                
$\mbox{\boldmath $p$}^{(1)}=-\mbox{\boldmath $p$}^{(2)}=\mbox{\boldmath $p$}$.                                
The semi-relativistic hamiltonian of the system is then given by                                
\begin{equation}                                
E_{12}~~ \Phi=\sum_{i=1}^{2}\sqrt{(m_{i}+\frac{1}{2}br)^{2}+                              
\mbox{\boldmath $p$}^{2}}~~\Phi.                                
\end{equation}                                
Taking $m_{1}=m_{2}=m$ for the quark-antiquark system, we have                                
\begin{equation}                                
E_{12}~~\Phi= 2\sqrt{(m+\frac{1}{2}br)^{2}+p^{2}_{r}+                              
\frac{\ell(\ell+1)}{r^{2}}} ~~\Phi,           \label{eq:e}                                
\end{equation}                                
where we have written the momentum part in spherical coordinates.                              
                              
Putting                              
\begin{equation}                              
b=\mu^{2},~~~~~\rho = \mu ~~r,                              
\end{equation}                              
for the $q-\bar{q}$ system we find $E_{12}$ by minimizing the function                              
\begin{equation}                                
E_{q \bar{q}}~~= 2\sqrt{(m+\frac{1}{2} \mu \rho)^{2}+                              
\frac{\mu^{2}}{\rho^{2}} (\ell+\frac{1}{2})^{2}} .                                
\end{equation}                                
                                
For $u$ and $d$ quarks, $m$ is small and can be neglected so that                                
\begin{equation}                              
E^{2} = \mu^{2} [\rho^{2} + \rho^{-2} (2 \ell + 1)^{2}]                              
\end{equation}                              
which has a minimum for                              
\begin{equation}                              
\rho^{2} = \rho_{0}^{2} = 2 \ell + 1,                              
\end{equation}                              
giving                              
\begin{equation}                              
E_{min}^{2} = E^{2} (\rho_{0}) = 4 \mu^{2} (\ell+\frac{1}{2}).                       \label{eq:min}       
\end{equation}                              
Thus, we obtain a linear Regge trajectory with                              
\begin{equation}                              
\alpha^{'} = \frac{1}{4} \mu^{-2} = \frac{b}{4}.                       \label{eq:mina}       
\end{equation}                              
Also $ \mbox{\boldmath $J$}= \mbox{\boldmath $\ell$} +\mbox{\boldmath $S$},$ where $\mbox{\boldmath $S$}$ arises from the quark spins. Experimentally                              
\begin{equation}                              
\alpha^{'} = 0.88 (GeV)^{-2}                              
\end{equation}                              
for mesons giving the value $0.54$ GeV for $\mu$. A more accurate calculation (see [4]) gives
\begin{equation}                              
\alpha^{'} = (2 \pi \mu^{2})^{-1}, ~~~~~ \mu \sim 0.43 GeV.   \label{eq:alp}                              
\end{equation}      
                              
The constituent quark mass can be defined in two ways                              
\begin{equation}                              
M_{c}(\ell) = \frac{1}{2} E_{min} = \mu \sqrt{\ell + \frac{1}{2}} ,                              
\label{eq:mc}                              
\end{equation}                              
or                              
\begin{equation}                              
m_{c}^{'}(\ell)=S_{0} = \frac{1}{2} \mu \rho_{0} =                               
\frac{\mu}{\sqrt{2}} \sqrt{\ell + \frac{1}{2}}.                    \label{eq:mca}          
\end{equation}                              
The first definition gives for $\ell = 0$, 
\begin{equation}
M_{c} = 0.31 GeV ~~~for~~~\mu= 0.43    \label{eq:mcb}                              
\end{equation}                              
in the case of $u$ and $d$ quarks.                              
                              
When the Coulomb like terms are introduced in the simplified Hamiltonian                              
(\ref{eq:e}) with negligible quark masses one obtains                              
\begin{equation}                              
E=\frac{\mu}{\rho} [-\bar{\alpha}  + \sqrt{\rho^{4} +      
(2\ell + 1)^{2}}]                              
\end{equation}                              
with                              
\begin{equation}                              
\bar{\alpha} =\frac{4}{3} \alpha_{s} ~~~{\rm for}~~ (q\bar{q}),~~~                              
\bar{\alpha} =\frac{2}{3} \alpha_{s} ~~~{\rm for}~~ (q q).                              
\end{equation}                              
In the energy range around $1$ GeV, $\alpha_{s}$ is of order of unity.                              
Estimates range from $0.3$ to $3$. Minimization of $E$ gives                              
\begin{equation}                              
E_{0}=\mu u_{0}^{\frac{-1}{4}} (-\bar{\alpha} + \sqrt{u_{0} +(2\ell + 1)^{2}})                              
\end{equation}                              
where                              
\begin{equation}                              
u_{0}(\epsilon)=\rho_{0}^{4}= (2 \ell + 1)^{2} (1+\frac{1}{2} \beta^{2} +                              
\epsilon \sqrt{2} \beta                              
\sqrt{\ell+\frac{1}{8} \beta^{2}}),                              
\end{equation}                              
\begin{equation}                              
\epsilon = \pm{1},~~~~\beta=\frac{\bar{\alpha}}{(2 \ell + 1)}.                              
\end{equation}                              
                              
The minimum $E_{0}$ is obtained for $\epsilon = -1$, giving to second                              
order in $\beta$:                              
\begin{equation}                              
E_{0} = \mu \sqrt{2(2 \ell + 1 )} (1 - \frac{\beta}{\sqrt{2}} -                              
3 \frac{\beta^{2}}{8}).       \label{eq:ez}                              
\end{equation}                              
                              
Linear Regge trajectories are obtained if $\beta^{2}$ is negligible.                              
Then for mesons                              
\begin{equation}                              
E_{0}^{2} = 4 \mu^{2} \ell + 2 \mu^{2} (1-\sqrt{2} \bar{\alpha}).                              
\end{equation}                              
The $\beta^{2}$ is negligible for small $\ell$ only if we take the lowest                              
estimate for $\alpha_{s}$, giving $0.4$ for $\bar{\alpha}$ in the $q \bar{q}$                              
case. For mesons with $u$, $d$ constituents, incorporating their spins through                              
the Breit term we obtain approximately                              
\begin{equation}                              
m_{\rho} \simeq m_{\omega} = E_{0} + \frac{c}{4},~~                                                             
m_{\pi}= E_{0} - \frac{3c}{4}, ~~c = K~~ \frac{\Delta V}{M_{q}^{2}}
\end{equation}
where $M_{q}$ is the constituent quark mass. This gives                              
\begin{equation}                              
E_{0}= \frac{(3 m_{\rho} + m_{\pi})}{4} = 0.61 GeV.                               
\end{equation}                              
                              
The Regge slope being of the order of $1 GeV$ an average meson mass                              
of the same order is obtained from Eq.(\ref{eq:ez}) in the linear 
trajectory                               
approximation. To this approximation $\bar{\alpha}$ should be treated like                               
a parameter rather than be placed by its value derived from QCD under                               
varying assumptions. Using Eq.(\ref{eq:alp}) for $\mu$ one gets a better                               
fit to the meson masses by taking $\alpha_{s} \sim 0.2$.                              

Turning now to baryon masses, we must first estimate the diquark mass. We                              
have for the $qq$ system                              
\begin{equation}                              
M_{D}= \mu (\sqrt{2} - \frac{2}{3} \alpha_{s}),                \label{eq:dq}           
\end{equation}                              
that is slightly higher than the average meson mass                              
\begin{equation}                              
\tilde{m}= \mu (\sqrt{2} - \frac{4}{3} \alpha_{s}).                           \label{eq:con}      
\end{equation}                              
                              
Here we note that $E$ is not very sensitive to the precise value of the QCD running coupling constant in the $GeV$ range. Taking $\alpha_{s} \sim 0.3$ changes $E^{qq}$ from $0.55$ to $0.56 GeV$.      
      
Note that Eq.(\ref{eq:dq}) gives $m_{D} = 0.55 GeV$. For excited $q-\bar{q}$ and $q-D$ systems if the rotational excitation energy is large compared with $\mu$, then both the $m_{D}$ and the Coulomb term $- \frac{4}{3} \frac{\alpha_{s}}{r}$ (same for $q-D$ and $q-\bar{q}$ systems) can be neglected. Thus, for both ($q-D$) [excited baryon] and $q-\bar{q}$ [excited meson] systems we have Eq.(\ref{eq:min}), namely      
\begin{equation}      
(E^{q-D})^{2} \sim (E^{q-\bar{q}})^{2} \sim 4 \mu^{2} \ell + 2 \mu^{2}      
\end{equation}      
giving again Eq.(\ref{eq:mina}), i.e.      
\begin{equation}      
(\alpha^{'})_{q-D} = (\alpha^{'})_{q-\bar{q}} \cong \frac{1}{4 \mu^{2}}~~~~{\rm or} ~~(\frac{1}{2 \pi \mu^{2}})      
\end{equation}      
as an explanation of hadronic supersymmetry in the nucleon and meson Regge spectra. We also have, extrapolating to small $\ell$:      
\begin{equation}      
\Delta (M^{2})^{q-D} = \Delta (m^{2})^{q-\bar{q}} = 4 \mu^{2} \Delta \ell = \frac{1}{\alpha^{'}} \Delta \ell .      
\end{equation}        
For $\Delta \ell = 1$ we find      
\begin{equation}      
m_{\Delta}^{2} - m_{N}^{2} = m_{\rho}^{2} - m_{\pi}^{2} .      
\end{equation}      
      
This relationship is same as the one proved in our earlier paper$^{\cite{bidi}}$ through the assumption that $U(6/21)$ symmetry is broken by an operator that behaves like $s = 0$, $I = 0$ member of $35 \times 35$ representations of $SU(6)$, which is true to $5\%$. It corresponds to a confined quark approximation with $\alpha_{s} = 0$.      
      
The potential model gives a more accurate symmetry breaking ($\alpha_{s} \sim 0.2$):      
\begin{equation}      
\frac{9}{8} (m_{\rho}^{2} - m_{\pi}^{2}) = m_{\Delta}^{2} - m_{N}^{2}      
\end{equation}      
with an accuracy of $1\%$.       
      
This mass squared formula arises from the second order iteration of the $q-D$, $q-\bar{q}$ Dirac equation. The factor $\frac{9}{8}$ comes from      
\begin{equation}      
\frac{1}{2} (\frac{4}{3} \alpha_{s})^{2} = \frac{8}{9} \alpha_{s}^{2}       
\end{equation}

At this point it is more instructive to derive a first order mass formula. Since the constituent quark mass $M_{q}$ is given by Eq.(\ref{eq:mc})                              
$(\ell = 0)$, we have                              
\begin{equation}                              
M_{q} = \frac{\mu}{\sqrt{2}}                              
\end{equation}                              
so that                              
\begin{equation}                              
\bar{m} = 2 M_{q} ( 1 -\frac{\sqrt{2}}{3} \alpha_{s}) \simeq 1.9 M_{q}                               
\end{equation}                              
                              
When the baryon is regarded as a $q-D$ system, each constituent gains an                               
effective mass $\frac{1}{2} \mu \rho_{0}$ which was approximately the                              
effective mass of the quark in the meson. Hence, the effective masses of                              
$q$ and $D$ in the baryon are                               
\begin{equation}                              
m_{q}^{'} \simeq M_{q}, ~~~~~m_{D}^{'} = M_{D} + M_{q} \simeq 3 M_{q}                              
\end{equation}                              
                              
The spin splittings for the nucleon $N$ and the $\Delta$ are given by                              
the Breit term                              
\begin{equation}                              
\Delta M = K \Delta V \frac{\mbox{\boldmath $S$}_{q}                                
\mbox{\boldmath $\cdot S$}_{D}}{m_{q}^{'} m_{D}^{'}}                              
\end{equation}                              
                              
For the nucleon with spin $\frac{1}{2}$ the term ${\bf S}_{q} \cdot {\bf S}_{D}$                              
gives $-1$ while it has the value $\frac{1}{2}$ for $\Delta$ with spin                              
$\frac{3}{2}$. Using the same $K$ for mesons and baryons which are                              
both considered to be a bound state of a color triplet with a color                               
antitriplet we can relate the baryon splitting $\Delta M$ to the meson                              
splitting $\Delta m$ for which ${\bf S}_{q} \cdot {\bf S}_{\bar{q}}$ takes the                              
values $\frac{1}{4}$ and $\frac{-3}{4}$. Hence we find                              
\begin{equation}                              
\Delta M = M_{\Delta} - M_{N} = \frac{3}{2}\cdot \frac{K \Delta V}{m_{q}^{'} m_{D}^{'}} = 
\frac{1}{2} \cdot \frac{K \Delta V}{ M_{q}^{2}},~~~~{\rm and}~~~~~\Delta m = \frac{K \Delta V}{M_{q}^{2}}                              
\end{equation}                              
which leads to a linear mass formula                              
\begin{equation}                              
\Delta M = \frac{1}{2} \Delta m                              
\end{equation}                              
which is well satisfied, and has been verified before using the three                               
quark constituents for the baryon$^{\cite{bidi}}$.                              
                              
The formation of diquarks which behave like antiquarks as far as QCD is                               
concerned is crucial to hadronic supersymmetry and to quark dynamics for                               
excited hadrons. The splittings in the mass spectrum are well understood                              
on the basis of spin-dependent terms derived from QCD. This approach to                               
hadronic physics has led to many in depth investigations recently. For extensive references we refer to recent papers by Lichtenberg and collaborators$^{\cite{lichty}}$ and by Klempt$^{\cite{klempt}}$.                
To see the symmetry breaking effect, note that the mass of a hadron will take the approximate form
\begin{equation}
m_{12}=m_1+m_2+K\frac{{\bf S}_1\cdot{\bf S}_2}{m_1m_2}
\end{equation}
where $m_i$ and ${\bf S}_i$ ($i=1,2$) are respectively the constituent mass and the spin of a quark or a diquark. The spin-dependent Breit term will split the masses of hadrons of different spin values. If we assume $m_q=m_{\bar q}=m$, where $m$ is the constituent mass of $u$ or $d$ quarks, and denote the mass of a diquark as $m_D$, then this approximation gives
\begin{equation}
m_\pi= (m_{q\bar{q}})_{s=0}=2m-K\frac{3}{4m^2},~~~~m_\rho= (m_{q\bar{q}})_{s=1}=2m+K\frac{1}{4m^2}
\end{equation} 
\begin{equation}
m_\Delta= (m_{qD})_{s=3/2}=m+m_D+K\frac{1}{2mm_D}
\end{equation}
\begin{equation}
m_N= (m_{qD})_{s=1/2}=m+m_D-K\frac{1}{mm_D},
\end{equation}
Eliminating $m$, $m_D$ and $K$, we obtain a mass relation 
\begin{equation}
\frac{8}{3}\cdot \frac{2m_{\Delta}+m_N}{3m_{\rho}+m_{\pi}}=1+\frac{3}{2} \cdot \frac{m_\rho-m_\pi}{m_\Delta-m_N}
\end{equation} 
which agrees with experiment to 13\%.
\section*{Relativistic Case}
Let us consider a quark-antiquark system in the approximation that the                                
potential is only a scalar.                                
In the center of mass system,                                
$\mbox{\boldmath $p$}^{(1)}+\mbox{\boldmath $p$}^{(2)}=0$, or                                
$\mbox{\boldmath $p$}^{(1)}=-\mbox{\boldmath $p$}^{(2)}=\mbox{\boldmath $p$}$                                
The semi-relativistic hamiltonian of the system is then given by                                

\begin{equation}                                
H=\sum_{i=1}^{2}\sqrt{(m_{i}+\frac{1}{2}br)^{2}+\mbox{\boldmath $p$}^{2}}                                
\end{equation}                                
Taking $m_{1}=m_{2}=m$ for the quark-antiquark system, we have                                

\begin{equation}                                
H=2\sqrt{(m+\frac{1}{2}br)^{2}+p^{2}_{r}+\frac{\ell(\ell+1)}{r^{2}}}                                
\end{equation}                                
where we have written the momentum part in spherical coordinates.                                
For $u$ and $d$ quarks, $m$ is small and can be neglected so that                                

\begin{equation}                                
H^{2}=4[\frac{1}{4}b^{2}r^{2}+p^{2}_{r}+\frac{\ell(\ell+1)}{r^{2}}]                                
\end{equation}                                
Similar equations for a $q-\bar{q}$ meson system were already proposed                                 
and solved numerically$^{\cite{lichtenberg},\cite{martin}}$ or exactly$^{\cite{g3}}$ by several authors.                                
The eigenfunction $\Psi$ for $H^{2}$ has eigenvalue $E^{2}$ such that                                

\begin{equation}                                
4[\frac{1}{4}b^{2}r^{2}-\frac{1}{r}\frac{d^{2}}{dr^{2}}r                                
+\frac{\ell(\ell+1)}{r^{2}}]\Psi=E^{2}\Psi                                
\end{equation}                                
The differential equation can be solved exactly, and                                
and the normalized eigenfunction is found to be                                

\begin{eqnarray}                                
\Psi(r,\theta,\phi)&=&\left[\frac{2(\frac{b}{2})^{2\ell+3}                                
\Gamma(\ell+\frac{3}{2}+n_{r}-1)}                                
{(n_{r}-1)!\Gamma^{2}(\ell+\frac{3}{2})}\right]^{1/2}                                
r^{\ell}\exp[-\frac{b}{4}r^{2}]\times  \nonumber \\                                
&&F(-n_{r}+1,{\textstyle \ell+\frac{3}{2},\frac{b}{2}r^{2}})                                
Y_{\ell}^{m}(\theta,\phi)                                
\end{eqnarray}                                
where $F(-n_{r}+1,\ell+\frac{3}{2},\frac{b}{2}r^{2})$ is the confluent                                
hypergeometric function, and $n_{r}=1,2,3,\ldots$, is the radial quantum                                
number. The eigenvalue is given by                                

\begin{equation}                                
E^{2}=4b[2(n_{r}-1)+\ell+\frac{3}{2}]  \label{eq:regge}                                
\end{equation}                                
therefore we obtain linear Regge trajectories of slope $1/4b$ when we make                                
plots of $\ell$ versus $M^{2}$.  The case of                                
$n_{r}=1$ corresponds to the leading Regge trajectory, and cases of                                
$n_{r}=2,3,\ldots$, correspond to the parallel daughter trajectories.                              

Note that Iachello and his collaborators$^{\cite{i1}}$ have obtained a similar mass                                
formula based on algebraic methods. Starting from a spectrum generating                                
algebra $G$, we can write a chain of subalgebras                                

\begin{equation}                                
G \supset G' \supset G'' \supset \cdots .                                
\end{equation}                                
They proposed that the Hamiltonian can be expanded in terms of                                
invariants of the chain of subalgebras                                

\begin{equation}                                
H=\alpha C(G)+\beta C(G')+\gamma C(G'')+\cdots,                                
\end{equation}                                
where $C(G)$ denotes one of the invariants of $G$. In a two-body problem, the spectrum generating algebra is $U(4)$, and in the total symmetric representation, one of the two chains of subalgebras is $U(3)$ followed by $SO(3)$ with respective Casimir invariants $n(n+2)$ and $\ell(\ell+1)$, where $n$ corresponds to the vibrational mode.  In order to derive the relativistic mass formula like Eq.(\ref{eq:regge}), one has to write a formula for the square of the mass, rather                                
than the mass itself.  Furthermore, without violating the dynamic symmetry, the Hamiltonian can be written in terms of non-linear functions of the Casimir invariants. Therefore, we can have                                

\begin{eqnarray}                                
M^{2}&=&\alpha+\beta\sqrt{n(n+2)+1}+\gamma\sqrt{\ell(\ell+1)+1/4} \nonumber \\                                
&=&\alpha '+\beta n+\gamma \ell                         \label{eq:iachello}                                
\end{eqnarray}                                
Comparing this mass formula with Eq.(\ref{eq:regge}), we find that the relativistic                                
quark model suggests a relation of $\beta=2\gamma$ in Eq.(\ref{eq:iachello}).                                
However, the above algebraic method treats $\beta$ and $\gamma$ as two                                
independent parameters, and so the extra degree of freedom allows Eq.(\ref{eq:iachello}) a better fit to the experimental data. The fitted parameters for mesons are found to be: $\beta\approx 1.5$ (GeV)$^{2}$ and $\gamma\approx 1.1$ (GeV)$^{2}$.                                
                                
The inclusion of the Coulomb-like term causes deviations from the linear                                
trajectories at low energies and changes the relation between the parameters                                
$\beta$ and $\gamma$. It is important to carry out a calculation based on a                                
better approximation to see if the empirical values of these parameters are                                
compatible with our model.                                

\section*{Bilocal Approximation to Hadronic Structure and Inclusion of Color}

The supergroup $SU(6/21)$ acts on a                        
quark and antidiquark situated at the same point $r_{1}$.  At the                       
point $r_{2}$ we can consider                        
the action of this supergroup with the same parameters, or one with                       
different parameters.  In the first case we have a global symmetry.  In the                        
second case, if we only deal with bilocal fields the symmetry will be                        
represented by $SU(6/21)\times SU(6/21)$, doubling the supergroup.                        
On the other hand, if any number of points are considered, with                       
different parameters                        
attached to each point, we are led to introduce a local supersymmetry                       
$SU(6/21)$ to which we should add the local color group $SU(3)^{c}$.                        
Since it is not a                        
fundamental symmetry, we shall not deal with the local $SU(6/21)$ group here.                         
However, the doubling the supergroup is useful for bilocal fields since the                        
decomposition of the adjoint representation of the 728-dimensional $SU(6/21)$                        
group with respect to $SU(6)\times SU(21)$ gives                       

\begin{equation}                      
728= (35,1)+(1,440)+(6,21)+ (\bar{6}, \bar{21}) +(1,1)                      
\end{equation}                       
                       
A further decomposition of the double supergroup into its field with                        
respect to its center of mass coordinates, as will be seen below                       
leads to the decomposition of the 126-dimensional cosets $(6,21)$ and                       
$(21,6)$ into $56^{+} + 70^{-}$  of the diagonal $SU(6)$. We would have a much tighter and more elegant scheme if we could perform such a decomposition from the start and be able to identify $(1,21)$                       
part of the fundametal representation of $SU(6/21)$ with the $21$-dimensional                       
representation of the $SU(6)$ subgroup, which means going beyond the                       
$SU(6/21)$ supersymmetry to a smaller supergroup having $SU(6)$ as a subgroup. Full description of a bilocal treatment and a minimal scheme where the bilocal treatment gets carried over unchanged into it will be a subject of another publication.                        
                      
Setting up of an effective relativistic theory based on the Dirac equation for the quark and the Klein-Gordon equation for the antiquark, exhibiting invariance under relativistic supersymmetry is needed. This can be achieved by means of an effective Wess-Zumino type Lagrangian constructed out of a gluon field interacting with a vector superfield formed by a quark and an antidiquark. In what follows we will present some preliminary work in this direction.                     

Low-lying baryons occur in the symmetric $56$ representation$^{ \cite{gr}}$                       
of $SU(6)$, whereas the Pauli principle would have led to the antisymmetrical $20$                        
representation.  This was a crucial fact for the introduction of color degree                        
of freedom$^{\cite{gre}}$ based on $SU(3)^{c}$. Since the quark field                       
transforms like a color triplet and the diquark like a color                       
antitriplet under $SU(3)^{c}$, the color                        
degrees of freedom of the constituents must be included correctly in order                        
to obtain a correct representation of the q-D system.  Hadronic states must                        
be color singlets.  These are represented by bilocal operators                       
$O({\bf r}_{1},{\bf r}_{2})$ in the bilocal 
approximation$^{\cite{tak}}$                       
that gives $\bar{q}(1)q(2)$ for mesons and $D(1) q(2)$ for                        
baryons.  Here $ \bar{q}(1)$ represents the antiquark situated at ${\bf r}_{1}$, $q(2)$ the quark situated at ${\bf r}_{2}$, and $D(1) = q(1)q(1)$ the                       
diquark situated at ${\bf r}_{1}$.  If we denote                        
the c.m. and the relative coordinates of the consituents by ${\bf R}$ and                       
${\bf r}$, where ${\bf r} ={\bf r}_{2}-{\bf r}_{1}$ and                       

\begin{equation}                      
{\bf R} = \frac{(m_{1} {\bf r}_{1}+m_{2} {\bf r}_{2})}{(m_{1}+m_{2})}                      
\end{equation}                       
with $m_{1}$ and $m_{2}$ being their masses, we can                        
then write $O({\bf R},{\bf r})$ for the operator that creates hadrons                       
out of the vacuum.                         
The matrix element of this operator between the vacuum and the hadronic state                        
$h$ will be of the form                        

\begin{equation}                      
<h| O({\bf R},{\bf r}) |0>= \chi({\bf R}) \psi({\bf r})                      
\end{equation}                       
where $\chi ({\bf R})$ is the free wave function of the hadron as a                       
function of the c.m. coordinate and $\psi ({\bf r})$ is the bound-state                       
solution of the $U(6/21)$ invariant Hamiltonian describing the                       
$q-\bar{q}$ mesons, $q-D$ baryons, $\bar{q}-\bar{D}$ antibaryons and                           
$D-\bar{D}$ exotic mesons, given by                       

\begin{equation}                      
i \partial_{t} \psi_{\alpha \beta} =                       
[\sqrt{(m_{\alpha} + \frac{1}{2} V_{s})^{2} + {\bf p}^{2}} +                      
\sqrt{(m_{\beta} + \frac{1}{2} V_{s})^{2} + {\bf p}^{2}} -\frac{4}{3}                      
\frac{\alpha_{s}}{r} + k \frac{{\bf s}_{\alpha} \cdot{\bf s}_{\beta}}                      
{m_{\alpha} m_{\beta}}] \psi_{\alpha \beta}                      
\end{equation}

Here ${\bf p}= -i {\bf \nabla}$ in the c.m. system and $m$ and $s$ denote                       
the masses and spins of the constituents, $\alpha_{s}$ the                       
strong-coupling constant, $V_{s} = br$ is the scalar potential with $r$                       
being the distance between the constituents in the bilocal object, and                        
$k= |\psi (0)|^{2}$.                      
                       
 The operator product expansion$^{\cite{wilson2}}$ will give a singular part depending only                        
on ${\bf r}$ and proportional to the propagator of the field binding the two                        
constituents.  There will be a finite number of singular coefficients $c_{n}                      
({\bf r})$ depending on the dimensionality of the constituent fields.                        
For example, for a                        
meson, the singular term is proportional to the progagator of the gluon field                        
binding the two constituents.  Once we subtract the singular part, the                        
remaining part $\tilde{O}({\bf R}, {\bf r})$ is analytic in r and thus                       
we can write

\begin{equation}                      
\tilde{O}({\bf R}, {\bf r})= O_{0}({\bf R}) + {\bf r} \cdot{\bf O}_{1}({\bf R})                      
+O(r^{2}).                      
\end{equation}                       
                       
   Now $O_{0}({\bf R})$ creates a hadron at its c.m. point ${\bf R}$                       
equivalent to a $\ell=0$, s-state of                        
the two constituents.  For a baryon this is a state associated with $q$ and                       
$D \sim qq$ at the same point ${\bf R}$, hence it is essentially a 3-quark                       
state when the three quarks are at a common location. The                        
${\bf O}_{1}({\bf R})$ can create three $\ell=1$ states with                        
opposite parity to the state created by $O_{0}({\bf R})$.  Hence, if                       
we denote the nonsingular parts of $\bar{q}(1)q(2)$ and $D(1)q(2)$ by                       
$[\bar{q}(1)q(2)]$ and $[D(1)q(2)]$, respectively, we have                       

\begin{equation}                      
[\bar{q}(1)q(2)]|0> = |M({\bf R})> + {\bf r} \cdot |{\bf M}^{'}({\bf R})> + O(r^{2})                       
\end{equation}                       

\begin{equation}                      
[D(1)q(2)]|0> = |B({\bf R})> + {\bf r} \cdot |{\bf B}^{'}({\bf R})> + O(r^{2})                       
\end{equation}
and similarly for the exotic meson states $D(1) \bar{D}(2)$.                      
                       
   Here M belongs to the $(35+1)$-dimensional representation of $SU(6)$                        
corresponding to an $\ell=0$ bound state of the quark and the antiquark.                        
The $M^{'}({\bf R})$ is an orbital excitation $(\ell=1)$ of opposite                       
parity, which are in the $(35+1,3)$ representation of the group                       
$SU(6) \times O(3)$, $O(3)$ being associated with the relative                        
angular momentum of the constituents.  The $M^{'}$ states contain                       
mesons like $B$, ${\bf A}_{1}$, ${\bf A}_{2}$ and scalar particles.                        
On the whole, the $\ell=0$ and $\ell=1$ part $\bar{q}(1)q(2)$ contain                       
$4 \times(35+1) = 144$ meson states.                      
                       
   Switching to the baryon states, the requirement of antisymmetry in color                       
and symmetry in spin-flavor indices gives the $(56)^{+}$                       
representation for $B({\bf R})$. The $\ell=1$ multiplets have negative                       
parity and have mixed spin-flavor symmetry.                       
They belong to the representation $(70^{-},3)$ of $SU(6) \times O(3)$ and                       
are represented by the states $|{\bf B}^{'}({\bf R})>$ which are $210$ in                       
number.  On the whole, these $266$ states                       
account for all the observed low-lying baryon states obtained from                        
$56 + 3 \times 70 = 266$.  A similar analysis can be carried out for the                       
exotic meson states $D(1) \bar{D}(2)$, where the diquark and the                       
antidiquark can be bound in an $\ell=0$ or $\ell=1$ state with                       
opposite parities.

\section*{Color Algebra and Octonions}

The exact, unbroken color group
$SU(3)^c$ is the
backbone of the strong interaction. It is worthwhile to understand its role in the
diquark picture more clearly.

Two of the colored quarks in the baryon combine into an anti-triplet $ {\bf 3\times 3
=\bar{3}+(6)} $, $ {\bf 3\times \bar{3} = 1+(8)} $. The ${\bf (6)} $ partner of the diquark and the $ {\bf (8)} $ partner 
of the nucleon do not
exist. In hadron dynamics the only color combinations to consider
are $ {\bf 3\times 3 \rightarrow \bar{3}} $ and $ {\bf \bar{3}\times 3 \rightarrow 1} $. These relations imply 
the existence of split octonion units through
a representation of the Grassmann algebra $\{u_i,u_j\}=0$, $i= 1,2,3$. What is a bit strange is that 
operators $u_i$, unlike ordinary fermionic operators, are not associative. We also have 
$\frac{1}{2}[u_i,u_j]=\epsilon_{ijk}~u_k^{*}$. The Jacobi identity
does not hold since $[u_i,[u_j,u_k]]=- i e_7 \neq 0$, where $e_7$, anticommute with 
$u_i$ and $u_i^{*}$. 

~~~~  The behavior of various states under the color group are best                       
seen if we use split octonion units defined by$^{\cite{gunaydin}}$                       

\begin{equation}                      
u_{0} = \frac{1}{2} (1 +i e_{7}) ,                      
~~~~~~~u_{0}^{*} = \frac{1}{2} (1 -i e_{7})                       
\end{equation}                        

\begin{equation}                      
u_{j} = \frac{1}{2} (e_{j} +i e_{j+3}) ,                      
~~~~~u_{j}^{*} = \frac{1}{2} (e_{j} -i e_{j+3}) , ~~~j=1,2,3                      
\end{equation}                       
                      
The automorphism group of the octonion algebra is the 14-parameter                       
exceptional group $G_{2}$.  The imaginary octonion units                       
$e_{\alpha} (\alpha  =1,...,7)$                       
fall into its 7-dimensional representation.                      
                       
   Under the $SU(3)^{c}$ subgroup of $G_{2}$ that leaves $e_{7}$                       
invariant, $u_{0}$ and $u_{0}^{*}$ are singlets, while $u_{j}$ and                       
$u_{j}^{*}$ correspond, respectively, to the                       
representations ${\bf 3}$ and $\bar{\bf 3}$.  The multiplication table can now be                       
written in a manifestly $SU(3)^{c}$ invariant manner (together with the                       
complex conjugate equations):                       

\begin{equation}                      
u_{0}^{2} = u_{0},~~~~~u_{0}u_{0}^{*} = 0                      
\end{equation}                       

\begin{equation}                      
u_{0} u_{j} = u_{j} u_{0}^{*} = u_{j},~~~~~                      
u_{0}^{*} u_{j} = u_{j} u_{0} = 0                          
\end{equation}                       

\begin{equation}                      
u_{i} u_{j}  = - u_{j} u_{i} = \epsilon_{ijk} u_{k}^{*}      \label{eq:oct}                      
\end{equation}                       

\begin{equation}                      
u_{i} u_{j}^{*} =  - \delta_{ij} u_{0}         \label{eq:octa}                      
\end{equation}                       
where $\epsilon_{ijk}$ is completely antisymmetric with  $\epsilon_{ijk} =1$                      
for  $ijk$  = $123$, $246$, $435$, $651$, $572$, $714$, $367$.                        
Here, one sees the                       
virtue of octonion multiplication.  If we consider the direct products

\begin{equation}                      
C:~~~~~{\bf 3} \otimes \bar{\bf 3} = {\bf 1} + {\bf 8}                       
\end{equation}

\begin{equation}                      
G:~~~~~{\bf 3} \otimes {\bf 3} = \bar{\bf 3} + {\bf 6}                      
\end{equation}                       
for $SU(3)^{c}$, then these equations show that octonion multiplication                       
gets rid of ${\bf 8}$ in ${\bf 3} \otimes \bar{\bf 3}$, while it gets rid                       
of ${\bf 6}$ in ${\bf 3} \otimes {\bf 3}$.  Combining  Eq.(\ref{eq:oct}) and                      
Eq.(\ref{eq:octa}) we find                       

\begin{equation}                      
(u_{i} u_{j}) u_{k} = - \epsilon_{ijk} u_{0}^{*}                       
\end{equation}                       
                       
   Thus the octonion product leaves only the color part in                       
${\bf 3} \otimes \bar{\bf 3}$ and ${\bf 3} \otimes {\bf 3} \otimes {\bf 3}$,                      
so that it is a natural algebra for colored quarks. 

For convenience we now produce the following multiplication table for the split octonion units:

\begin{center}
\begin{tabular}{|l|c|c|c|c|}\hline
 & $u_0$ & $u_{0}^{*}$ & $u_{k} $ & $u_{k}^{*}$  \\ \hline 
$u_0$  &  $ u_0 $  & $0$ & $u_k$ & $0$ \\ \hline

$u_{0}^{*}$  &  $ 0 $  & $u_{0}^{*}$ & $0$ & $u_{k}^{*}$ \\ \hline

$u_j$  &  $ 0 $  & $u_{j}$ & $\epsilon_{jki}u_{i}^{*}$ & $-\delta_{jk}u_{0}$ \\ \hline

$u_{j}^{*}$  &  $ u_{j}^{*} $  & $0$ & $-\delta_{jk}u_{0}^{*}$ & $\epsilon_{jki}u_{i}$ \\ \hline

\end{tabular}
\end{center}

It is worth noting that $u_i$ and $u_{j}^{*}$ behave like fermionic annihilation and creation operators:

\begin{equation}
\{u_i,u_j\}=\{u_{i}^{*},u_{j}^{*}\}=0,~~~\{u_i,u_{k}^{*}\}=-\delta_{ij}
\end{equation}                      

For more recent reviews on octonions and nonassociative algebras we refer to papers by Okubo$^{\cite{okubo}}$, and Baez$^{\cite{baez}}$.
                      
   The quarks, being in the triplet representation of the color                       
group $SU(3)^{c}$, are represented by the local fields                      
$q_{\alpha}^{i}(x)$, where $i = 1,2,3$ is the color index and $\alpha$                        
the combined spin-flavor index. Antiquarks at point $y$ are color                       
antitriplets $q_{\beta}^{i}(y)$.  Consider the two-body systems                       

\begin{equation}                      
C_{\alpha j}^{\beta i} = q_{\alpha}^{i} (x_{1}) \bar{q}_{\beta}^{j} (x_{2})                       
\label{eq:c}                      
\end{equation}                       

\begin{equation}                      
G_{\alpha \beta}^{i j} = q_{\alpha}^{i} (x_{1}) q_{\beta}^{j} (x_{2})                       
\label{eq:g}                      
\end{equation}                       
so that $C$ is either a color singlet or color octet, while $G$ is a                       
color antitriplet or a color sextet.  Now $C$ contains meson states                       
that are color singlets and hence observable.  The octet $q-\bar{q}$ state                       
is confined and not observed as a scattering state.  In the case of                       
two-body $G$ states, the antitriplets are diquarks which, inside a                       
hadron can be combined with another triplet quark to give                       
observable, color singlet, three-quark baryon states.  The color                       
sextet part of $G$ can only combine with a third quark to give                       
unobservable color octet and color decuplet three-quark states.                        
Hence the hadron dynamics is such that the ${\bf 8}$ part of $C$                       
and the ${\bf 6}$                       
part of $G$ are suppressed.  This can best be achieved by the use of the                       
above octonion algebra$^{ \cite{dom}}$.  The dynamical suppression of the                       
 octet and sextet states in Eq.(\ref{eq:c}) and Eq.(\ref{eq:g}) is , therefore,                       
automatically achieved.  The split octonion units can be contracted                       
with color indices of triplet or antitriplet fields.  For quarks                       
and antiquarks we can define the "transverse" octonions (calling $u_{0}$                       
and $u_{0}^{*}$ longitidunal units)

\begin{equation}                      
q_{\alpha} = u_{i} q_{\alpha}^{i} = {\bf u} \cdot{\bf q}_{\alpha} ,~~~~~                      
\bar{q}_{\beta} = u_{i}^{\dagger} \bar{q}_{\beta}^{j} = -{\bf u}^{*}                       
\cdot \bar{\bf q}_{\beta}        \label{eq:qal}                      
\end{equation}                       
                       
We find                       

\begin{equation}                      
q_{\alpha}(1) \bar{q}_{\beta}(2) = u_{0} {\bf q}_{\alpha}(1)                       
\cdot{\bf q}_{\beta}(2)                      
\end{equation}                       

\begin{equation}                      
\bar{q}_{\alpha}(1) q_{\beta}(2) = u_{0}^{*} \bar{\bf q}_{\alpha}(1)                       
\cdot{\bf q}_{\beta}(2)                      
\end{equation}

\begin{equation}                      
G_{\alpha \beta}(12) = q_{\alpha}(1) q_{\beta}(2) = {\bf u}^{*}                       
\cdot{\bf q}_{\alpha}(1) \times {\bf q}_{\beta}(2)                      
\end{equation}                       

\begin{equation}                      
G_{\beta \alpha}(21) = q_{\beta}(2) q_{\alpha}(1) = {\bf u}^{*}                       
\cdot{\bf q}_{\beta}(2) \times {\bf q}_{\alpha}(1)                      
\end{equation}                       
                       
Because of the anticomutativity of the quark fields, we have 
                      
\begin{equation}                      
G_{\alpha \beta}(12) = G_{\beta \alpha}(21) =                       
\frac{1}{2}  \{q_{\alpha}(1), q_{\beta}(2)\}                      
\end{equation}                       
                       
If the diquark forms a bound state represented by a field $D_{\alpha \beta}(x)$                      
at the center-of-mass location $x$

\begin{equation}                      
x = \frac{1}{2} (x_{1} +x_{2})                       
\end{equation}                       
when $x_{2}$ tends to $x_{1}$ we can replace the argument by $x$, and we obtain                       

\begin{equation}                      
D_{\alpha \beta}(x) = D_{\beta \alpha}(x)                      
\end{equation}                       
so that the local diquark field must be in a symmetric                       
representation of the spin-flavor group.  If the latter is taken to                       
be $SU(6)$, then $D_{\alpha \beta}(x)$ is in the 21-dimensional symmetric                       
representation, given by                        

\begin{equation}                      
({\bf 6} \otimes {\bf 6})_{s} = {\bf 21}                       
\end{equation}                       
                       
If we denote the antisymmetric $15$ representation by $\Delta_{\alpha \beta}$,                      
we see that the octonionic fields single out the $21$ diquark representation                       
at the expense of $\Delta_{\alpha \beta}$.  We note that without this                       
color algebra supersymmetry would give antisymmetric configurations as noted by                       
Salam and Strathdee$^{\cite{salam}}$ in their possible supersymmetric                       
generalization of hadronic supersymmetry.  Using the nonsingular                       
part of the operator product expansion we can write                       

\begin{equation}                      
\tilde{G}_{\alpha \beta}({\bf x}_{1}, {\bf x}_{2}) =                      
D_{\alpha \beta}(x) + {\bf r} \cdot {\bf \Delta}_{\alpha \beta}(x)                      
\label{eq:qam}                      
\end{equation}                       
The fields $\Delta_{\alpha \beta}$ have opposite parity to $D_{\alpha \beta}$;                      
${\bf r}$ is the relative                       
coordinate at time $t$ if we take $t$ = $t_{1}$ = $t_{2}$.  They play no role in                       
the excited baryon which becomes a bilocal system with the 21-                       
dimensional diquark as one of its constituents.                       
                      
   Now consider a three-quark system at time $t$.  The c.m. and                       
relative coordinates are

\begin{equation}                      
{\bf R} = \frac{1}{\sqrt{3}}({\bf r}_{1} + {\bf r}_{2} + {\bf r}_{3})                      
\end{equation}                       

\begin{equation}                      
{\bf  \rho} = \frac{1}{\sqrt{6}}(2 {\bf r}_{3} - {\bf r}_{1} - {\bf r}_{2})                     
\end{equation}                       

\begin{equation}                      
{\bf r} = \frac{1}{\sqrt{2}}({\bf r}_{1} - {\bf r}_{2})                      
\end{equation}                        
giving                       

\begin{equation}                      
{\bf r}_{1} = \frac{1}{\sqrt{3}} {\bf R} - \frac{1}{\sqrt{6}} {\bf \rho}                      
+ \frac{1}{\sqrt{2}} {\bf r}                      
\end{equation}                       

\begin{equation}                      
{\bf r}_{2} = \frac{1}{\sqrt{3}} {\bf R} - \frac{1}{\sqrt{6}} {\bf \rho}                      
- \frac{1}{\sqrt{2}} {\bf r}                      
\end{equation}                       

\begin{equation}                      
{\bf r}_{3} = \frac{1}{\sqrt{3}} {\bf R} + \frac{2}{\sqrt{6}} {\bf \rho}                      
\end{equation}                       
                       
The baryon state must be a color singlet, symmetric in the three                       
pairs ($\alpha$, $x_{1}$), ($\beta$, $x_{2}$), ($\gamma$, $x_{3}$).  We find                       

\begin{equation}                      
(q_{\alpha}(1) q_{\beta}(2)) q_{\gamma}(3) = -u_{0}^{*}                       
F_{\alpha \beta \gamma}(123)                       
\end{equation}                        

\begin{equation}                      
q_{\gamma}(3) (q_{\alpha}(1) q_{\beta}(2))  = -u_{0}                       
F_{\alpha \beta \gamma}(123)                      
\end{equation}                        
so that                       

\begin{equation}                      
- \frac{1}{2} \{\{q_{\alpha}(1), q_{\beta}(2)\}, q_{\gamma}(3)\} =                        
F_{\alpha \beta \gamma}(123)                       
\end{equation}                       
                       
The operator $F_{\alpha \beta \gamma}(123)$ is a color singlet and is                       
symmetrical in the three pairs of coordinates.  We have                       

\begin{equation}                      
F_{\alpha \beta \gamma}(123) = B_{\alpha \beta \gamma} ({\bf R}) +                      
{\bf \rho} \cdot {\bf B}'({\bf R}) + {\bf r} \cdot {\bf B}''({\bf R}) + C                      
\label{eq:fat}                      
\end{equation}                      
where $C$ is of order two and higher in ${\bf \rho}$ and ${\bf r}$.  Because                       
${\bf R}$ is symmetric in ${\bf r}_{1}$, ${\bf r}_{2}$ and ${\bf r}_{3}$,                       
the operator $B_{\alpha \beta \gamma}$    that creates a baryon                       
at ${\bf R}$ is totally symmetrical in its flavor-spin indices.  In the                       
$SU(6)$ scheme it belongs to the ($56$) representation.  In the bilocal                       
$q-D$ approximation we have ${\bf r}=0$ so that $F_{\alpha \beta \gamma}$ is                       
a function only of ${\bf R}$ and ${\bf \rho}$ which are both symmetrical in                       
${\bf r}_{1}$ and ${\bf r}_{2}$.  As before, ${\bf B}'$                        
belongs to the orbitally excited $70^{-}$ represenation of $SU(6)$.  The                       
totally antisymmetrical ($20$) is absent in the bilocal                       
approximation.  It would only appear in the trilocal treatment that                       
would involve the 15-dimensional diquarks.  Hence, if we use local                       
fields, any product of two octonionic quark fields gives a ($21$)                       
diquark                       

\begin{equation}                      
q_{\alpha}({\bf R}) q_{\beta}({\bf R}) = D_{\alpha \beta}({\bf R})                      
\end{equation}                       
and any nonassociative combination of three quarks, or a diquark                       
and a quark at the same point give a baryon in the $56^{+}$ representation:                       

\begin{equation}                      
(q_{\alpha}({\bf R}) q_{\beta}({\bf R})) q_{\gamma}({\bf R}) = - u_{0}^{*}                       
B_{\alpha \beta \gamma}({\bf R})                      
\end{equation}                       

\begin{equation}                      
q_{\alpha}({\bf R}) (q_{\beta}({\bf R}) q_{\gamma}({\bf R})) = - u_{0}                       
B_{\alpha \beta \gamma}({\bf R})                      
\end{equation}                       

\begin{equation}                      
q_{\gamma}({\bf R}) (q_{\alpha}({\bf R}) q_{\beta}({\bf R})) = - u_{0}                       
B_{\alpha \beta \gamma}({\bf R})                      
\end{equation}                       

\begin{equation}                      
(q_{\gamma}({\bf R}) q_{\alpha}({\bf R})) q_{\beta}({\bf R}) = - u_{0}^{*}                       
B_{\alpha \beta \gamma}({\bf R})                      
\end{equation}

The bilocal approximation gives the ($35+1$) mesons and the $70^{-}$                       
baryons with $\ell=1$ orbital excitation.                       
                      
In order to go beyond the $SU(6/21)$ symmetry we could go to a smaller supergroup and build a colored supersymmetry scheme based on $SU(6)\times SU(6/1)$, for example. Applications of such extended algebraic structures and their implication for the construction of supersymmetric meson-baryon lagrangians will be deferred to another publication.

We can now consider the $28\times 28$ octonionic matrix

\begin{equation}
Z= \left(
\begin{array}{ccc}
u_0 M & u_0 B & \mbox{\boldmath$u$} \cdot \mbox{\boldmath$Q$}  \\
u_0 B^\dag & u_0 N & \mbox{\boldmath$u$} \cdot \mbox{\boldmath$D$}^{*}   \\
\epsilon \mbox{\boldmath$u$}^{*} \cdot \mbox{\boldmath$Q$}^\dag   & \epsilon \mbox{\boldmath$u$}^{*} \cdot 
\mbox{\boldmath$D$}^T   & u_0 L 
\end{array} \right)
\end{equation} 
Here $\epsilon$ can be given values $1$, $-1$ or $0$. $M$ and $N$ are respectively $6\times 6$ and $21\times 21$ hermitian matrices, $B$ a rectangular $6\times 21$ matrix, $\mbox{\boldmath$u$} \cdot \mbox{\boldmath$Q$}$ a $6\times 1$ column matrix, $\mbox{\boldmath$u$} \cdot \mbox{\boldmath$D$}^{*}$ a $21\times 1$ column matrix, and $L$ a $1\times 1$ scalar. Such matrices close under anticommutator operations for $\epsilon=1$. Matrices $iZ$ close under commutator operations. In either case, they do not satisfy the Jacobi identity. But for $\epsilon=0$, when the algebra is no longer semi-simple, the Jacobi identity is satisfied and we obtain hadronic superalgebra which is an extension of the algebra $SU(6/21)$. Its automorphism group includes $SU(6)\times SU(21)\times SU(3)^c$. Thus color is automatically incorporated.    
 \section*{Relativistic Formulation Through the Spin Realization of the 
Wess-Zumino Algebra}

It is possible to use a spin representation of the Wess-Zumino algebra 
to write first order relativistic equations for quarks and diquarks that are  
invariant under supersymmetry transformations. In this section we  
briefly deal 
with such Dirac-like supersymmetric equations and with a short discussion of  
experimental possibilities for the observation of the diquark structure 
and exotic $\bar{D}-D = (\bar{q}\bar{q})(qq)$ mesons. For an extensive discussion of the experimental situation we refer to recent papers by Anselmino, et al.$^{\cite{lichty}}$, and by Klempt$^{\cite{klempt}}$. 
 
There is a spin realization of the Wess-Zumino super-Poincar\'{e} algebra 

\begin{equation} 
[p_{\mu}, p_{\nu}] = 0 ,~~~~~[D_{\alpha}, p_{\mu}] = 0 
\end{equation}
 
\begin{equation} 
[\bar{D}_{\dot{\beta}}, p_{\mu}] = 0, ~~~~~ 
[D^{\alpha}, \bar{D}^{\dot{\beta}}] = \sigma_{\mu}^{\alpha \dot{\beta}} 
p^{\mu} 
\end{equation} 
with $p_{\mu}$ transforming like a $4$-vector and $D^{\alpha}$, 
$\bar{D}^{\dot{\beta}}$ like the left and right handed spinors under the  
Lorentz group with generators $J_{\mu \nu}$. 
 
We also note that 

\begin{equation} 
[J_{\mu \nu}, p_{\lambda}] = \delta_{\mu \nu} p_{\lambda} - 
\delta_{\nu \lambda} p_{\mu} 
\end{equation} 
and 

\begin{equation} 
[J, J] = J  
\end{equation}  
 
The finite non unitary spin realization is in terms of $4 \times 4$ 
matrices for $J_{\mu \nu}$ and $p_{\nu}$ 

\begin{equation} 
J_{\mu \nu} = \frac{1}{2} \sigma_{\mu \nu} = 
\frac{1}{4i} [\gamma_{\mu}, \gamma_{\nu}]  
\end{equation} 

\begin{equation} 
J_{\mu \nu}^{L} = \frac{1-\gamma_{5}}{2} \frac{1}{2} \sigma_{\mu \nu} = 
\Sigma_{\mu \nu}^{L}    
\end{equation} 

\begin{equation} 
p_{\mu} = \Pi_{\mu}^{L} = \frac{1-\gamma_{5}}{2} \gamma_{\mu}   \label{eq:va} 
\end{equation} 
 
Introducing two Grassmann numbers $\theta_{\alpha}$ ($\alpha = 1, 2$) 
that transforms 
like the components of a left handed spinor and commute with the Dirac 
matrices $\gamma_{\mu}$, we have the representation 

\begin{equation} 
D_{\alpha} = \Delta_{\alpha} = \frac{\partial}{\partial \theta_{\alpha}} 
\end{equation}
 
\begin{equation} 
\bar{D}^{\dot{\beta}} = \bar{\Delta}^{\dot{\beta}} = \theta_{\alpha} 
\sigma_{\mu}^{\alpha \dot{\beta}} \Pi_{\mu}^{L}  
\end{equation} 
 
Such a representation of the super-Poincar\'{e} algebra acts on a Majorana  
chiral superfield 

\begin{equation} 
S(x, \theta) = \psi (x) + \theta_{\alpha} B^{\alpha} (x) + \frac{1}{2} 
\theta_{\alpha} \theta^{\alpha} \chi (x) . 
\end{equation} 
 
Here $\psi$ and $\chi$ are Majorana superfields associated with fermions and 
$B^{\alpha}$ has an unwritten Majorana index and a chiral spinor index 
$\alpha$, so that it represents a boson. 
 
Note that the sum of the two representations we wrote down is also a realization 
of the Wess-Zumino algebra. 
 
On the other hand we have the realization of $p_{\mu}$ in terms of the  
differential operator $-i \partial_{\mu} = -i \frac{\partial}{\partial  
x^{\mu}}$. In the Majorana representation, the operator $\gamma_{\mu} 
\partial_{\mu} = i \gamma_{\mu} p_{\mu}$ is real, and $\psi  = \psi ^{c} = 
\psi ^{*}$. Let us now define $\psi_{L}$ and $\psi_{R}$ by 

\begin{equation} 
\psi_{L} = \frac{1}{2} (1 + \gamma_{5}) \psi        
\end{equation}
and 

\begin{equation} 
\psi_{R} = \frac{1}{2} (1 - \gamma_{5}) \psi = \psi_{L}^{*}   
\end{equation} 
The free particle Dirac equation can now be written as 

\begin{equation} 
\Pi_{\mu}^{L} \partial_{\mu} \psi_{L} = m \psi_{L}^{*}    \label{eq:pi} 
\end{equation} 
or 

\begin{equation} 
\Pi_{\mu} p^{\mu} \psi_{L} = - i m \psi_{L}^{*} 
\end{equation} 
We can introduce left and right handed component fields 

\begin{equation} 
S_{L} = \frac{1}{2} (1 + \gamma_{5}) S  
\end{equation}  

\begin{equation} 
S_{R} = \frac{1}{2} (1 - \gamma_{5}) S = S_{L}^{*} 
\end{equation}
so that Eq.(\ref{eq:pi}) generalizes to the superfield equation 

\begin{equation} 
\Pi_{\mu}^{L} \partial_{\mu} S_{L} = m S_{L}^{*}    \label{eq:pia} 
\end{equation} 
or, 

\begin{equation} 
\Pi_{\mu} p^{\mu} S_{L} = - i m S_{L}^{*} 
\end{equation} 
Now consider the supersymmetry transformation 

\begin{equation} 
\delta S_{L} = (\xi^{\alpha} \Delta_{\alpha} + \bar{\xi}_{\dot{\beta}} 
\bar{\Delta}^{\dot{\beta}}) S_{L} = \Xi S_{L}    \label{eq:pib} 
\end{equation} 
This transformation commutes with the operator $\Pi_{\mu}^{L} \partial_{\mu}$ 
so that 

\begin{equation} 
\Pi_{\mu}^{L} \partial_{\mu} (S_{L} + \delta S_{L}) = m 
(S_{L} + \delta S_{L})^{*}. 
\end{equation} 
 
If $\psi_{L}$ is a left handed quark and $B^{\alpha}(x)$ an antidiquark 
with the same mass as the quark, Eq.(\ref{eq:pib}) provides a  
relativistic form of the quark antidiquark symmetry which is in fact broken 
by the quark-diquark mass difference. The scalar supersymmetric potential  
is introduced through $m \longrightarrow m + V_{s}$ as before and    
Eq.(\ref{eq:pia}) remains supersymmetric. By means of this formalism, it is 
possible to reformulate the treatments given in the earlier sections in  
first order relativistic form. 
 
To write equations in the first order form, we consider $V$ and $\Phi$ 
given in terms of the boson fields by 
\begin{equation} 
V = i \gamma_{\mu} V_{\mu} -\frac{1}{2}   \sigma_{\mu \nu} 
\end{equation} 
and 

\begin{equation} 
\Phi = i \gamma_{5} \phi + i \gamma_{5} \gamma_{\mu} \phi_{\mu} . 
\end{equation} 
In the Majorana representation 
 
\begin{equation} 
V^{*} = - V , ~~~~~and ~~~~~~ \Phi = \Phi^{*} 
\end{equation} 
 
We now define the left and right handed component fields by 
\begin{equation} 
V_{L} = \frac{1- \gamma_{5}}{2} V
\end{equation}
and

\begin{equation}
V_{R}= \frac{1+\gamma_{5}}{2} V 
\end{equation} 
so that

\begin{equation} 
V_{R} = - V_{L}^{*} 
\end{equation}  
with 

\begin{equation} 
V_{L} = i \frac{1- \gamma_{5}}{2} \gamma_{\mu} V_{\mu} - 
 \frac{1- \gamma_{5}}{2} \frac{1}{2} \sigma_{\mu \nu} V_{\mu \nu} 
\end{equation} 
and making use of  Eq.(\ref{eq:va}) we have
 
\begin{equation} 
V_{L} = i \Pi_{\mu}^{L} V_{\mu} - 
 \frac{1- \gamma_{5}}{2} \frac{1}{2} \sigma_{\mu \nu} V_{\mu \nu} 
\label{eq:vi} 
\end{equation} 
 
Noting that for any $a_{\mu}$ and $b_{\nu}$ we can write

\begin{equation} 
\Pi_{\mu} \Pi_{\nu}^{*} a_{\mu} b_{\nu} =  
\frac{1- \gamma_{5}}{2} \gamma_{\mu} \frac{1+ \gamma_{5}}{2} \gamma_{\nu} 
a_{\mu} b_{\nu} = \frac{1- \gamma_{5}}{2} \gamma_{\mu} \gamma_{\nu} 
a_{\mu} b_{\nu} 
\end{equation} 
and incorporating  the definition of $\sigma_{\mu \nu}$  

\begin{equation} 
\frac{1}{2} (\Pi_{\mu} \Pi_{\nu}^{*} - \Pi_{\nu} \Pi_{\mu}^{*}) = 
\frac{1- \gamma_{5}}{2} i \sigma_{\mu \nu} 
\end{equation} 
in Eq.(\ref{eq:vi}) leads to 

\begin{equation} 
V_{L} = i \Pi_{\mu}^{L} V_{\mu} + \frac{i}{2}  
(\Pi_{\mu} \Pi_{\nu}^{*} - \Pi_{\nu} \Pi_{\mu}^{*}) V_{\mu \nu}  
\label{eq:vl} 
\end{equation} 
Letting $\Sigma_{\mu \nu}^{L} = \Pi_{\mu} \Pi_{\nu}^{*} -  
\Pi_{\nu} \Pi_{\mu}^{*}$, Eq.(\ref{eq:vl}) reads 

\begin{equation} 
V_{L} = i \Pi_{\mu}^{L} V_{\mu} + \frac{i}{2} \Sigma_{\mu \nu}^{L} V_{\mu \nu}  
\end{equation} 

We now have a first order equation:
 
\begin{equation} 
\Pi_{\mu}^{L} \partial_{\mu} V_{R} = m V_{L}   \label{eq:vc} 
\end{equation} 
Similarly 

\begin{equation} 
\Pi_{\mu}^{R} \partial_{\mu} V_{L} = m V_{R} 
\end{equation} 
Therefore 

\begin{equation} 
\Pi_{\mu}^{L} \Pi_{\mu}^{R} \partial_{\mu} \partial_{\nu} V_{L} =  
 m \Pi_{\mu}^{L} \partial_{\mu} V_{R} = m^{2} V_{L} 
\end{equation} 
which after substitution of Eq.(\ref{eq:vc}) gives 

\begin{equation} 
\Box{V_{L}} = m^{2} V_{L} 
\end{equation} 

For the $\Phi$ part we can write 

\begin{equation} 
\Pi_{\mu}^{L} \partial_{\mu} \Phi_{R} = m \Phi_{L} 
\end{equation} 
with 

\begin{equation} 
\Phi_{L} = \frac{1- \gamma_{5}}{2} \Phi
\end{equation}
and

\begin{equation}
\Phi_{R} = \frac{1+ \gamma_{5}}{2} \Phi 
\end{equation} 

Clearly,

\begin{equation} 
\Phi_{R} = \Phi_{L}^{*}. 
\end{equation} 
and above procedure can now be repeated for the $\Phi$ fields.

\section{Skyrmion Model}

The Lagrangian density

\begin{equation}
{\cal{L}}=-\frac{1}{16} F_{\pi}^2 ~tr(\partial_\mu U^{\dag} \partial^{\mu} U) + \frac{1}{32e^2}~tr([U^{\dag} \partial_{\mu} U,U^{\dag} \partial_{\nu} U]^2)    \label{eq:1s}
\end{equation}
consists of the original chiral model term (first term) plus the Skyrme (second) term. Here $F_\pi$ is the pion decay constant whose experimental value is $186$ MeV, and $U$ is an $SU(2)$ matrix given by

\begin{equation}
U= exp{\{ 2i\mbox{\boldmath${\tau}$}\cdot \frac{\mbox{\boldmath$\phi$}(t,\mbox{\boldmath$x$})}{F_\pi} \}}
\end{equation}

In above equations $\mbox{\boldmath$\phi$}$ is the pseudoscalar pion field ($I=1, S=0)$, and $e$ is an unspecified dimensionless parameter of the model. $U$ is subject to the constraint

\begin{equation}
U U^\dag =1
\end{equation}
and ${\cal{L}}$ is invariant under $SU_L(2)\times SU_R(2)$ under which

\begin{equation}
U\rightarrow LUR^\dag
\end{equation}
where $LL^\dag=1$, $RR^\dag=1$ and $\mbox{\boldmath$\tau$}$ are the $2\times2$ Pauli matrices. $F(r)$ is the shape function to be determined by the variation of the ground state energy of the soliton. In the large $F_\pi$ limit we can write

\begin{equation}
U\approx 1+ 2i\mbox{\boldmath${\tau}$}\cdot \frac{\mbox{\boldmath$\phi$}}{F_\pi} 
\end{equation}
so that

\begin{equation}
\partial_{\mu} U^\dag \partial^{\mu} U = 4 F_\pi^{-2} \partial_{\mu} \mbox{\boldmath$\phi$}\cdot \partial^{\mu} \mbox{\boldmath$\phi$}
\end{equation}
which gives

\begin{equation}
-\frac{1}{16}~ tr~\partial_{\mu} U^\dag \partial^{\mu} U = -\frac{1}{2} \partial_{\mu} \mbox{\boldmath$\phi$}\cdot \partial^{\mu} \mbox{\boldmath$\phi$} 
\end{equation}
the kinetic term for a massless pion.

A mass term for the pion would come from

\begin{equation}
{\cal{L}}_m= -\frac{1}{16} F_{\pi}^2 m_{\pi}^2 ~tr(U+U^{\dag} -2) \approx -\frac{1}{2} m_{\pi}^2 \mbox{\boldmath$\phi$}^2 +O(F_\pi)
\end{equation}
which would break the symmetry of ${\cal{L}}$ to the diagonal $SU(2)$ (isospin) under which 
\begin{equation}
U\rightarrow MUM^\dag~~~~~~ (L=R=M)
\end{equation}
or 
\begin{equation}
\mbox{\boldmath$\tau$}\cdot \mbox{\boldmath$\phi$}\rightarrow M\mbox{\boldmath$\tau$}\cdot \mbox{\boldmath$\phi$}M^\dag 
\end{equation}

Going to the next order

\begin{equation}
U= 1+ 2i \mbox{\boldmath$\tau$}\cdot\frac{\mbox{\boldmath$\phi$}}{F_\pi} - 2\frac{\phi^2}{F_\pi^2} +\cdots
\end{equation}
where $\phi^2=\mbox{\boldmath$\phi$}\cdot\mbox{\boldmath$\phi$}$, and
\begin{equation}
U^\dag= 1- 2i \mbox{\boldmath$\tau$}\cdot\frac{\mbox{\boldmath$\phi$}}{F_\pi} - 2\frac{\phi^2}{F_\pi^2} +\cdots
\end{equation}
we have

\begin{equation}
\partial_\mu U = 2i \mbox{\boldmath$\tau$}\cdot F_\pi^{-1} \partial_\mu \mbox{\boldmath$\phi$}-4F_\pi^{-2} \mbox{\boldmath$\phi$}\cdot \partial_\mu \mbox{\boldmath$\phi$}
\end{equation}

We can now write the right handed current

\begin{eqnarray}
J_\mu^R&=&\frac{F_\pi^2}{4} ~U^\dag \partial_\mu U^\dag  =i\mbox{\boldmath$\tau$}\cdot(\mbox{\boldmath$\phi$}\times \partial_\mu \mbox{\boldmath$\phi$}+ \frac{F_\pi}{2} ~\partial_\mu \mbox{\boldmath$\phi$} + O(F_\pi^{-1})  
\\ \nonumber &=& i \mbox{\boldmath$\tau$} \cdot (\mbox{\boldmath$V$}_\mu +\mbox{\boldmath$A$}_\mu) 
\end{eqnarray}
where

\begin{equation}
\mbox{\boldmath$V$}_\mu =\mbox{\boldmath$\phi$}\times \partial_\mu \mbox{\boldmath$\phi$}+\cdots
\end{equation}
is the isospin current, and

\begin{equation}
\mbox{\boldmath$A$}_\mu =\frac{F_\pi}{2}~\partial_\mu \mbox{\boldmath$\phi$}+\cdots
\end{equation}
is the axial vector current. Also

\begin{equation}
J_\mu^L=\frac{F_\pi^2}{4}~U\partial_\mu U^\dag = -\frac{F_\pi^2}{4} ~(\partial_\mu U)U^\dag = 
i\mbox{\boldmath$\tau$}\cdot ( \mbox{\boldmath$V$}_\mu -\mbox{\boldmath$A$}_\mu )
\end{equation}
with additional terms coming from the Skyrme term.

Under $SU_L(2)\times SU_R(2)$ we have

\begin{equation}
U\rightarrow LUR^\dag,~~~~~U^\dag \rightarrow RU^\dag L^\dag
\end{equation}
so that

\begin{equation}
J_\mu^L \rightarrow LJ_\mu^L L^\dag,~~~~~J_\mu^R \rightarrow RJ_\mu^R R^\dag
\end{equation}
Both currents are conserved from the equation of motion ($m_\pi=0$). It is the charged components of $J_\mu^L$ that couple to the $W_\mu^+$ weak vector bosons in weak interactions. Both occur ($J_\mu^{L,R}$) in strong interactions governed by QCD.

${\cal{L}}$ is assumed to arise from the QCD theory of gluons and quarks when these colored fields are integrated out in the Feynman path integral to leave a skyrmion like effective theory in the color singlet sector. Note that using

\begin{equation}
\partial_\mu U^\dag=-U^\dag (\partial_\mu U)U^\dag
\end{equation}
${\cal{L}}$ can be cast in the Sommerfield-Sugawara form

\begin{eqnarray}
{\cal{L}}&=& \frac{F_\pi^2}{16}~tr(J_\mu^L J^{L\mu}) +\frac{1}{32 e^2}~tr([J_\mu^L, J_\nu^L]^2) 
\nonumber \\ &=& \frac{F_\pi^2}{16}~tr(J_\mu^R J^{R\mu}) +\frac{1}{32 e^2}~tr([J_\mu^R, J_\nu^R]^2) 
\end{eqnarray}

These currents obey the $SU_L(2)\times SU_R(2)$ current algebra of Gell-Mann, so that we have the correspondence with quark bilinears

\begin{equation}
J_\mu^L\sim \overline{q}~\frac{1-\gamma_5}{2}~\gamma_\mu~q=V_\mu-A_\mu
\end{equation}
and

\begin{equation}
J_\mu^R \sim \overline{q}~\frac{1+\gamma_5}{2}~\gamma_\mu~q=V_\mu+A_\mu
\end{equation}

Another conserved current is the "topological" baryon current$^{\cite{witten}}$

\begin{equation}
B^\mu=\frac{4}{3\pi^2}~\frac{\epsilon^{\mu\nu\alpha\beta}}{F_\pi^6}~tr\{J_\nu^R J_\alpha^R J_\beta^R \}
\end{equation}
which is conserved identically without use of the equation of motion. The same current is obtained by using $J_\nu^L$ instead of $J_\nu^R$. Integration over the baryonic current $B^0$ gives the baryonic charge.

Lagrangian in Eq.(\ref{eq:1s}) should also contain the Wess-Zumino action$^{\cite{wz}}$ which can not be written in closed form in four dimensions but in five dimensions

\begin{equation}
\Gamma_{WZ}= -\frac{iN_c}{240 \pi^2}~\int_{M^5}~tr[\alpha^5]
\end{equation}
where the 1-form $\alpha$ is defined as $\alpha=U^\dag \partial_\mu U ~dx^\mu$ and $M^5$ is a five-dimensional manifold with boundary $\partial M^5=M^4$, the Minkowski space. In the case of $SU(2)$ the Wess-Zumino term vanishes. We can now proceed to show this.

${\cal{L}}$ has extra symmetries that QCD does not have. For example, $K^+ K^- \rightarrow \pi^+ \pi^0 \pi^-$ violates parity $P$ which is a symmetry of ${\cal{L}}$. The Euler-Lagrange equation for ${\cal{L}}$ is

\begin{equation}
\partial_\mu (\frac{1}{8} F_\pi^2~U^\dag \partial_\mu U)=0
\end{equation}
We can try to add an extra term that violate $P$. Levi-Civita symbol $\epsilon_{\mu\nu\alpha\beta}$ is a perfect choice. The Euler-Lagrange equation is generalized to

\begin{equation}
\partial_\mu (\frac{1}{8} F_\pi^2~U^\dag \partial_\mu U)+ \lambda \epsilon^{\mu\nu\alpha\beta}~ (U^\dag \partial_\mu U) (U^\dag \partial_\nu U) (U^\dag \partial_\alpha U) (U^\dag \partial_\beta U)=0 
\end{equation}
We can now try to get the term in the Lagrangian that will yield the above equation of motion. In the equation above, the term that multiplies $\lambda$ is an obvious choice but it vanishes identically. To solve the problem, we write down the Wess-Zumino term which is a five-dimensional integral

\begin{equation}
{\cal{L}}_{WZ}= \int_{D5} d^5x~\epsilon^{\mu\nu\alpha\beta}~tr[L_\mu L_\nu L_\alpha L_\beta L_\gamma] 
\end{equation}
where $L_\mu=U^\dag \partial_\mu U$, and $D_5$ is a five-dim disc whose boundary $M_4$ is space-time. Then ${\cal{L}}+{\cal{L}}_{WZ}$ would give the required equation of motion.

Let $Q$ be the generator of the gauge group. The $U$ transforms as $U\rightarrow U+i\alpha(x)[Q,U]$. Now introduce gauge field $A_\mu$ which transforms as

\begin{equation}
A_\mu\rightarrow A_\mu - \partial_\mu \alpha(x)
\end{equation}
${\cal{L}}$ is not gauge invariant, so we have to add terms to it. The gauge invariant W-Z term is

\begin{eqnarray}
{\cal{L}}_{WZ}^{\rm inv} & =& {\cal{L}}_{WZ} -\int d^4x A_\mu J_\mu \nonumber \\  & & +\frac{i}{2}\int~d^4x ~\epsilon^{\mu\nu\alpha\beta}~(\partial_\mu A_\nu)A_\alpha~tr[ Q^2(\partial_\beta U) U^\dag \nonumber \\
& & ~~~~~~+Q^2 U^\dag(\partial_\beta U)+Q UQU^\dag (\partial_\beta U) U^\dag]
\end{eqnarray}
where

\begin{eqnarray}
J^\mu &=&\frac{1}{48\pi^2}~ \epsilon^{\mu\nu\alpha\beta}~tr[ Q(\partial_\nu UU^\dag)(\partial_\alpha UU^\dag)(\partial_\beta UU^\dag)   \nonumber \\
& & +Q(U^\dag \partial_\nu U)(U^\dag \partial_\alpha U)(U^\dag \partial_\beta U)
\end{eqnarray}
is the current coupled to gauge field $A_\mu$.

At quark level, the baryon number operator is

\begin{equation}
Q=\frac{1}{N_c}~I
\end{equation}
Therefore, we have the baryonic current

\begin{equation}
B^\mu= \frac{1}{24\pi^2}~ \epsilon^{\mu\nu\alpha\beta}~tr[(U^\dag \partial_\nu U)(U^\dag \partial_\alpha U)(U^\dag \partial_\beta U)]
\end{equation}
${\cal{L}_{WZ}}$ vanishes identically when there are only two flavors, i.e. $U \in SU(2)$. This can be seen easily if we write

\begin{equation}
U= 1+i\mbox{\boldmath$\tau$}\cdot \mbox{\boldmath$a$}
\end{equation}
so that

\begin{equation}
{\cal{L}_{WZ}}\sim \int d^4x~\epsilon^{\mu\nu\alpha\beta}~a_a \partial_\mu a_b \partial_\nu a_c \partial_\alpha a_d \partial_\beta a_e~tr[\tau_a \tau_b\tau_c \tau_d \tau_e]
\end{equation}
In $SU(2)$ there are only three independent generators $\tau$. However, the $\epsilon^{\mu\nu\alpha\beta}$ has four indices which require $b$, $c$, $d$ and $e$ be completely antisymmetric. This is impossible, so this forces ${\cal{L}}\equiv 0$. The baryonic current, although it is obtained from Wess-Zumino term, is non-vanishing.

\section*{The Skyrme Kink Solution; Rotating Skyrmion and its Quantization}

The equations of motion obtained for ${\cal{L}}$ have a static solution of the hedgehog form
 
\begin{equation}
U=U_0(\mbox{\boldmath$r$})= exp\{ {i F(r) \frac{\mbox{\boldmath$\tau$}\cdot \mbox{\boldmath$r$}}{r}}\}  \label{eq:ansatz}
\end{equation}
where $r=\sqrt{x^2+y^2+z^2}$, and $\mbox{\boldmath$r$}$ is aligned along isospin giving $I+J$ invariance and    

\begin{equation}
U_0(0)=-1,~~~~U_0(\infty)=1
\end{equation} 
corresponds to the following boundary conditions on the function $F(r)$:
 
\begin{equation}
F(0)=\pi,~~~~~F(\infty)=0
\end{equation}

Here $F(r)$ is the shape function to be determined by the variation of the ground state of the soliton. A realistic model would necessarily contain the dynamical quantities of spins and isospins. The time dependence of the soliton is obtained by rotating it with a time dependent $SU(2)$ matrix $A(t)$

\begin{equation}
U(\mbox{\boldmath$r$}, t)=A(t) U_0(\mbox{\boldmath$r$}) A^\dag(t)  \label{eq:u}
\end{equation}
where 

\begin{equation}
A(t) A^\dag (t) =1
\end{equation}
Thus $A(t)$ is a rotation matrix which makes the soliton rotate in time.

Sustituting Eq.(\ref{eq:u}) into Eq.(\ref{eq:1s}) we obtain the Lagrangian

\begin{equation}
L(t)=\int d^3x~{\cal{L}}(t,\mbox{\boldmath$x$})= -M+\lambda~tr({\dot{A}} {\dot{A}}^\dag)
\end{equation}
where

\begin{equation}
\dot{A}=\frac{dA(t)}{dt},~~~~~\lambda=\frac{2\pi}{3}~\frac{\Lambda}{e^3 F_\pi}
\end{equation}
with

\begin{equation}
\Lambda=\int_0^\infty d\rho~\rho^2~sin^2F (1+4(F'^2 + \frac{sin^2F}{\rho^2}))
\end{equation}
Calculation gives $\Lambda\sim 50.9$ for the Skyrme solution. Now if
 
\begin{equation}
A=a_0+i\mbox{\boldmath$a$}\cdot \mbox{\boldmath$\tau$}=\left(
\begin{array}{cc}
a_0+ia_3 & ia_1+a_2 \\
ia_1-a_2 & a_0-ia_3
\end{array}
\right)
\end{equation}
($a_0^2+{\mbox{\boldmath$a$}}^2=1)$, then the Hamiltonian is

\begin{equation}
H=M+\frac{1}{8\lambda} \sum_{i=0}^3(-\frac{\partial^2}{\partial a_i^2})
\end{equation}
with 

\begin{equation}
\sum_{i=0}^3 a_i^2=1
\end{equation}
constraints. This constrains $a_i$ on the sphere $S^3$ and $(H-M)$ is the Laplacian on $S^3$. Its eigenfunctions are spherical harmonics, polynomials in $a_0\pm ia_3$ and $a_1\pm ia_2$ of degree $\ell$.

One obtains a tower of states with (spin=isospin) 
 
\begin{equation}
I=J=\frac{\ell}{2}    \label{eq:201}
\end{equation}
$\ell=1$ and $\ell=3$ are identified with $N$ and $\Delta$ respectively with the corresponding $E_{\ell}$ eigenvalues of $H$ giving their masses.

The states for $\ell >3$ are not found in nature. This is one of the weaknesses of the model.

The tower can be chopped off by quantizing $A(t)$ fermionically, so that

\begin{equation}
\{a_i,\Pi_j\}=\delta_{ij}
\end{equation}
In this case one has only $\ell$ odd solutions. To have a non vanishing cubic polynomials in $a_i$ one has also to introduce a color degree of freedom for $a_i$ and choose color singlet cubic expressions for $\Delta$. 

The wave functions are traceless symmetric polynomials in $a_i$ and that they have spin and isospin given by Eq.(\ref{eq:201}) can be immediately seen by considering the operators

\begin{equation}
J_k= \frac{1}{2} i(-a_0 \frac{\partial}{\partial a_k} + a_k \frac{\partial}{\partial a_0} - \epsilon_{klm} a_l \frac{\partial}{\partial a_m})
\end{equation}
and 

\begin{equation}
I_k= \frac{1}{2} i( a_0 \frac{\partial}{\partial a_k} - a_k \frac{\partial}{\partial a_0} -\epsilon_{klm} a_l \frac{\partial}{\partial a_m})
\end{equation}

The baryon charge density and the baryon currents are given by the components $B^0$ and $\mbox{\boldmath$B$}$ of the baryon current $B^\mu$. They are

\begin{equation}
B^0=-\frac{1}{2\pi^2}~\frac{sin^2F}{r^2}~F'
\end{equation}
and

\begin{equation}
B^i= \frac{\epsilon^{ijk}}{\pi^2}~\frac{sin^2F}{r}~F'~{\hat{r}}_j s_k
\end{equation}

Note that the expressions for the spin and isospin are

\begin{equation}
\mbox{\boldmath$K$}=-\frac{i}{2}~tr(\mbox{\boldmath$\tau$}A^\dag \dot{A})  
\end{equation}
and

\begin{equation}
\mbox{\boldmath$S$}= \frac{i}{2}~tr(A^\dag \mbox{\boldmath$\tau$}\dot{A}) 
\end{equation}
with the baryon number

\begin{equation}
B=\int_0^\infty 4\pi r^2 B^0(r)=\frac{[F(0)-F(\infty)]}{\pi}
\end{equation}
Note that $B=1$ when $F(0)=\pi$ and $f(\infty)=0$.

We also note that integrating ${\cal{L}}$ we obtain

\begin{equation}
L=\int d^3r~{\cal{L}}=-M+2\lambda \mbox{\boldmath$S$}^2 = -M+2\lambda \mbox{\boldmath$K$}^2
\end{equation} 
where $M$ is given by

\begin{equation}
M= 4\pi \int_0^\infty dr~r^2\{ \frac{F_\pi^2}{8} [ F'^2 +2\frac{sin^2F}{r^2} ] +\frac{1}{2e^2} ~\frac{sin^2F}{r^2}~[\frac{sin^2F}{r^2} + 2F'^2]\}   \label{eq:M}
\end{equation}
and the moment of inertia $\lambda$ is

\begin{equation}
\lambda= \frac{4 \pi}{3} \int_0^\infty dr~r^2 [ \frac{F_\pi^2}{2} sin^2F +\frac{2}{e^2} sin^2F (F'^2+\frac{sin^2F}{r^2})] \label{eq:2s}
\end{equation}

Let us now show these results explicitly. First we add a term $\frac{1}{2} \gamma^2 B^\mu B_\mu$ to ${\cal{L}}$ in Eq.(\ref{eq:1s}). Since

\begin{equation}
B^\mu= \frac{1}{24\pi^2}~\epsilon^{\mu\nu\alpha\beta}~tr[ U^\dag \partial_\nu U~ U^\dag \partial_\alpha U~
U^\dag \partial_\beta U ]
\end{equation}
and writing $U$ in component form, we have

\begin{equation}
U= \frac{1}{D}~A(cosF(r) + i \mbox{\boldmath$\tau$} \cdot \mbox{\boldmath${\hat{x}}$}~sinF(r))A^\dag =\frac{1}{D}~A(c+i
\mbox{\boldmath$\tau$} \cdot \mbox{\boldmath${\hat{x}}$}~s)A^\dag
\end{equation}
where we defined $c=cosF(r)$, $s=sinF(r)$ and $D=det~A$. Now

\begin{equation}
K_a= -i~tr(\tau_a A^\dag \dot{A})=-i~tr(\dot{A} \tau_a A^\dag)=-i~tr(A^\dag \dot{A} \tau_a)
\end{equation}
and 

\begin{equation}
S_a = -i~tr(A^\dag \tau_a \dot{A})=-i~tr(\tau_a \dot{A} A^\dag)=-i~tr(\dot{A} A^\dag \tau_a)
\end{equation}

Therefore

\begin{eqnarray}
{\cal{L}}&=& \frac{F_\pi^2}{8}~[-(F'^2 + 2\frac{s^2}{r^2}) + s^2~\frac{K_a}{D}~\delta_{ab}^T~\frac{K_b}{D}] \nonumber \\
& &+ \frac{1}{2e^2}[ -\frac{s^2}{r^2}~(2 F'^2 + \frac{s^2}{r^2}) + s^2 (F'^2 + \frac{s^2}{r^2})~\frac{K_a}{D}~\delta_{ab}^T~\frac{K_b}{D}] \nonumber\\
& & +\frac{\gamma^2}{2} [ -(\frac{1}{2\pi^2}~\frac{s^2}{r^2}~F')^2 + (\frac{1}{2\pi^2}~\frac{s^2}{r}~F')^2 ~\frac{K_a}{D}~\delta_{ab}^T~\frac{K_b}{D}] 
\end{eqnarray}
where we defined $\delta_{ab}^T= \delta_{ab}-{\hat{x}}_a {\hat{x}}_b$. Integrating ${\cal{L}}$ yields

\begin{eqnarray}
L&=&\int {\cal{L}}~d^3r \nonumber \\ 
& &= -4\pi \int_0^\infty r^2 dr [ \frac{F_\pi^2}{8}~(F'^2 +2~\frac{s^2}{r^2}) + \frac{1}{2e^2}~\frac{s^2}{r^2} (2F'^2 +\frac{s^2}{r^2}) \nonumber \\ 
& & + \frac{\gamma^2}{2} (\frac{1}{2\pi^2}~\frac{s^2}{r^2}~F')^2] +\frac{8\pi}{3} \int_0^\infty r^2 dr [\frac{F_\pi^2}{8}~s^2 \nonumber \\
& & + \frac{1}{2e^2}~s^2 (F'^2 + \frac{s^2}{r^2}) + \frac{\gamma^2}{2}~(\frac{1}{2\pi^2}~\frac{s^2}{r^2}~F')^2] \frac{\mbox{\boldmath$K$} \cdot \mbox{\boldmath$K$}}{D^2} \nonumber \\
& & = -M +2 \lambda ~\mbox{\boldmath$K$} \cdot \mbox{\boldmath$K$}= -M +2 \lambda \mbox{\boldmath$S$} \cdot 
\mbox{\boldmath$S$} \nonumber \\
& & =-M + \lambda~(\mbox{\boldmath$K$} \cdot \mbox{\boldmath$K$} + \mbox{\boldmath$S$} \cdot 
\mbox{\boldmath$S$})  
\end{eqnarray}
giving the desired expression.

Now with the use of boundary conditions $F(0)=\pi$ and $F(\infty)=0$ the shape function $F(r)$ is solved numerically by minimizing $M$. The differential equation for $F(\rho)$ is (where $\rho$ is a dimensionless parameter defined as $\rho=eF_\pi r$)

\begin{equation}
\rho^2 (\rho^2 +8 sin^2F) F'' +2\rho^3 F' + sin2F(4\rho^2 F'^2 -4 sin^2F-\rho^2)=0  \label{eq:shape}
\end{equation}
Behavior of $F(\rho)$ is plotted against $\rho$ in Fig.(1) below.

We further note that using Eq.(\ref{eq:1s}) the equation of motion is written as

\begin{equation}
\partial^\mu(U^\dag\partial_\mu U) -\frac{1}{e^2 F_\pi}~ \partial^\mu (U^\dag  \partial^\nu U [ U^\dag \partial_mu U , U^\dag \partial_\nu U])=0
\end{equation}
and in the static limit we have

\begin{equation}
\partial_i(U^\dag \partial_i U) -\frac{1}{e^2 F_\pi^2}  \partial_i (U^\dag  \partial_j U[U^\dag \partial_i U, U^\dag \partial_j U])=0     \label{eq:stat}
\end{equation}
Substituting the soliton ansatz Eq.(\ref{eq:ansatz}) into Eq.(\ref{eq:stat}) we obtain

\begin{equation}
r^4 F'' + 2r^3 F' -r^2 sin2F +\frac{4}{e^2 F_\pi^2} [ sin2F(sin^2F - r^2 F'^2) -2r^2 sin^2F~F'']=0
\end{equation}
which is identical to Eq.(\ref{eq:shape}) when we change the variable $r$ to $\rho=eF_\pi r$. therefore the hedgehog soliton is obtained from solving Eq.(\ref{eq:shape}) is consistent with the above equation of motion.

To find the mass spectrum of the solitons, we first have to write the system in the Hamiltonian form and then quantize the collective coordinates $A$. The matrix $A$ has a canonical conjugate matrix ${\cal{A}}$ with matrix elements 

\begin{equation}
{\cal{A}}^{\alpha\beta}= \partial L/\partial \dot{A}_{\alpha\beta}
\end{equation}
where $\dot{A}_{\alpha\beta}$ is the matrix element of $\dot{A}$. The quantization prescription is given by

\begin{equation}
[A_{\alpha\beta}, {\cal{A}}^{\mu\nu}]=i \delta_\alpha^\mu \delta_\beta^\nu   \nonumber
\end{equation} 

\begin{equation}
[A_{\alpha\beta}, A_{\mu\nu}]=[ {\cal{A}}^{\alpha\beta}, {\cal{A}}^{\mu\nu}]=0
\end{equation}
The Hamiltonian is obtained from the Legendre transformation

\begin{equation}
H= M+\mbox{\boldmath$K$}^2/2\lambda = M+\mbox{\boldmath$S$}^2/2\lambda    \label{eq:3s}
\end{equation}

The baryon charge density and the baryon currents were given by the components $B^0$ and $\vec{B}$ of the baryon current $B_\mu$ above. From these the electric charge radius and the magnetic moments can be computed.

Eigenvalues of the Hamiltonian are 

\begin{equation}
E=M+\frac{1}{8\lambda}~\ell (\ell+2)
\end{equation}
where $\ell=2J$. Therefore the nucleon ($N$) and the $\Delta$ masses are given by

\begin{equation}
M_N= M+\frac{1}{2\lambda}~\frac{3}{4}
\end{equation}

\begin{equation}
M_\Delta= M+\frac{1}{2\lambda}~\frac{15}{4}
\end{equation}
where $M$ is obtained numerically from Eq.(\ref{eq:M}) which comes out to be

\begin{equation}
M= 36.416 ~\frac{F_\pi}{e}
\end{equation}
and similarly $\lambda$ is numerically evaluated from Eq.(\ref{eq:2s}) to be

\begin{equation}
\lambda= \frac{2\pi}{3}~ (\frac{1}{e^3 F_\pi})\times 50.42
\end{equation}

Above values are, for $M=865.15~ MeV$, and for $\lambda= 5.11 \times 10^{-3}$. Experimentally 
$M_N=938.5$ MeV, and $M_\Delta=1232$ MeV. Best fit to these nucleon and delta masses yield $e=5.43$ and $F_\pi=129$ MeV which is much smaller than the experimental value of $F_\pi=186$ MeV. Improvement of $F_\pi$ value will be shown later in this paper.

The isoscalar mean radius is

\begin{eqnarray}
<r^2>_{E,I=0} &=& \int_0^\infty 4\pi r^2~dr B_0(r) r^2 = -\frac{2}{\pi} \int_0^\infty r^2~sin^2F~F'~dr \nonumber \\
  & & -\frac{2}{\pi} \int_0^\infty r^2~dr~F'~sin^2F = \frac{4.47}{e^2 F_\pi^2}
\end{eqnarray}
gives

\begin{equation}
<r^2>_{E,I=0}^{1/2} = 0.597
\end{equation}

The isoscalar magnetic mean radius is given by

\begin{equation}
<r^2>_{M,I=0}  = \frac{3}{5}~ \frac{\int_0^\infty 4 \pi r^2~dr~B_0(r)~r^4}{<r^2>_{E,I=0} }=0.837 {\rm fm}^2
\end{equation}
so that $<r^2>_{M,I=0}  = 0.915$ fm. 

Magnetic moments are given by

\begin{equation}
\mu_p = 2 M_N (\frac{<r^2>_{I=0} }{12 \lambda} + \frac{\lambda}{6})
\end{equation}
and

\begin{equation}
\mu_n = 2 M_N (\frac{<r^2>_{I=0} }{12 \lambda} - \frac{\lambda}{6})
\end{equation}
Numerical calculations give (in terms of Bohr magneton) $\mu_p=1.871$ and $\mu_n=-1.34$. The ratio $|\frac{\mu_p}{\mu_n}|=1.4$ is in excellent agreement with the experimental value $1.46$.

The axial coupling is given by

\begin{equation}
g_A = -\frac{\pi}{3e^2} D
\end{equation}
where 

\begin{equation}
D= \int_0^\infty r^2 ~dr \{\frac{sin2F}{r} [1+4(F'^2 +\frac{sin^2F}{r^2})]+F'(1+\frac{8 sin^2F}{r^2})\}
\end{equation}
We find $D=-17.2$ and $g_A= 0.611$.

Strong couplings are given by

\begin{equation}
g_{\pi NN} =\frac{2M_N}{F_\pi} ~g_A
\end{equation}
which is the Goldberger-Treiman relation. We get $g_{\pi NN}=9.327$ and $g_{\pi N \Delta} = \frac{3}{2} g_{\pi NN}$ gives $g_{\pi N \Delta}= 13.99$. Experimental values are $g_{\pi NN}=13.5$ and $g_{\pi N \Delta} = 20.3$.

From the axial vector current $A_\mu$ and the vector current $V_\mu$ one can compute $g_{\pi N N}$ and $ \frac{g_A}{g_V}$. One finds

\begin{equation}
g_A=\frac{1}{2}~g_A^{\rm exp}
\end{equation}
and $g_{\pi N N}$ smaller than $g_{\pi N N}^{\rm exp}$ by $40\%$, as well as the magnetic moments $\mu_p$ and $\mu_N$ being smaller than the experimental values by $40\%$. We also fin $ |\frac{\mu_p}{\mu_N}| $ is good to $2\%$, and $<r^2>$ is off by $30\%$. Thus the agreement is only qualitative. Similar calculations performed by Nam and Workman$^{\cite{nam}}$ and Hahm, et.al.$^{\cite{hahm}}$ agree with these results. Magnetic moments are improved by adding vector mesons to the model$^{\cite{4s}}$, but $F_\pi$ resists all improved models so far. With many other contributing factors not yet taken into account, such as the effect of the pion mass and the interaction of vector mesons, this level of agreement is remarkable.

\section*{Possible Strategy for an Improved Skyrmion Model for Hadrons}

We have seen that the two-body potential models for the $q-\overline{q}$ and $q-D$ bound state systems give a good description of hadrons when the potentials are extracted from QCD. In this case gluons are integrated out but the quark degrees of freedom remain. The model reflects the basic symmetries of QCD: its flavor independence, its approximate spin independence of the confining forces, the vector nature of gluons (Coulomb force at short distances) and its chiral invariance broken by $q \overline{q}$ condensates. In addition we have $q- \overline{q}$ structure for mesons and $qqq$ or $q-D$ structure for baryons giving the corresponding representations of $SU(6)$ for low lying hadrons and its supersymmetric extension $SU(6/21)$.

The skyrmion model, to be successful must reflect these symmetries. It does not. We now make the following observations toward the implementation of such a programme:

(1). Only $I=J$ states are present. The approximate $SU(4)$ symmetry of non strange hadrons has for the multiplet $\mbox{\boldmath$\pi$}$, $\omega_\mu$ and ${\mbox{\boldmath$\rho$}}_\mu$ (15 rep) and for the baryons $N$ and $\Delta$ the 20 dimensional symmetric multiplet (submultiplet of 56 of $SU(6)$). This tells us that $N$ and $\Delta$ are cubic functions of the quark wave function.

In the skyrmion theory $\omega_\mu$ ($I\neq J$) is missing. One must decouple isospin from spin. For that one should transform the skyrmion rotation

\begin{equation}
U_0(\mbox{\boldmath$r$}) \rightarrow A(t)U_0(\mbox{\boldmath$r$}) A^\dag(t), ~~~~(AA^\dag=1)  
\end{equation}
but also by deformation

\begin{equation}
U_0(\mbox{\boldmath$r$}) \rightarrow B(t)U_0(\mbox{\boldmath$r$}) B^\dag(t), ~~~~(BB^\dag=1)  
\end{equation}
so that one has

\begin{equation}
U(t,\mbox{\boldmath$x$})= M(t) U_0(\mbox{\boldmath$r$}) N^\dag(t) 
\end{equation}
with

\begin{equation}
M(t)=B(t)~ A(t),~~~~~N(t)=B^\dag(t)~ A(t).
\end{equation}
Then one has $6$ degrees of freedom instead of $4$ and $I$ and $J$ can be decoupled.

(2). In the skyrmion theory wave functions of $N$ are linear in $A(t)$ and $\Delta$ is cubic in $A(t)$. The $SU(4)$ symmetry between $N$ and $\Delta$ is lost.

It can be recovered by having both $N$ and $\Delta$ being cubic in  $M(t)$ and $N(t)$. Then we can ask what do the $\ell=1$ (linear in $M$ and $N$) and the $\ell=2$ (quadratic in $M$ and $N$) represent? If one has the color degree of freedom, necessary for fermionic quantization, these states will be colored. Then it is natural to associate $\ell=1$ linear functions with the quark (cubic root of the Skyrme soliton) and $\ell=2$ with the diquark ($\frac{2}{3}$th root of the skyrmion).

A simple modification of the skyrmion Lagrangian was proposed$^{\cite{chg}}$ with the aim of reflecting the cubic nature of the baryon. $U_0(\mbox{\boldmath$x$})$ is replaced by $V_0(\mbox{\boldmath$x$})=U_0(\mbox{\boldmath$x$})$ where

\begin{equation}
U(t,\mbox{\boldmath$x$})= exp \{ 2i~\mbox{\boldmath$\tau$} \cdot \frac{\mbox{\boldmath$\phi$}}{F_\pi} \}
\end{equation}
so that

\begin{equation}
V(t,\mbox{\boldmath$x$})= exp \{ 6i~\mbox{\boldmath$\tau$} \cdot \frac{\mbox{\boldmath$\phi$}}{F_\pi} \}
\end{equation}

The new Lagrangian ${\cal{L}}'$ is adjusted so that for $F_\pi$ large it gives

\begin{equation}
{\cal{L}}'\simeq -\frac{1}{2}~\partial_\mu \mbox{\boldmath$\phi$} \cdot \partial^\mu \mbox{\boldmath$\phi$}
\end{equation}

It reads

\begin{equation}
{\cal{L}}_n= - \frac{F_\pi^2}{16 n^2}~ tr(\partial_\mu U^{\dag n} \partial^\mu U^n) + \frac{1}{32e^2n^4}~tr([U^{\dag n} \partial_\mu U^n, U^{\dag n} \partial_\nu U^n]^2)      \label{eq:ln}
\end{equation}
with $n=3$.  We have ${\cal{L}}= {\cal{L}}_1$ for the usual skyrmion. Under $SU(2) \times SU(2)$ $U^n$ transforms as

\begin{equation}
U^n \rightarrow LU^n R^\dag
\end{equation}

The hedgehog solution is obtained by having $U_0(0)=-1$ , $U_0(\infty)=1$ so that 

\begin{equation}
U_0^3(0)=-1,~~~~~~~~U_0^3(\infty)=1
\end{equation}
is unchanged. Again

\begin{equation}
U_0(\mbox{\boldmath$r$})=exp~\{ iF(r) ~\frac{\mbox{\boldmath$\tau$} \cdot \mbox{\boldmath$r$}}{r} \}
\end{equation}
with $F(0)=\pi$, $F(\infty)=0$.

$F(r)$ now satisfies a new linear equation derived from the new Lagrangian ${\cal{L}}_3$. The constants $M$ and $\lambda$ have now new values as functions of $F_\pi$ and $e$. We can now proceed to calculate the static properties of composite solitons. Rotating composite configuration is given by

\begin{equation}
U^n(\mbox{\boldmath$r$},t)=A(t) U_0^n(\mbox{\boldmath$r$}) A^\dag(t)
\end{equation} 
and its substitution into Eq.(\ref{eq:ln}) gives the Lagrangian

\begin{equation}
L_n=-M_N + 2 \lambda_n {\mbox{\boldmath$s$}}^2
\end{equation}

The mass of the composite static soliton is

\begin{equation}
M_n= 4\pi \int_0^\infty dr~r^2\{ \frac{F_\pi^2}{8} [ F'^2 +2\frac{sin^2nF}{n^2 r^2} ] +\frac{1}{2e^2} ~\frac{sin^2nF}{r^2}~[\frac{sin^2nF}{n^2 r^2} + 2F'^2]\}   
\end{equation}
and the corresponding moment of inertia is

\begin{equation}
\lambda_n= \frac{4 \pi}{3} \int_0^\infty dr~r^2 [ \frac{F_\pi^2}{2n^2} sin^2nF +\frac{2}{e^2} \frac{sin^2nF}{n^2} (2F'^2+\frac{sin^2nF}{n^2 r^2})] 
\end{equation}

The normalized baryonic current

\begin{equation}
B^\mu= \frac{1}{24\pi^2 n^2}~\epsilon^{\mu\nu\alpha\beta}~tr[ (U^{-n} \partial_\nu U^n)~( U^{-n} \partial_\alpha U^n)~
(U^{-n} \partial_\beta U^n) ]
\end{equation}
has a time component

\begin{equation}
B_0 = -\frac{1}{2\pi^2}~\frac{sin^2nF}{r^2}~F'
\end{equation}
and spatial component

\begin{equation}
B^i=\frac{\epsilon^{ijk}}{\pi^2}~\frac{sin^2nF}{r}~F'~{\hat{r}}_j~s_k
\end{equation}

The baryon number is equal to unity when $F(0)=\pi$ and $F(\infty)=0$. Note that all odd powers of $U$ obey the same "kink" boundary conditions $U(0)=U^3(0)=U^5(0)=\cdots=-1$, and $U(\infty)=U^3(\infty)=U^3(\infty)=\cdots=1$. This is not true for $U^2$ or other even powers of $U$.

The equation of motion for the shape function $F(\rho)$ now takes the form

\begin{equation}
\rho^2 (n^2 \rho^2 +8 sin^2nF) F'' +2n^2 \rho^3 F' + \frac{1}{n} sin2nF(4n^2\rho^2 F'^2 -4 sin^2nF- n^2\rho^2)=0  
\end{equation}
Once again this is solved numerically and its behaviour is plotted for various odd values of $n$ in figure $1$ below.

\begin{center}
\includegraphics[width=150mm]{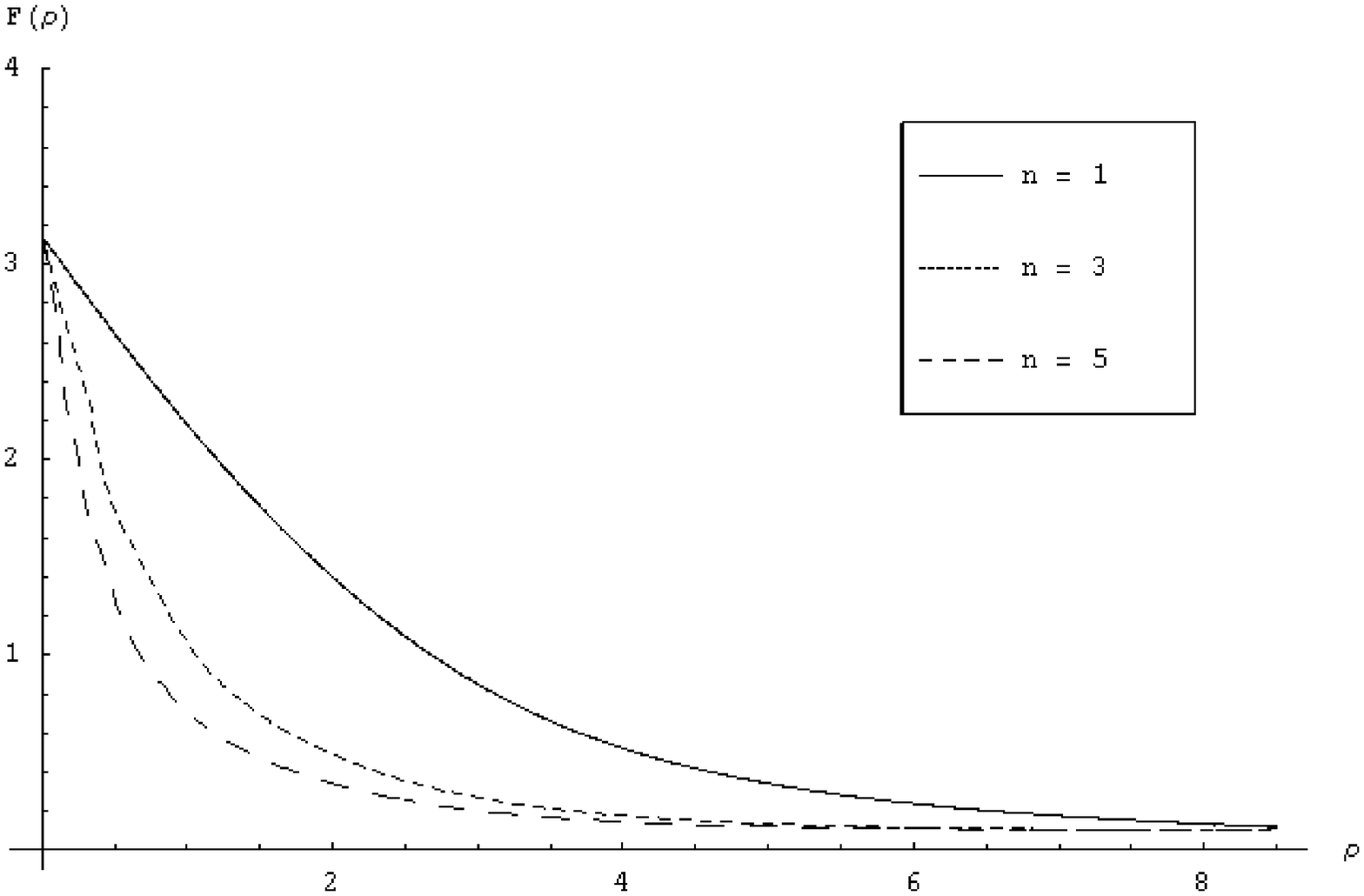}
\end{center} 
\begin{center}
{\bf Fig.1 Shape function $F(\rho)$ vs. $\rho$ }
\end{center}

Now the Hamiltonian is given by

\begin{equation}
H_n= {\cal{A}}^{\alpha\beta} {\dot{A}}_{\alpha\beta} -L_n= M_n + \frac{{\mbox{\boldmath$S$}}^2}{2 \lambda_n}
\end{equation}

In the usual manner, when they are fitted to the $N$ and $\Delta$ masses one finds $e=1.72$ (instead of 5.43 for $n=1$), and $F_\pi= 185$ MeV (instead of 129 MeV for $n=1$). 

Isoscalar mean radius calculated in the same manner as before turns out $<r^2>^{1/2}_{I=0}=0.74$ fm, and the isoscalar magnetic mean radius is $1.185$ fm. Similarly the magnetic moment calculation yields (where $\lambda$ in above $\mu_p$ and $\mu_n$ is now replaced by $\lambda_n$) yields the values $\mu_p=2.0343$ and $\mu_n=-1.17$. The ratio $|\frac{\mu_p}{\mu_n}|=1.97$ ($1.46$ experimentally).

The axial coupling is given by

\begin{equation}
g_A=-\frac{\pi}{3e^2}~D_n
\end{equation}
where

\begin{equation}
D_n=\int_0^\infty r^2~dr \{ \frac{sin2nF}{nr} [ 1+ 4(F'^2 +\frac{sin^2nF}{n^2 r^2}) +F' (1+ \frac{8 sin^2nF}{n^2 r^2}) \}
\end{equation}

We find $D_3=-3.71$ and hence $g_A=1.31$, which is close to the experimental value $1.23$. The strong couplings for the $n=3$ case turn out to be $g_{\pi NN}=13.2$ and $g_{\pi N \Delta}=19.8$. 

Results of the calculations are compiled in the table below

\begin{center}
\begin{tabular}{|c|c|c|c|} \hline
Quantity & Old Prediction ($n=1$) &Prediction ($n=3$) & Experiment  \\ \hline
$M_N$   & input &  input & 938.5 MeV  \\ \hline
$M_\delta$ & input & input & 1232 MeV \\ \hline
$F_\pi$ & 129 & 185 MeV & 186 MeV \\ \hline
${\sqrt{<r^2>}}_{E,I=0}$ & 0.597 & 0.74 fm  & 0.72 fm \\  \hline
${\sqrt{<r^2>}}_{M,I=0}$ & 0.915 & 1.185 fm  & 0.81 fm \\  \hline
$\mu_p$ & 1.871 & 2.034 & 2.79 \\  \hline
$\mu_N$ & -1.34 & -1.17 & -1.91 \\ \hline
$g_A$ & 0.611 & 1.31 & 1.23 \\  \hline
$g_{\pi NN}$ & 9.327 & 13.2 & 13.5 \\  \hline
$g_{\pi N \Delta}$ & 13.99 & 19.8 & 20.3 \\ \hline
\end{tabular}
\end{center} 
\vspace{0.5cm}

We see that most everything is improved (some spectacularly). Except for the magnetic properties of the nucleon, the predictions derived from the model are in excellent agreement with the experimental values. Inclusion of the effect of vector mesons could change and improve the results significantly since the virtual photon can make a transition to the $\omega$ and $\rho^0$ mesons in the vector dominance scenarios. Possible improvement of the predicted values of the magnetic properties will be discussed below. 

In the generalized model (as is also in $n=1$ case) we find that the calculated nucleon charge densities oscillate within the nucleons, a peculiar future of nucleon substructure. It is well known, from the deep inelastic scattering experiments, that nucleons have point-like constituents. Since the quark degrees of freedom are integrated out in the Skyrme model, we expect the substructure of the nucleons not to be point-like. Therefore the oscillatory behavior of the charge densities may be due to interference effect of the constituents, which is similar to the quark interference effect found in the bag model$^{\cite{chodos}}$.  

Excellent agreement with the experimental value of $F_\pi$ (=186 MeV) suggests that $U^3$ model could serve as a better description of the nucleons. The reason might be that composite model associated with the Lagrangian ${\cal{L}}_n$ describes an effective hadronic theory based on QCD with the color group $SU(n)$. The baryon consisting of $n$ quarks are in a spatially symmetric ground state. In that case the current $nB^\mu$ could be interpreted as a quark number current, so that the baryon has quark number 3 for $n=3$. Under spin and isospin transformations, the skyrmion matrix

\begin{equation}
U^3(\mbox{\boldmath$r$},t) = A(t)~U_0^3(\mbox{\boldmath$r$})~A^\dag(t)
\end{equation}
has same transformation properties as the nucleon matrix of spin states $n_\alpha$ and $p_\alpha$ ($\alpha=1,2$)

\begin{equation}
N= \left(
\begin{array}{cc}
-n_1 & p_1 \\
-n_2 & p_2 
\end{array}  \right)
\end{equation}
with left and right multiplications being respectively associated with spin and isospin rotations. Now $N$ and $\Delta$ which form the 20-dimensional symmetric representation of the spin-isospin group $SU(4)$ are cubic functions of the elements of the underlying quark matrix

\begin{equation}
Q= \left(
\begin{array}{cc}
-d_1 & u_1 \\
-d_2 & u_2 
\end{array}  \right)
\end{equation}
that can be taken to be proportional to $A(t)$. When the color group is $SU(3)$ and $Q$ is quantized fermionically, it is possible to obtain the symmetric ${\bf 20}$ representation for the $(N,\Delta)$ system that is cubic in $Q$. Four of these twenty cubic functions, which are eigenstates of the Hamiltonian, are associated with $N$. They have both spin and isospin $1/2$ and can be put in a $2\times 2$ matrix form $Q\bar{Q}Q$, where $Q\bar{Q}= Det~Q$ with 

\begin{equation}
\bar{Q}= \left(
\begin{array}{cc}
u_2 & -u_1 \\
d_2 & -d_1 
\end{array}  \right)
\end{equation}
$Q\bar{Q}$ being a singlet in spin and isospin. The remaining 16 cubic functions correspond to the wave functions of the $\Delta$ resonances. 

Since the quark has a baryon number $1/3$ and the matrix $U_0^3$ gives baryon number $1$, it is natural to interpret the $\mbox{\boldmath$r$}$ dependence of the quark in the nucleon as being given by $U_0(\mbox{\boldmath$r$})$. Then the quark matrix could be split into the form

\begin{equation}
Q=A(t) U_0(\mbox{\boldmath$r$})
\end{equation}
in a static approximation. The $N$ and $\Delta$ wave functions would then be associated with the cubic functions of the elements of $U_0$ as well as being cubic in $A(t)$. The fermionic quantization is preferable for eliminating the higher excited states of the skyrmion represented by higher order polynomials in $A(t)$ that are eigenstates of the Hamiltonian. A convincing derivation of $n=3$ composite Skyrme Lagrangian from QCD theory of quarks and gluons with color group $SU(3)$ is still lacking. For the moment one can regard it as a phenomenological model that gives reasonable quantitative predictions. We shall deal with this issue and the wave functions of the $\Delta$ resonances written in terms of the cubic functions in a forthcoming publication$^{\cite{sultan}}$. For now we give a brief introduction to this subject.

The baryon states can be written in terms of $A$, for example

\begin{equation}
<A|\Delta^{++},~S_3=\frac{3}{2}> \sim A_{21}^3
\end{equation}
 
\begin{equation}
<A|p,~S_3=\frac{1}{2}> \sim A_{21}
\end{equation}

Note that we can write
\begin{equation}
u\uparrow= a_1 +ia_2=\psi_1, ~~~~~~d\uparrow= i(a_0 + ia_3)=\psi_2
\end{equation}
so that

\begin{equation}
u\downarrow=-i(a_0 -ia_3)=\psi_2^*,~~~~~~d\downarrow= -(a_1 -ia_2)=-\psi_1^*
\end{equation}
giving

\begin{equation}
Q=\left( \begin{array}{cc}
\psi_1  & -\psi_2^* \\
\psi_2 & \psi_1^*
\end{array}  \right)
\end{equation}
with 

\begin{equation}
|\psi_1|^2 + |\psi_2|^2=1
\end{equation}

Note that the left column of $Q$ is

\begin{equation}
\psi = \left( \begin{array}{c}
\psi_1 \\  \psi_2 \end{array} \right)
\end{equation}
while the right column is

\begin{equation}
\hat{\psi} = -i \sigma_2~\psi^*
\end{equation}
so that $Q$ can be written in compact form as

\begin{equation}
Q = \left( \begin{array}{cc}
\psi & \hat{\psi}
\end{array}  \right)
\end{equation}

Now the $\Delta^{++}$ state is

\begin{equation}
\Delta^{++}=\frac{\sqrt{2}}{\pi}~ \psi_1^3
\end{equation}
and $\Delta^{+}$ state is given by

\begin{equation}
\Delta^+= \frac{-\sqrt{2}}{\pi}~ \psi_1 (1-3 \psi_2^{*} \psi_2)=\frac{\sqrt{2}}{\pi}~ \psi_1 (2-3\psi_1^{*} \psi_1)
\end{equation}

In general we can construct the $\Delta$ resonances in this way. We can write down the $Q\bar{Q}Q$ ($s=\frac{3}{2}$) part in a matrix form

\begin{equation}
Q\bar{Q} Q= \left( \begin{array}{cc}
 -u_1d_1d_2+u_2 d_1^2 & -u_1u_2d_1 + u_1^2 d_2  \\
u_2d_1d_2 -u_1d_2^2  & u_1u_2d_2 - u_2^2 d_1 
\end{array} \right)    \label{eq:central}
\end{equation}
This term arises as part of a general quark matrix (in terms of $u$'s and the $d$'s) as follows: Let us write down the following matrices  
 
\begin{equation}
Q K_\alpha Q K_\beta Q + Q K_\beta Q K_\alpha Q
\end{equation}
and

\begin{equation}
Q \bar{K}_\alpha Q \bar{K}_\beta Q + Q \bar{K}_\beta Q \bar{K}_\alpha Q
\end{equation}
each expression giving $3\times 4-2=10$ triplet quark combinations as expected. In component form 
 
\begin{equation}
K_1= \frac{1+\tau_3}{2},~~~~~~K_2=\frac{1-\tau_3}{2}
\end{equation}
and

\begin{equation}
\bar{K}_1=\frac{\tau_1+i\tau_2}{2},~~~~~~\bar{K}_2=\frac{\tau_1-i\tau_2}{2}
\end{equation}

We can now write the full matrix $\Omega$ as

\begin{equation}
\Omega= \left( \begin{array}{cc}
-Q K_1 Q K_1 Q & Q \bar{K}_2Q \bar{K}_2 Q  \\
-Q \bar{K}_1 Q \bar{K}_1 Q & Q K_2 Q K_2 Q
\end{array} \right)
\end{equation}
where
 
\begin{equation}
\Omega_{11}= -Q K_1 Q K_1 Q = \left( \begin{array}{cc}
d_1^3  & -u_1d_1^2 \\
d_1^2 d_2 & -u_1 d_1 d_2
\end{array} \right)
\end{equation}
 
\begin{equation}
\Omega_{12}= Q \bar{K}_2Q \bar{K}_2 Q  = \left(  \begin{array}{cc}
-u_1^2 d_1  &  u_1^3  \\
-u_1 u_2 d_1 & u_1^2 u_2
\end{array}  \right)
\end{equation}
 
\begin{equation}
\Omega_{21}=Q \bar{K}_1 Q \bar{K}_1 Q = \left(  \begin{array}{cc}
-d_1 d_2^2 & u_2 d_1 d_2  \\
-d_2^3  & u_2 d_2^2
\end{array}  \right)
\end{equation}
 
\begin{equation}
\Omega_{22}= -Q K_2 Q K_2 Q = \left(  \begin{array}{cc}
u_1 u_2 d_2  & -u_1 u_2^2 \\
u_2^2 d_2  & -u_2^3
\end{array}  \right)
\end{equation}

We can now embed $s=\frac{3}{2}$ part into the center of the $\Omega$ matrix. We get the $s=\frac{3}{2}$ part from
 
\begin{equation}
Q[\bar{K}_2 Q \bar{K}_1 + \bar{K}_1 Q \bar{K}_2 -K_2 Q K_1 - K_1 Q K_2]Q=  \nonumber
\end{equation}
\begin{equation}
~~=\left( \begin{array}{cc}
u_2 d_1^2 - u_1 d_1 d_2  & -u_1 u_2 d_1 + u_1^2 d_2   \\
u_2 d_1 d_2 - u_1 d_2^2  & u_1 u_2 d_2 - u_2^2 d_1
\end{array}  \right)   \label{eq:centrali}
\end{equation}

Similarly we can write down the $s=\frac{1}{2}$ parts as

\begin{equation}
Q[\bar{K}_2 Q \bar{K}_1 - \bar{K}_1 Q \bar{K}_2]Q= [u_1 d_2-u_2 d_1] \left( \begin{array}{cc}
d_1  & 0  \\
0  & u_2
\end{array}  \right)
\end{equation}
and
 
\begin{equation}
Q[K_1 Q K_2 -K_2 Q K_1]Q= [u_1 d_2 -u_2 d_1] \left( \begin{array}{cc}
0  & u_1  \\
d_2  & 0
\end{array}  \right)
\end{equation}

Finally the full matrix containing all $20$ terms become:
 
\begin{equation}
\Xi= \left( \begin{array}{cccc}
d_1^3  & -u_1 d_1^2  & -u_1^2 d_1  & u_1^3  \\
d_1^2 d_2  & -u_1 d_1 d_2 +u_2 d_1^2 & -u_1 u_2 d_1 + u_1^2 d_2 & u_1^2 u_2  \\
-d_1 d_2^2  & u_2 d_1 d_2 -u_1 d_2^2  & u_1 u_2 d_2 - u_2^2 d_1  & -u_1 u_2^2  \\
-d_2^3  & u_2 d_2^2  & u_2^2 d_2  & -u_2^3
\end{array}   \right)
\end{equation}
The expression shown in Eq.(\ref{eq:centrali}) then corresponds to the following term:

\begin{equation}
\left( \begin{array}{cc}
\Xi_{22} & \Xi_{23} \\
\Xi_{32}  & \Xi_{33}
\end{array}  \right)
\end{equation}
 
Note that the first column of $\Xi$ contains all the $S_3=\frac{3}{2}$ $\Delta^{++}$, $\Delta^+$, $\Delta^0$, and $\Delta^-$ terms, the second column contains $S_3=\frac{1}{2}$, third column contains $S_3=\frac{-1}{2}$, and the last column contains $S_3=\frac{-3}{2}$ expressions. 

We now give these terms explicitly. For the $\Delta^{++}$ states we have

\begin{equation}
<A|~\Delta^{++},~ S_3=\frac{3}{2} >= \frac{\sqrt{2}}{\pi} ~\psi_1^3
\end{equation}

\begin{equation}
<A|~\Delta^{++}, ~S_3=\frac{1}{2} >= -\frac{\sqrt{6}}{\pi} ~\psi_1^2~\psi_2^*
\end{equation}

\begin{equation}
<A|~\Delta^{++},~ S_3=\frac{-1}{2} >= -\frac{\sqrt{6}}{\pi} ~\psi_1 ~ \psi_2^{*2}
\end{equation}

\begin{equation}
<A|~\Delta^{++}, ~S_3=\frac{-3}{2} >= \frac{\sqrt{2}}{\pi} ~\psi_2^{*3}
\end{equation}

For the $\Delta^+$ state we have

\begin{equation}
<A|~\Delta^{+},~ S_3=\frac{3}{2} >= \frac{\sqrt{6}}{\pi} ~\psi_1^2~\psi_2
\end{equation}

\begin{equation}
<A|~\Delta^{+},~ S_3=\frac{1}{2} >= - \frac{\sqrt{2}}{\pi} ~\psi_1~(1-3\psi_2^*~\psi_2)
\end{equation}

\begin{equation}
<A|~\Delta^{+},~ S_3=\frac{-1}{2} >= \frac{\sqrt{2}}{\pi} ~\psi_2^*~(1-3\psi_1^*~\psi_1)
\end{equation}

\begin{equation}
<A|~\Delta^{+},~ S_3=\frac{-3}{2} >=- \frac{\sqrt{6}}{\pi} ~\psi_2^{*2}~\psi_1^*
\end{equation}

For the $\Delta^0$ states we have

\begin{equation}
<A|~\Delta^{0},~ S_3=\frac{3}{2} >= \frac{\sqrt{6}}{\pi} ~\psi_1~\psi_2^2
\end{equation}

\begin{equation}
<A|~\Delta^{0},~ S_3=\frac{1}{2} >= \frac{\sqrt{2}}{\pi} ~\psi_2~(1-\psi_1^*~\psi_1)
\end{equation}

\begin{equation}
<A|~\Delta^{0},~ S_3=\frac{-1}{2} >= \frac{\sqrt{2}}{\pi} ~\psi_1^*~(1-\psi_2^*~\psi_2)
\end{equation}

\begin{equation}
<A|~\Delta^{0},~ S_3=\frac{-3}{2} >= \frac{\sqrt{6}}{\pi} ~\psi_1^{*2}~ \psi_2^*
\end{equation}

Finally, for the $\Delta^-$ states we have

\begin{equation}
<A|~\Delta^{-},~ S_3=\frac{3}{2} >= \frac{\sqrt{2}}{\pi} ~\psi_2^3
\end{equation}

\begin{equation}
<A|~\Delta^{-},~ S_3=\frac{1}{2} >=- \frac{\sqrt{6}}{\pi} ~\psi_1^*~\psi_2^2
\end{equation}

\begin{equation}
<A|~\Delta^{-},~ S_3=\frac{-1}{2} >= \frac{\sqrt{6}}{\pi} ~\psi_1^{*2}~\psi_2
\end{equation}

\begin{equation}
<A|~\Delta^{-}, ~S_3=\frac{-3}{2} >=- \frac{\sqrt{2}}{\pi} ~\psi_1^{*3}
\end{equation}

Action of the Hamiltonian operator given above on the $\Delta^0$, $\Delta^+$, $\Delta^-$ and $\Delta^{++}$ states  will be dealt with in detail in a forthcoming publication.

(3). Restoration of symmetries.

One can now have a supersymmetry algebra generated by the creation operators for $\ell=1$ (fermionic) and those for $\ell=2$ (bosonic). The $\ell=3$ cubic functions will be a representation of this superalgebra. 

(4). Construction of the enlarged Hamiltonian that will be approximately invariant under $SU(4)$ and its supersymmetric extension $SU(4/10)$ (subalgebra of $SU(6/21)$) is the next step.

Finding its eigenvalues and linear (quark), quadratic (diquark) and cubic ($N$ and $\Delta$) eigenstates is the following step. 

Final step will be recalculation of masses, charge radii, meson-nucleon coupling constants and magnetic moments, pion decay constant, meson-meson couplings, etc.

\section*{Chiral Lagrangians with Vector Mesons}

In the large $N_c$ expansion baryons emerge as solitons within an effective theory of infinitely many weakly interacting mesons. A more realistic model should include not just the pseudoscalar meson field but also other heavier vector mesons. If we assume the $\pi$'s, $\omega$'s and $\rho$'s form an $SU(4)$ multiplet, they should be treated in a unified manner. Accordingly, we now consider an $8\times 8$ matrix configuration of these mesons

\begin{equation}
W= exp\{i\gamma_5 \mbox{\boldmath$\tau$}\cdot \mbox{\boldmath$ \phi$} + \frac{1}{2} \sigma_{\mu\nu} \omega^{\mu\nu} + \frac{1}{2} \sigma_{\mu\nu} \mbox{\boldmath$\tau$}\cdot \mbox{\boldmath$\rho$}^{\mu\nu}\}   \label{eq:v1}
\end{equation}
where 
\begin{equation}
\mbox{\boldmath$\phi$} \rightarrow \frac{2}{F_\pi}~\mbox{\boldmath$\phi $} ,~~~~\omega \rightarrow \frac{g_\omega}{M_\omega^2}~\omega,~~{\rm and}~~ \mbox{\boldmath$\rho$} \rightarrow \frac{g_\rho}{M_\rho^2}~ \mbox{\boldmath$\rho$}  \label{eq:v2}
\end{equation}
will be substituted back into Eq.(\ref{eq:v1}) later. Here $g_\omega$ and $g_\rho$ are the couplings, and $M_\omega$ and $M_\rho$ are the masses of $\omega$ and $\rho$ mesons.  The $8\times 8$ Pauli and Dirac matrices are formed by making the direct products
 
\begin{equation}
\tau_a \rightarrow \tau_a \otimes I_4, ~~~~{\rm and} ~~~~\gamma_\mu \rightarrow I_2 \otimes \gamma_\mu  \nonumber
\end{equation}
also

\begin{equation}
\sigma_{\mu\nu}= \frac{1}{2i} [\gamma_\mu, \gamma_\nu]
\end{equation}
 
First let us write down the simplest chiral model

\begin{eqnarray}
{\cal{L}}_0&=& tr[(\partial^\mu W)W^{-1} (\partial_\mu W)W^{-1}] + {\rm h.c.} \nonumber \\
& =& -8~ \partial_\mu \mbox{\boldmath$\phi$} \cdot \partial^\mu \mbox{\boldmath$\phi$} + 4~(\partial_\mu \omega_{\alpha\beta} ~\partial^\mu \omega^{\alpha\beta}+ \partial_\mu \mbox{\boldmath$\rho$}_{\alpha\beta} \cdot \partial^\mu \mbox{\boldmath$\rho$}^{\alpha\beta}) + O(4) 
\end{eqnarray}
which is not sophisticated enough to contain much interesting physics. Since isospin space and spin space do not mix, there is no interaction among mesons. To connect the isospin space and the spin space while keeping the isospin symmetry intact we can insert $\tau$- and $\sigma$-matrices between the configuration $W$. To write down the more realistic Lagrangian we first evaluate the following terms:
 
\begin{equation}
tr[ \gamma_\theta (\partial^\mu W) W^{-1} \gamma^\theta (\partial_\mu W) W^{-1} +{\rm h.c.}] = 32~ \partial_\mu \mbox{\boldmath$\phi$} \cdot \partial^\nu \mbox{\boldmath$\phi$} +O(4)    \label{eq:v3}
\end{equation}
and
 
\begin{eqnarray}
& & tr[\gamma_\mu (\gamma^\mu W) W^{-1} \gamma_\nu (\partial^\nu W)W^{-1} ]+{\rm h.c.}~~~~~~~~~~~~~~~~~~~~~~~ \nonumber \\
& &~~~~ =8~ \partial_\mu \mbox{\boldmath$\phi$}  \cdot \partial^\mu \mbox{\boldmath$\phi$}  + 4~(\partial_\mu \omega_{\alpha\beta} ~\partial^\mu  \omega^{\alpha\beta} +\partial_\mu \mbox{\boldmath$\rho$}_{\alpha\beta} \cdot   \partial^\mu  \mbox{\boldmath$\rho$}^{\alpha\beta}) \nonumber \\
& &~~~~-16 ~( \partial_\mu \omega^{\mu\nu} ~\partial^\lambda \omega_{\lambda\nu} +  \partial_\mu \mbox{\boldmath$\rho$}^{\mu\nu} \cdot  \partial^\lambda \mbox{\boldmath$\rho$}_{\lambda \nu}) \nonumber \\
& & ~~~~-48 ~\partial_\mu \mbox{\boldmath$\rho$}^{\mu\nu} \cdot (\mbox{\boldmath$\phi$} \times  \partial_\nu \mbox{\boldmath$\phi$} ) - 32 ~\epsilon_{\mu\nu\beta\kappa} ~\mbox{\boldmath$\phi$} \cdot \partial^\mu \mbox{\boldmath$\rho$}_\lambda^\beta ~\partial^\nu \omega^{\lambda\kappa} \nonumber \\
& & ~~~~-8~\mbox{\boldmath$\rho$}^{\alpha\beta} \cdot ( \partial_\mu \mbox{\boldmath$\rho$}_{\alpha\beta} \times \partial_\nu \mbox{\boldmath$\rho$}^{\mu\nu}) \nonumber \\
& & ~~~~-2 ~\epsilon_{\mu\nu\delta\epsilon}~ \epsilon_{\alpha \beta \lambda\kappa} \mbox{\boldmath$\rho$}^{\alpha\beta} \cdot ( \partial^\mu \mbox{\boldmath$\rho$}^{\lambda\kappa} \times \partial^\nu 
\mbox{\boldmath$\rho$}^{\delta\epsilon}) +O(4)  \label{eq:v4}
\end{eqnarray}
and
 
\begin{eqnarray}
& & tr~[\gamma_\theta ~\gamma_\mu (\partial^\mu W) W^{-1} \gamma^\theta ~\gamma_\nu (  \partial^\nu W)W^{-1} ] +{\rm h.c.} \nonumber \\
& & ~= 16 ~ \partial_\mu \mbox{\boldmath$\phi$} \cdot  \partial^\mu \mbox{\boldmath$\phi$} + 8~( \partial_\mu \omega_{\alpha\beta}~ \partial^\mu \omega^{\alpha\beta} +   \partial_\mu \mbox{\boldmath$\rho$}_{\alpha\beta} \cdot \partial^\mu \mbox{\boldmath$\rho$}^{\alpha\beta}) \nonumber \\
& & ~~~ -32~  \partial_\mu \mbox{\boldmath$\rho$}^{\mu\nu} \cdot (\mbox{\boldmath$\phi$} \times  \partial_\nu \mbox{\boldmath$\phi$}) -64~ \epsilon_{\mu\nu\beta\kappa} \mbox{\boldmath$\phi$} \cdot \partial^\mu \mbox{\boldmath$\rho$}_\lambda^\beta ~\partial^\nu \omega^{\lambda\kappa} \nonumber \\
& & ~~~+16~\mbox{\boldmath$\rho$}^{\alpha\beta} \cdot ( \partial_\mu \mbox{\boldmath$\rho$}_{\alpha\beta} \times \partial_\nu \mbox{\boldmath$\rho$}^{\mu\nu}) \nonumber \\
& & ~~~ -4~ \epsilon_{\mu\nu\delta\epsilon}~ \epsilon_{\alpha \beta \lambda\kappa} \mbox{\boldmath$\rho$}^{\alpha\beta} \cdot ( \partial^\mu \mbox{\boldmath$\rho$}^{\lambda\kappa} \times \partial^\nu 
\mbox{\boldmath$\rho$}^{\delta\epsilon}) +O(4)     \label{eq:v5}
\end{eqnarray}
 
We can now combine Eqs.(${\ref{eq:v3}-\ref{eq:v5}}$):

\begin{eqnarray}
& & tr[(\partial^\mu W)W^{-1} (\partial_\mu W)W^{-1}] +\frac{1}{4}
tr[ \gamma_\theta (\partial^\mu W) W^{-1} \gamma^\theta (\partial_\mu W) W^{-1} ] \nonumber \\
& &+~tr[\gamma_\mu (\gamma^\mu W) W^{-1} \gamma_\nu (\partial^\nu W)W^{-1} ] \nonumber \\
& & - ~tr~[\gamma_\theta ~\gamma_\mu (\partial^\mu W) W^{-1} \gamma^\theta ~\gamma_\nu (  \partial^\nu W)W^{-1} ] +{\rm h.c.}= \nonumber \\
& & ~= -8 ~ \partial_\mu \mbox{\boldmath$\phi$} \cdot  \partial^\mu \mbox{\boldmath$\phi$} -16~( \partial_\mu \omega^{\mu\nu}~ \partial^\lambda \omega_{\lambda\nu} +   \partial_\mu \mbox{\boldmath$\rho$}^{\mu\nu} \cdot \partial^\lambda \mbox{\boldmath$\rho$}_{\lambda\nu}) \nonumber \\
& & ~~~ -16~  \partial_\mu \mbox{\boldmath$\rho$}^{\mu\nu} \cdot (\mbox{\boldmath$\phi$} \times  \partial_\nu \mbox{\boldmath$\phi$}) +32~ \epsilon_{\mu\nu\beta\kappa} \mbox{\boldmath$\phi$} \cdot \partial^\mu \mbox{\boldmath$\rho$}_\lambda^\beta ~\partial^\nu \omega^{\lambda\kappa} \nonumber \\
& & ~~~-24~\mbox{\boldmath$\rho$}^{\alpha\beta} \cdot ( \partial_\mu \mbox{\boldmath$\rho$}_{\alpha\beta} \times \partial_\nu \mbox{\boldmath$\rho$}^{\mu\nu}) \nonumber \\
& & ~~~ +2~ \epsilon_{\mu\nu\delta\epsilon}~ \epsilon_{\alpha \beta \lambda\kappa} \mbox{\boldmath$\rho$}^{\alpha\beta} \cdot ( \partial^\mu \mbox{\boldmath$\rho$}^{\lambda\kappa} \times \partial^\nu 
\mbox{\boldmath$\rho$}^{\delta\epsilon}) +O(4)     \label{eq:v6}
\end{eqnarray}

Now using Eq.(${\ref{eq:v2}}$) in $W$ and Eq.($\ref{eq:v6}$) we have

\begin{equation}
W= exp\{\frac{2i}{F_\pi} \gamma_5 \mbox{\boldmath$\tau$}\cdot \mbox{\boldmath$ \phi$} + \frac{1}{2}\frac{g_\omega}{M_\omega^2} \sigma_{\mu\nu} \omega^{\mu\nu} + \frac{1}{2} \frac{g_\rho}{M_\rho^2}\sigma_{\mu\nu} \mbox{\boldmath$\tau$}\cdot \mbox{\boldmath$\rho$}^{\mu\nu}\}   
\end{equation}
and

\begin{eqnarray}
& &\frac{F_\pi^2}{64}\{ tr[(\partial^\mu W)W^{-1} (\partial_\mu W)W^{-1}] +\frac{1}{4}
tr[ \gamma_\theta (\partial^\mu W) W^{-1} \gamma^\theta (\partial_\mu W) W^{-1} ] \nonumber \\
& &+~tr[\gamma_\mu (\gamma^\mu W) W^{-1} \gamma_\nu (\partial^\nu W)W^{-1} ] \nonumber \\
& & - ~tr~[\gamma_\theta ~\gamma_\mu (\partial^\mu W) W^{-1} \gamma^\theta ~\gamma_\nu (  \partial^\nu W)W^{-1} ] +{\rm h.c.}\} \nonumber \\
& & ~= -\frac{1}{2} ~ \partial_\mu \mbox{\boldmath$\phi$} \cdot  \partial^\mu \mbox{\boldmath$\phi$} -\frac{F_\pi^2~g_\omega^2}{4 M_\omega^4}~( \partial_\mu \omega^{\mu\nu}~ \partial^\lambda \omega_{\lambda\nu}) 
-\frac{F_\pi^2~g_\rho^2}{4M_\rho^4}~ \partial_\mu \mbox{\boldmath$\rho$}^{\mu\nu} \cdot \partial^\lambda \mbox{\boldmath$\rho$}_{\lambda\nu} \nonumber \\
& & ~~~ -\frac{g_\rho}{M_\rho^2}~  \partial_\mu \mbox{\boldmath$\rho$}^{\mu\nu} \cdot (\mbox{\boldmath$\phi$} \times  \partial_\nu \mbox{\boldmath$\phi$}) +\frac{F_\pi~g_\omega~g_\rho}{M_\omega^2~M_\rho^2}~ \epsilon_{\mu\nu\beta\kappa} \mbox{\boldmath$\phi$} \cdot \partial^\mu \mbox{\boldmath$\rho$}_\lambda^\beta ~\partial^\nu \omega^{\lambda\kappa} \nonumber \\
& & ~~~-\frac{3}{8}~\frac{F_\pi^2~g_\rho^3}{M_\rho^6}~\mbox{\boldmath$\rho$}^{\alpha\beta} \cdot ( \partial_\mu \mbox{\boldmath$\rho$}_{\alpha\beta} \times \partial_\nu \mbox{\boldmath$\rho$}^{\mu\nu}) \nonumber \\
& & ~~~ +\frac{F_\pi^2~g_\rho^3}{32~M_\rho^6}~ \epsilon_{\mu\nu\delta\epsilon}~ \epsilon_{\alpha \beta \lambda\kappa} \mbox{\boldmath$\rho$}^{\alpha\beta} \cdot ( \partial^\mu \mbox{\boldmath$\rho$}^{\lambda\kappa} \times \partial^\nu 
\mbox{\boldmath$\rho$}^{\delta\epsilon}) +O(4)     \label{eq:v7}
\end{eqnarray}

Since the masses of vector mesons are significant in comparison to the baryonic masses, we must include the meson masses in a more realistic Lagrangian. Two mass terms comes from

\begin{eqnarray}
& &tr[\gamma_\mu W \gamma^\mu W -4] + {\rm h.c.} \nonumber \\
 & &~~~= 16~\omega_{\alpha\beta}~\omega^{\alpha\beta} +16~\mbox{\boldmath$\rho$}_{\alpha\beta} \cdot \mbox{\boldmath$\rho$}^{\alpha\beta} - 16~ \epsilon_{\alpha\beta\lambda\kappa} \mbox{\boldmath$\phi$} \cdot \mbox{\boldmath$\rho$}^{\alpha\beta}~\omega^{\lambda\kappa} + O(4)
\end{eqnarray}
which gives masses to the (spin triplets) vector mesons $\omega$ and $\rho$ while leaving the $\pi$ massless (spin singlet). Using Eq.(${\ref{eq:v2}}$) we get

\begin{eqnarray}
& &-\frac{F_\pi^2~M^2}{128} ~tr[\gamma_\mu W \gamma^\mu W -4] + {\rm h.c.} \nonumber \\
 & &~~~= - \frac{F_\pi^2~M^2~g_\omega^2}{8~M_\omega^4}~\omega_{\alpha\beta}~\omega^{\alpha\beta} -\frac{F_\pi^2~ M^2~g_\rho^2}{8~M_\rho^4}~\mbox{\boldmath$\rho$}_{\alpha\beta} \cdot \mbox{\boldmath$\rho$}^{\alpha\beta} \nonumber \\      & & + \frac{F_\pi~M^2~g_\rho~g_\omega}{4~M_\rho^2~M_\omega^2}~ \epsilon_{\alpha\beta\lambda\kappa} \mbox{\boldmath$\phi$} \cdot \mbox{\boldmath$\rho$}^{\alpha\beta}~\omega^{\lambda\kappa} + O(4)
\end{eqnarray}

We now evaluate the term

\begin{equation}
tr[W^{-1}~\tau_i~W~\tau_i -3]= -32 ~\mbox{\boldmath$\phi$} \cdot \mbox{\boldmath$\phi$} +16~\mbox{\boldmath$\rho$}^{\alpha\beta} \cdot  \mbox{\boldmath$\rho$}_{\alpha\beta} + O(4)
\end{equation}
and again using Eq.($\ref{eq:v2}$) we arrive at 

\begin{equation}
\frac{F_\pi^2~ m_\pi^2}{256}~tr[W^{-1}~\tau_i~W~\tau_i -3]= -\frac{1}{2}~m_\pi^2~  ~\mbox{\boldmath$\phi$} \cdot \mbox{\boldmath$\phi$} +\frac{F_\pi^2~ g_\rho^2~ m_\pi^2}{16~M_\rho^4}~\mbox{\boldmath$\rho$}^{\alpha\beta} \cdot  \mbox{\boldmath$\rho$}_{\alpha\beta} + O(4)
\end{equation}
a term which shifts the masses of the isospin triplets $\pi$ and $\rho$ with respect to the isospin singlet $\omega$. We now combine all these into an approximate meson Lagrangian, expanded to third order, and containing $\rho \pi\pi$, $\omega\rho\pi$ and $\rho\rho\rho$ interactions:

\begin{eqnarray}
{\cal{L}}&=& -\frac{1}{2} ~ \partial_\mu \mbox{\boldmath$\phi$} \cdot  \partial^\mu \mbox{\boldmath$\phi$} -\frac{F_\pi^2~g_\omega^2}{4 M_\omega^4}~( \partial_\mu \omega^{\mu\nu}~ \partial^\lambda \omega_{\lambda\nu}) 
-\frac{F_\pi^2~g_\rho^2}{4M_\rho^4}~ \partial_\mu \mbox{\boldmath$\rho$}^{\mu\nu} \cdot \partial^\lambda \mbox{\boldmath$\rho$}_{\lambda\nu} \nonumber \\
& & ~~~ -\frac{g_\rho}{M_\rho^2}~  \partial_\mu \mbox{\boldmath$\rho$}^{\mu\nu} \cdot (\mbox{\boldmath$\phi$} \times  \partial_\nu \mbox{\boldmath$\phi$}) +\frac{F_\pi~g_\omega~g_\rho}{M_\omega^2~M_\rho^2}~ \epsilon_{\mu\nu\beta\kappa} \mbox{\boldmath$\phi$} \cdot \partial^\mu \mbox{\boldmath$\rho$}_\lambda^\beta ~\partial^\nu \omega^{\lambda\kappa} \nonumber \\
& & ~~~-\frac{3}{8}~\frac{F_\pi^2~g_\rho^3}{M_\rho^6}~\mbox{\boldmath$\rho$}^{\alpha\beta} \cdot ( \partial_\mu \mbox{\boldmath$\rho$}_{\alpha\beta} \times \partial_\nu \mbox{\boldmath$\rho$}^{\mu\nu})-\frac{1}{2}~m_\pi^2~\mbox{\boldmath$\phi$} \cdot \mbox{\boldmath$\phi$} \nonumber \\
& & ~~~ +\frac{F_\pi^2~g_\rho^3}{32~M_\rho^6}~ \epsilon_{\mu\nu\delta\epsilon}~ \epsilon_{\alpha \beta \lambda\kappa} \mbox{\boldmath$\rho$}^{\alpha\beta} \cdot ( \partial^\mu \mbox{\boldmath$\rho$}^{\lambda\kappa} \times \partial^\nu 
\mbox{\boldmath$\rho$}^{\delta\epsilon}) \nonumber \\
& &
- \frac{F_\pi^2~M^2~g_\omega^2}{8~M_\omega^4}~\omega_{\alpha\beta}~\omega^{\alpha\beta} -\frac{F_\pi^2~ g_\rho^2}{16~M_\rho^4}~(2M^2-m_\pi^2)~\mbox{\boldmath$\rho$}_{\alpha\beta} \cdot \mbox{\boldmath$\rho$}^{\alpha\beta} \nonumber \\      & & + \frac{F_\pi~M^2~g_\rho~g_\omega}{4~M_\rho^2~M_\omega^2}~ \epsilon_{\alpha\beta\lambda\kappa} \mbox{\boldmath$\phi$} \cdot \mbox{\boldmath$\rho$}^{\alpha\beta}~\omega^{\lambda\kappa} + O(4)  \label{eq:v8}
\end{eqnarray}
Note that some of the total divergences have been subtracted. 

From the kinetic and mass terms of the Lagrangian we get

\begin{equation}
F_\pi^2 ~g_\omega^2 =2~M_\omega^2
\end{equation}
and

\begin{equation}
F_\pi^2~g_\rho^2 =2~M_\rho^2
\end{equation}
which is the KSRF$^{\cite{KSRF}}$ relation, and substituting

\begin{equation}
M=M_\omega
\end{equation}
we have mass relations for the mesons

\begin{equation}
M_\omega^2 -\frac{1}{2}~m_\pi^2=M_\rho^2   \label{eq:v9}
\end{equation}
in perfect agreement with experiment.

If we identify the $\omega$ and $\rho$ fields as 

\begin{equation}
\partial_\mu~\omega_{\mu\nu}=M_\omega^2~\omega^\nu
\end{equation}
and

\begin{equation}
\partial_\mu~\mbox{\boldmath$\rho$}^{\mu\nu}= M_\rho^2 ~\mbox{\boldmath$\rho$}^\nu
\end{equation}
up to second order, Lagrangian only of free fields ($\omega$ mesons)

\begin{eqnarray}
{\cal{L}}&=& -\frac{1}{2}~\partial_\mu \mbox{\boldmath$\phi$} \cdot \partial_\mu \mbox{\boldmath$\phi$}- \frac{1}{2}~m_\pi^2  \mbox{\boldmath$\phi$} \cdot \mbox{\boldmath$\phi$} \nonumber \\
& & -\frac{1}{2M_\omega^2}~ \partial_\mu \omega^{\mu\nu}~\partial^\lambda \omega_{\lambda\nu} -\frac{1}{4}~ \omega_{\alpha\beta}~\omega^{\alpha\beta} \nonumber \\
& & -\frac{1}{2M_\rho^2}~ \partial_\mu \mbox{\boldmath$\rho$}^{\mu\nu} \cdot \partial^\lambda \mbox{\boldmath$\rho$}_{\lambda\nu} -\frac{1}{4}~ \mbox{\boldmath$\rho$}_{\alpha\beta} \cdot \mbox{\boldmath$\rho$}^{\alpha\beta}      
\end{eqnarray}

Take

\begin{equation}
\partial_\mu \omega^{\mu\nu}= M_\omega^2~\omega^\nu,~~~~~~\partial_\mu \mbox{\boldmath$\rho$}^{\mu\nu}=M_\rho^2~ \mbox{\boldmath$\rho$}^\nu   
\end{equation}

So the equations of motion from ${\cal{L}}$ are

\begin{equation}
\partial^\mu \omega^\nu -\partial^\nu \omega^\mu = \omega^{\mu\nu} 
\end{equation}

\begin{equation}
\partial^\mu \mbox{\boldmath$\rho$}^\nu -  \partial^\nu \mbox{\boldmath$\rho$}^\mu= \mbox{\boldmath$\rho$}^{\mu\nu} 
\end{equation}

\begin{equation}
\Box \mbox{\boldmath$\phi$} = m_\pi^2~ \mbox{\boldmath$\phi$} 
\end{equation}

\section*{Vector Mesons as Gauge Bosons}
Gauge potentials can be obtained from the decomposition of a larger algebra into a subalgebra and a coset. Gauge bosons can be combined with the fundamental mesons to form a single multiplet so that all mesons are treated equally in the larger symmetry group. 

We can write $U$ in terms of the new variables $\Xi_L$ and $\Xi_r$ 

\begin{equation}
U=\Xi_L~\Xi_R
\end{equation}
Under the group

\begin{equation}
SU(2)_{L,{\rm global}} \times SU(2)_{R, {\rm global}} \times SU(2)_{\rm local}
\end{equation}
$\Xi_{L,R}$ transform as

\begin{equation}
\Xi_L \rightarrow \theta(\mbox{\boldmath$r$})~\Xi_L~A^\dag
\end{equation}

\begin{equation}
\Xi_R \rightarrow \theta(\mbox{\boldmath$r$})~\Xi_R~B^\dag
\end{equation}
so that the Lagrangian 

\begin{equation}
{\cal{L}}= -\frac{F_\pi}{16}~tr(\partial_\mu U^\dag~\partial^\mu U)  \label{eq:v12}
\end{equation}
is invariant under the global symmetry $A\rightarrow A~U~B^\dag$, with $A\in SU(2)_{L, {\rm global}}$, $B\in SU(2)_{R, {\rm global}}$, and $\theta(\mbox{\boldmath$r$}) \in SU(2)_{\rm local}$. Since $U$ transforms as $U \rightarrow A~U~B^\dag$ without any explicit $\theta(\mbox{\boldmath$r$})$ dependence, the symmetry is said to be "hidden". The associated gauge field 

\begin{equation}
V_\mu= \frac{1}{2} ~\mbox{\boldmath$\tau$} \cdot \mbox{\boldmath$V$}_\mu
\end{equation}
transforms as 

\begin{equation}
V_\mu \rightarrow \frac{i}{g} ~ \theta(\mbox{\boldmath$r$})~\partial_\mu ~\theta(\mbox{\boldmath$r$}) + \theta(\mbox{\boldmath$r$})~V_\mu~\theta^\dag (\mbox{\boldmath$r$})
\end{equation}
We can now construct a Lagrangian that respects this gauge symmetry. Writing the covariant derivative

\begin{equation}
D_\mu=\partial_\mu -i~g~V_\mu
\end{equation}
where $g$ is the gauge coupling, we can write two independent second order terms that are gauge invariant:

\begin{equation}
{\cal{L}}_V= -\frac{F_\pi}{16}~tr[(D_\mu \Xi_L)~\Xi_L^\dag + (D_\mu \Xi_R)~\Xi_R^\dag]^2
\end{equation}

\begin{equation}
{\cal{L}}_A= -\frac{F_\pi}{16}~tr[(D_\mu \Xi_L)~\Xi_L^\dag - (D_\mu \Xi_R)~\Xi_R^\dag]^2
\end{equation}
so that a combination of the form 

\begin{equation}
{\cal{L}}= {\cal{L}}_A~+ a ~{\cal{L}}_V
\end{equation}
is equivalent to the non-linear sigma model Eq.(\ref{eq:v12}). Here $a$ is an arbitrary parameter. Using unitary gauge

\begin{equation}
\Xi_L^\dag=\Xi_R=\Xi=U^{\frac{1}{2}}
\end{equation}
we have

\begin{equation}
{\cal{L}}_A= -\frac{F_\pi^2}{16}~tr(\partial_\mu U^\dag \partial^\mu U)
\end{equation}

\begin{equation}
{\cal{L}}_V= -\frac{F_\pi^2}{16}~tr \{g~V_\mu -\frac{i}{2}~[(\partial_\mu \Xi)~\Xi^\dag + (\partial_\mu \Xi^\dag)~\Xi]\}^2
\end{equation}
We see that when we use the equation of motion for ${\cal{L}}_V$

\begin{equation}
g~V_\mu^a = -i~tr\{ \frac{\tau^a}{2} ~[(\partial_\mu \Xi)~\Xi^\dag + (\partial_\mu \Xi^\dag)~\Xi]\}
\end{equation}
${\cal{L}}_V$ vanishes and ${\cal{L}}$ becomes identical to ${\cal{L}}_A$. Making the identification of vector fields $V_\mu$ with the $\rho$-meson fields $\rho_\mu$, we add by hand the kinetic term of $\rho$, so that

\begin{equation}
{\cal{L}}={\cal{L}}_A + a~{\cal{L}}_V -\frac{1}{4}~ \mbox{\boldmath$\rho$}_{\mu\nu}\cdot \mbox{\boldmath$\rho$}^{\mu\nu}
\end{equation}
Again, expanding to third term, using the weak field expansion we get the Lagrangian

\begin{eqnarray}
{\cal{L}}&=& -\frac{F_\pi^2}{16}~tr(\partial_\mu U^\dag~\partial^\mu U) -\frac{1}{4} ~\mbox{\boldmath$\rho$}_{\mu\nu} \cdot \mbox{\boldmath$\rho$}^{\mu\nu} \nonumber \\
& & -\frac{a}{8}~g^2~F_\pi^2~ \mbox{\boldmath$\rho$}_{\mu} \cdot \mbox{\boldmath$\rho$}^{\mu} - \frac{a}{2} ~g~\mbox{\boldmath$\rho$}_{\mu} \cdot (\mbox{\boldmath$\rho$}_{\phi} \times \partial^\mu \mbox{\boldmath$\rho$}_{\phi}) +\cdots
\end{eqnarray}
The mass of the $\rho$ meson is given by

\begin{equation}
M_\rho^2 =\frac{a}{4}~g^2~F_\pi^2
\end{equation}
which is identical to earlier value when $a=2$. Overall result of this approach agrees with experiment very well. 

Even though this description we just gave works very well with vector mesons, it would be more useful if we could find a way of including scalars and vectors in a unified manner. Toward achieving such a goal, we can now consider the genesis of covariant derivatives through a decomposition of a larger algebra involving spin or Dirac operators, into a subalgebra and a coset. The $\sigma$-model obtained from the decomposition is valued in the subalgebra of the larger algebra, while the gauge fields are valued in the coset space. Once we identify the derivative of the coset elements as vector gauge potentials, the derivative of the larger algebra configuration can be written as a covariant derivative, i.e., a derivative of the subalgebra element plus a gauge potential. Let us rewrite $W$ as

\begin{equation}
W(\phi,\omega,\rho)=V(\omega, \rho)~U(\phi) 
\end{equation}
where 

\begin{equation}
U(\phi)= exp\{\frac{2i}{F_\pi}~\gamma_5~\mbox{\boldmath$\tau$} \cdot \mbox{\boldmath$\phi$} \}
\end{equation}
is a configuration of pion fields only, and

\begin{equation}
V(\omega,\rho)= exp\{ \frac{g_\omega}{2M_\omega^2}~\sigma_{\alpha\beta}~\omega^{\alpha\beta} + \frac{g_\rho}{2M_\rho^2}~\sigma_{\alpha\beta}~ \mbox{\boldmath$\tau$} \cdot \mbox{\boldmath$\rho$}^{\alpha\beta} \}
\end{equation}
is a configuration of vector meson fields only. We also note that

\begin{eqnarray}
W^{-1} \partial_\mu W &=& U^\dag V^{-1} \partial_\mu (VU)= U^\dag V^{-1} [V\partial_\mu U +(\partial_\mu V)U] \nonumber \\
& & U^\dag (\partial_\mu + V^{-1} \partial_\mu V)U = U^\dag D_\mu U
\end{eqnarray}

Therefore, the covariant derivative
 
\begin{equation}
D_\mu= \partial_\mu + V^{-1} \partial_\mu V
\end{equation}
arises naturally with vector mesons being the gauge potentials $V^{-1} \partial_\mu V$.

Under this decomposition, the ${\cal{L}}$ of the non-linear $\sigma$-model becomes

\begin{equation}
{\cal{L}}\sim tr[W^{-1} (\partial_\mu W) W^{-1} (\partial^\mu W)] \rightarrow tr[U^\dag (D_\mu U) U^\dag (D^\mu U)]
\end{equation}
We see that the ordinary derivatives are replaced by the covariant derivatives, and the field configuration $W$ is replaced by the field configuration $U$ that is valued in the subalgebra. But there is one problem: we note that

\begin{equation}
V^{-1}\partial_\mu V \sim \sigma_{\alpha\beta}~\partial_\mu F^{\alpha\beta}
\end{equation}
where $F$ is a function of the vector meson fields, has a spin matrix $\sigma_{\alpha\beta}$ dependence but $U^\dag \partial_\mu U$ has none. Therefore the trace of the matrix with scalar-vector meson interaction terms is zero. To generate non-zero values of such terms it is necessary to add $\sigma$- and $\gamma$-matrices to ${\cal{L}}$. For simplicity, we now take

\begin{equation}
V(\rho)=exp\{ \frac{g_\rho}{2M_\rho^2}~\sigma_{\alpha\beta} ~\mbox{\boldmath$\tau$} \cdot \mbox{\boldmath$\rho$}^{\alpha\beta} \}
\end{equation}
to be a configuration of $\rho$-mesons only. The covariant derivative is

\begin{equation}
D_\mu= \partial_\mu + \frac{g_\rho}{2M_\rho^2}~\sigma_{\alpha\beta} ~\mbox{\boldmath$\tau$} \cdot \mbox{\boldmath$\rho$}^{\alpha\beta}+\ldots 
\end{equation}

If we now define 

\begin{equation}
\Gamma_\alpha= V^{-1} \gamma_\alpha V
\end{equation}
and

\begin{equation}
\Sigma_{\alpha\beta} = V^{-1} ~\sigma_{\alpha\beta} V
\end{equation}
such that they satisfy the same Dirac algebra as the original $\tau$- and $\sigma$-matrices, namely

\begin{equation}
\{ \Gamma_\alpha, \Gamma_\beta \}= \{ \gamma_\alpha, \gamma_\beta \}= \eta_{\alpha\beta}
\end{equation}
and

\begin{eqnarray}
[\Sigma_{\alpha\beta}, \Sigma_{\mu\nu} ] &=& V^{-1}~[\sigma_{\alpha\beta}, \sigma_{\mu\nu} ]~V \nonumber\\
& &=2i~V^{-1}~(\eta_{\alpha\mu} \Sigma_{\beta\nu} +\eta_{\beta\nu} \Sigma_{\alpha\mu} - \eta_{\alpha\nu} \Sigma_{\beta\mu}  - \eta_{\beta\mu} \Sigma_{\alpha\nu}) V \nonumber \\ 
& & =2i~V^{-1}~(\eta_{\alpha\mu} \Sigma_{\beta\nu} +\eta_{\beta\nu} \Sigma_{\alpha\mu} - \eta_{\alpha\nu} \Sigma_{\beta\mu}  - \eta_{\beta\mu} \Sigma_{\alpha\nu})  
\end{eqnarray}
Using these in Eq.(\ref{eq:v8}) after some lengthy algebra we arrive at a weak field expansion of the Lagrangian

\begin{eqnarray}
{\cal{L}}&=& -\frac{1}{2}~ \partial_\mu \mbox{\boldmath$\phi$} \cdot \partial^\mu \mbox{\boldmath$\phi$} -\frac{1}{4}~\mbox{\boldmath$\rho$}_{\mu\nu} \cdot \mbox{\boldmath$\rho$}^{\mu\nu} \nonumber \\
& & -\frac{1}{2}~M_\rho^2~ \mbox{\boldmath$\rho$}_\mu \cdot  \mbox{\boldmath$\rho$}^\mu - g_\rho~ \mbox{\boldmath$\rho$}_\mu \cdot ( \mbox{\boldmath$\phi$} \times \partial^\mu \mbox{\boldmath$\phi$}) +\ldots 
\end{eqnarray}
where we have made the identification $\partial_\mu \rho_\alpha^{\mu\nu} = M_\rho^2~\rho_\alpha^\nu$, and obtained the KSRF relation. It is easy to see that this result agrees with hidden symmetry approach we discussed. Therefore, a realistic connection between covariant derivatives and the coset elements of the larger algebra has been demonstrated through this correspondence.

Another example of obtaining covariant derivatives is from the consideration of a chiral Lagrangian that includes the meson nucleon interactions. The Lagrangian

\begin{equation}
{\cal{L}}= -\bar{\psi}~\gamma_\mu \partial^\mu \psi - m~\bar{\psi} W \psi +{\rm meson~~ terms}
\end{equation}
is invariant under the chiral transformation. There is no mass $m~\bar{\psi} \psi$ for the fermions because such a term is not chiral invariant, hence the fermion field $\psi$ can not represent a realistic physical nucleon, which has to be massive. However, by performing a unitary transformation involving the meson field we can define a field $\xi$ 

\begin{equation}
\xi=W^{\frac{1}{2}}~\psi =exp(\frac{2i}{F_\pi}~\gamma_5 \mbox{\boldmath$\tau$} \cdot \mbox{\boldmath$\phi$} + \frac{g_\omega}{2M_\omega^2}~\sigma_{\alpha\beta} \omega^{\alpha\beta} +\frac{g_\rho}{2M_\rho^2}~\sigma_{\alpha\beta} \mbox{\boldmath$\tau$} \cdot \mbox{\boldmath$\rho$}^{\alpha\beta})~\psi
\end{equation}  
such that the new nucleon mass term $m~\bar{\xi} \xi$ becomes chiral invariant. Since

\begin{equation}
\bar{W}^{\frac{1}{2}}=\gamma_4~W^{\dag\frac{1}{2}}~\gamma_4= W^{\frac{1}{2}}
\end{equation}
we have $\bar{\xi}=\bar{\psi}~W^{\frac{1}{2}}$. Then, the transformed Lagrangian becomes

\begin{equation}
{\cal{L}}= -(\bar{\xi}~W^{-\frac{1}{2}} ) \gamma_\mu \partial^\mu (W^{-\frac{1}{2}}~\xi) -m~\bar{\xi} \xi +\cdots = -\bar{\xi} \gamma_\mu^{'} D^\mu \xi - m~\bar{\xi} \xi + \cdots  \label{eq:v13}
\end{equation}
where $\gamma_\mu^{'}=W^{-\frac{1}{2}} \gamma_\mu W^{-\frac{1}{2}}$ and the covariant derivative $D_\mu=\partial_\mu-A_\mu$ with
\begin{equation}
A_\mu =(\partial_\mu W^{\frac{1}{2}}) W^{-\frac{1}{2}} = \frac{i}{F_\pi}~\gamma_5 \mbox{\boldmath$\tau$} \cdot \partial_\mu \mbox{\boldmath$\phi$} + \frac{g_\omega}{2}~\omega_\mu + \frac{g_\rho}{2}~\mbox{\boldmath$\tau$} \cdot \mbox{\boldmath$\rho$}_\mu +\cdots
\end{equation}
We see that the Lagrangian in Eq.(\ref{eq:v13}) gives a Dirac equation for the nucleons ($\gamma_\mu^{'}~D^\mu \xi= m~\xi$, which contains both the vector and the pseudoscalar meson-nucleon interactions through the covariant derivatives.

These examples show that gauge potentials can be obtained from the decompositions of larger algebra into a subalgebra and a coset. Gauge bosons, which take on special roles in the model originally, can be combined with the fundamental mesons to form a single multiplet so that all mesons are treated equally in the larger symmetry group. 

\section*{Fermionic Quantization}
Earlier we quantized the skyrmion by rotating the soliton solution by the $SU(2)$ time-dependent matrix $A(t)$. The canonical conjugate matrix of $A$ is given by

\begin{equation}
{\cal{A}}= \partial L/ \partial {\dot{A}}_{\alpha\beta}
\end{equation}
and the Hamiltonian is obtained from

\begin{equation}
H= {\cal{A}}^{\alpha\beta} A_{\alpha\beta} - L
\end{equation}
Although at the classical level $A_{\alpha\beta}$ and ${\cal{A}}^{\alpha\beta}$ are commuting variables, after quantization they become operators and are no longer commuting. We can either perform bosonic quantization
 
\begin{equation}
[A_{\alpha\beta}, {\cal{A}}^{\alpha\beta}] =i\delta_\alpha^\beta \delta_\beta^\nu
\end{equation}

\begin{equation}
[A_{\alpha\beta}, A_{\mu\nu}]=[{\cal{A}}, {\cal{A}}]=0
\end{equation}
or fermionic quantization

\begin{equation}
\{A_{\alpha\beta}, {\cal{A}}^{\alpha\beta}\} =i\delta_\alpha^\beta \delta_\beta^\nu
\end{equation}

\begin{equation}
\{A_{\alpha\beta}, A_{\mu\nu}\}=\{{\cal{A}}, {\cal{A}}\}=0
\end{equation}
that give the same algebra of the spin operators formed by the $A$'s and ${\cal{A}}$'s. In the bosonic quantization, operators $A_{\alpha\beta}$ commute among themselves, allowing particle states be described in terms of power functions in $A$ of arbitrary orders. On the other hand, in the case of fermionic quantization, the $A_{\alpha\beta}$ anticommute among themselves so that states with two or more identical $A_{\alpha\beta}$ do not exist, and the powers in $A$ of the state functions are limited. In the case when we take $A$ to be fermionic, then the cubic functions for $\Delta^{++}$ vanishes.

To overcome this difficulty, we consider the same problem which occurred in the quark model when the problem arose of putting three spin up $u$ quarks together to form $s=3/2$ $\Delta^{++}$ state. To conform with Pauli exclusion principle, three colors had to be introduced in the quark model to make the state antisymmetric in permuting any two quarks. In a similar sense, we now triple the number of variables in the Skyrme model by introducing two more matrices $B$ and $C$ in addition to the matrix $A$, and require the new Lagrangian to be invariant under the permutations of $A$, $B$ and $C$. The color group $SU(3)$ is replaced by the discrete subgroup $Z_3$. In general for $1/N$ expansion, the color group $SU(N)$ is replaced by $Z_N$, and there will be $N$ different matrices. We propose a new Lagrangian that accommodates the new degrees of freedom

\begin{eqnarray}
{\cal{L}}&=& \frac{1}{3!} \{ - \frac{F_\pi^2}{16}~tr (\partial_\mu U_{AB}^\dag~\partial^\mu U_{AB}) + \frac{1}{32 e^2} ~
tr([U_{AB}^\dag~\partial_\mu U_{AB} , U_{AB}^\dag~\partial_\nu U_{AB}]^2) \nonumber \\
& & ~~~~~~~~+{\rm cyclic~permutations~in}~ A, B, C \}
\end{eqnarray}

We note that the time dependent parts of ${\cal{L}}$ remain unchanged, so we only have to concentrate on the time derivatives of 

\begin{eqnarray}
U_{AB}({\mbox{\boldmath$r$}},t)=A(t) U_0({\mbox{\boldmath$r$}})B^\dag(t)   \nonumber \\
U_{BC}({\mbox{\boldmath$r$}},t)=B(t) U_0({\mbox{\boldmath$r$}})C^\dag(t)    \nonumber \\
U_{CA}({\mbox{\boldmath$r$}},t)=C(t) U_0({\mbox{\boldmath$r$}})A^\dag(t).
\end{eqnarray}

Writing 
\begin{eqnarray}
\mbox{\boldmath$s$}_A = - \frac{i}{2}~tr(\mbox{\boldmath$\tau$} A^\dag {\dot{A}}) \nonumber \\
\mbox{\boldmath$s$}_B = - \frac{i}{2}~tr(\mbox{\boldmath$\tau$} B^\dag {\dot{B}}) \nonumber \\
\mbox{\boldmath$s$}_C = - \frac{i}{2}~tr(\mbox{\boldmath$\tau$} C^\dag {\dot{C}}) 
\end{eqnarray}
in the same way as before, after some tedious algebra, we find that the time-dependent parts of the non-linear sigma model terms are given by

\begin{eqnarray}
tr(\partial_0U_{AB}~\partial_0U_{AB}^\dag ) &=& 2(\mbox{\boldmath$s$}_A -\mbox{\boldmath$s$}_B)^2 - 8 sinF~cosF~\epsilon_{abc}~s_A^a~s_B^b~{\hat{r}}^c  \nonumber \\
&  & ~+8 sin^2F (\mbox{\boldmath$s$}_A \cdot \mbox{\boldmath$s$}_B - \mbox{\boldmath$s$}_A \cdot {\hat{\mbox{\boldmath$r$}}}~\mbox{\boldmath$s$}_B \cdot {\hat{\mbox{\boldmath$r$}}})     
\end{eqnarray}

\begin{eqnarray}
tr(\partial_0U_{BC}~\partial_0U_{BC}^\dag ) &=& 2(\mbox{\boldmath$s$}_B -\mbox{\boldmath$s$}_C)^2 - 8 sinF~cosF~\epsilon_{abc}~s_B^a~s_C^b~{\hat{r}}^c  \nonumber \\
&  & ~+8 sin^2F (\mbox{\boldmath$s$}_B \cdot \mbox{\boldmath$s$}_C - \mbox{\boldmath$s$}_B \cdot {\hat{\mbox{\boldmath$r$}}}~\mbox{\boldmath$s$}_C \cdot {\hat{\mbox{\boldmath$r$}}})     
\end{eqnarray}

\begin{eqnarray}
tr(\partial_0U_{CA}~\partial_0U_{CA}^\dag ) &=& 2(\mbox{\boldmath$s$}_C -\mbox{\boldmath$s$}_A)^2 - 8 sinF~cosF~\epsilon_{abc}~s_C^a~s_A^b~{\hat{r}}^c  \nonumber \\
&  & ~+8 sin^2F (\mbox{\boldmath$s$}_C \cdot \mbox{\boldmath$s$}_A - \mbox{\boldmath$s$}_C \cdot {\hat{\mbox{\boldmath$r$}}}~\mbox{\boldmath$s$}_A \cdot {\hat{\mbox{\boldmath$r$}}})     
\end{eqnarray}
The time-dependent part of the Skyrme term can also be similarly calculated. After some more lengthy algebra, we obtain the Lagrangian

\begin{eqnarray}
L &=& -M + \frac{2}{3} \frac{F_\pi^2}{16}~ {\tilde{V}} [(\mbox{\boldmath$s$}_A -\mbox{\boldmath$s$}_B)^2 + (\mbox{\boldmath$s$}_B -\mbox{\boldmath$s$}_C)^2 + (\mbox{\boldmath$s$}_C -\mbox{\boldmath$s$}_A)^2   \nonumber \\
& & ~+ \frac{2}{3} \lambda [ (\mbox{\boldmath$s$}_A \cdot \mbox{\boldmath$s$}_B)+ (\mbox{\boldmath$s$}_B \cdot \mbox{\boldmath$s$}_C)+ (\mbox{\boldmath$s$}_C \cdot \mbox{\boldmath$s$}_A) ] 
\end{eqnarray}
where  $M$ and $\lambda$ are given by Eq.(\ref{eq:M}) and Eq.(\ref{eq:2s}), respectively, and ${\tilde{V}}=\int d^3r$ is the size of the $3$-space to be set equal to infinity after the Legendre transformation of the Hamiltonian. Defining the canonical conjugates

\begin{eqnarray}
{\cal{A}}^{\alpha\beta}= \frac{\partial L}{\partial {\dot{A}}_{\alpha\beta}}  \nonumber \\
{\cal{B}}^{\alpha\beta}= \frac{\partial L}{\partial {\dot{B}}_{\alpha\beta}}  \nonumber \\
{\cal{C}}^{\alpha\beta}= \frac{\partial L}{\partial {\dot{C}}_{\alpha\beta}}   
\end{eqnarray}
and the corresponding spin operators

\begin{eqnarray}
\mbox{\boldmath$S$}_A = - \frac{i}{2}~tr (\mbox{\boldmath$\tau$} {\cal{A}}^\dag A ) \nonumber \\
\mbox{\boldmath$S$}_B = - \frac{i}{2}~tr (\mbox{\boldmath$\tau$} {\cal{B}}^\dag B ) \nonumber \\
\mbox{\boldmath$S$}_C = - \frac{i}{2}~tr (\mbox{\boldmath$\tau$} {\cal{C}}^\dag C )
\end{eqnarray}
we obtain the Hamiltonian

\begin{eqnarray}
H &=& {\cal{A}}^{\alpha\beta} {\dot{A}}_{\alpha\beta} + {\cal{B}}^{\alpha\beta} {\dot{B}}_{\alpha\beta} + {\cal{C}}^{\alpha\beta} {\dot{C}}_{\alpha\beta} - L   \nonumber \\
&=& M + \frac{F_\pi^2}{16}~\tilde{V}~\frac{6}{(6 \tilde{V} - \lambda)^2} [(\mbox{\boldmath$S$}_A -\mbox{\boldmath$S$}_B)^2 + (\mbox{\boldmath$S$}_B -\mbox{\boldmath$S$}_C)^2 + (\mbox{\boldmath$S$}_C -\mbox{\boldmath$S$}_A)^2 ]  \nonumber \\
& & +\{[ (2 \tilde{V} +\lambda) \mbox{\boldmath$S$}_A + (2 \tilde{V} -\lambda) \mbox{\boldmath$S$}_B + (2 \tilde{V} -\lambda) \mbox{\boldmath$S$}_C] \times   \nonumber \\
& & ~~~~~[(2 \tilde{V} -\lambda) \mbox{\boldmath$S$}_A + (2 \tilde{V} +\lambda) \mbox{\boldmath$S$}_B + (2 \tilde{V} -\lambda) \mbox{\boldmath$S$}_C]   \nonumber \\
& & +[ (2 \tilde{V} +\lambda) \mbox{\boldmath$S$}_A + (2 \tilde{V} -\lambda) \mbox{\boldmath$S$}_B + (2 \tilde{V} -\lambda) \mbox{\boldmath$S$}_C] \times   \nonumber \\
& & ~~~~~[(2 \tilde{V} -\lambda) \mbox{\boldmath$S$}_A + (2 \tilde{V} -\lambda) \mbox{\boldmath$S$}_B + (2 \tilde{V} +\lambda) \mbox{\boldmath$S$}_C]   \nonumber \\
& & +[ (2 \tilde{V} -\lambda) \mbox{\boldmath$S$}_A + (2 \tilde{V} +\lambda) \mbox{\boldmath$S$}_B + (2 \tilde{V} -\lambda) \mbox{\boldmath$S$}_C] \times   \nonumber \\
& &~~~~~[(2 \tilde{V} -\lambda) \mbox{\boldmath$S$}_A + (2 \tilde{V} -\lambda) \mbox{\boldmath$S$}_B + (2 \tilde{V} +\lambda) \mbox{\boldmath$S$}_C]\} \frac{3}{2\lambda (6 \tilde{V} -\lambda)^2} 
\end{eqnarray}
Finally, letting $\tilde{V} \rightarrow \infty$, we arrive at the Hamiltonian

\begin{equation}
H= M+ \frac{1}{2 \lambda}~ (\mbox{\boldmath$S$}_A +\mbox{\boldmath$S$}_B + \mbox{\boldmath$S$}_C)^2   
\end{equation}
whose eigenstates will be functions of the variables $A$, $B$ and $C$. In analogy to the $SU(4)$ quark model, we can describe the nucleons and the $\Delta$'s as $SU(4)$ multiplets and write their states in terms of functions trilinear in $A$, $B$ and $C$. For example, we can write the state function of spin $3/2$ $\Delta^{++}$ as

\begin{eqnarray}
<A,B,C|\Delta^{++}, S_3=\frac{3}{2})> & \sim & A_{21}B_{21}C_{21} \nonumber  \\
& & ~+ {\rm antisymmetrization~in}~A,B,C   
\end{eqnarray}
and the spin up proton as

\begin{eqnarray}
<A,B,C|p, S_3=\frac{1}{2})> & \sim & 2A_{21}B_{21}C_{12} +2A_{12}B_{21}C_{21}+2A_{21}B_{12}C_{21}  \nonumber  \\
& & -A_{11} B_{22}C_{21} -A_{21}B_{11}C_{22}-A_{11}B_{21}C_{22}  \nonumber  \\
& & -A_{22} B_{11}C_{21} -A_{21}B_{22}C_{11}-A_{22}B_{21}C_{11}  \nonumber \\
& & ~+ {\rm antisymmetrization~in}~A,B,C   
\end{eqnarray}
The problem of the infinite towers of particle states disappears because the Pauli exclusion principle does not allow any more states with $I=S \geq 5/2$.

\section*{First Steps In the Implementation of $SU(4)$ and $SU(4/10)$ Symmetric Skyrmion Program}

In a recent paper Cheung and G\"ursey$^{\cite{3s}}$ showed that generators for $SU(4)$ are constructed from dynamical skyrmion operators. In their paper they gave a fermionic realization of $SU(2)\times SU(2)$ algebra of the Skyrme model. They imbedded the $SU(2)\times SU(2)$ algebra into $U(4)$ by introducing new generators which are simple modifications of the $SU(2) \times SU(2)$ generators. 

The particle wave functions are written in powers of the collective coordinates. The infinite tower of states, which produces one of the problems for the Skyrme model can now be eliminated by restricting the values of these powers. We observe that if the collective coordinates are quantized fermionically, arbitrary functions of $A$ (where $\psi \sim A^P$ with $2S=P$) can be written as polynomials in $A$ of finite degrees, and therefore the height of the tower of particle states will have an upper bound. Another motivation for introducing fermionic quantization is that the predicted baryon masses from the Skyrme model are too high if the experimental value of $F_\pi$ is taken as an input parameter.Modification of the model by addition of new terms will usually increase the particle mass further since usually the new terms give positive contributions to the Hamiltonian. Since in the renormalization procedure of a supersymmetric field theory the contributions coming from fermion loops are negative, such supersymmetric theory will have a better convergence. It is conceivable that the masses may be lowered if one can construct a supersymmetric Skyrme model, and fermionic quantization is a first step in this direction.

Note that the spin and the isospin operators we introduced form the $6$ generators of the $SU(2)\times SU(2)$ algebra. To imbed the algebra into $U(4)$, we introduce the operators

\begin{equation}
\mbox{\boldmath$N$}=i~tr({\cal{A}}^\dag A)
\end{equation}
and

\begin{equation}
E_{ij}= \frac{i}{2} ~tr(\tau_i {\cal{A}}^\dag \tau_j A)
\end{equation}

The form of the operators $\mbox{\boldmath$N$}$ and $E_{ij}$ are easy to obtain as they differ from the operators $\mbox{\boldmath$S$}$ and $\mbox{\boldmath$K$}$ only by a factor of a Pauli matrix. One can easily verify that the 15 operators $\mbox{\boldmath$S$}$, $\mbox{\boldmath$K$}$ and $E_{ij}$ form the generators of $SU(4)$. If we include the operator $\mbox{\boldmath$N$}$, which commutes with all the $SU(4)$ generators, we obtain a $U(4)$ algebra. The $SU(4)$ Casimir operator is given by

\begin{equation}
C= \mbox{\boldmath$S$}^2 + \mbox{\boldmath$K$}^2 + E_{ij} E^{ij}   \label{eq:5s}
\end{equation} 

The fermionic quantization prescription allows us to expand the algebra into superalgebra. The supersymmetric extension of the $U(N)$ algebra and the operators forming the $SU(2)\times SU(2)$ subalgebra can now be obtained. Let $A_\alpha$ be an $N$-vector, with $ {\cal{A}}_\alpha $ its canonical cojugate ($\alpha=1,\ldots,N)$. Together with their complex conjugates they obey the following commutation relations:

\begin{equation}
\{ A_\alpha, A_\beta \}=\{ A_\alpha^*, A_\beta^* \}=\{ {\cal{A}}_\alpha, {\cal{A}}_\beta \}=\{ {\cal{A}}_\alpha^*, {\cal{A}}_\beta^* \}=0
\end{equation}

\begin{equation}
\{A_\alpha, A_\beta^* \}=\{ {\cal{A}}_\alpha, {\cal{A}}_\beta^* \}=0
\end{equation}

\begin{equation}
\{A_\alpha, {\cal{A}}_\beta^* \}=\{A_\alpha^*, {\cal{A}}_\beta \}=0
\end{equation}

\begin{equation}
\{ A_\alpha, {\cal{A}}_\beta \}=\delta_{\alpha \beta},~~~\{ A_\alpha^*, {\cal{A}}_\beta^* \}=\delta_{\alpha \beta}
\end{equation}

Defining the operators $J_{\alpha\beta}$ as

\begin{equation}
J_{\alpha\beta}={\cal{A}}_\alpha A_\beta + {\cal{A}}_\alpha^* A_\beta^*
\end{equation}

They generate the $U(N)$ algebra

\begin{equation}
[J_{\alpha\beta}, J_{\mu\nu}]=\delta_{\beta\mu} J_{\alpha\nu} - \delta_{\alpha\nu} J_{\mu\beta}
\end{equation}

For the case of $N=4$, we keep the track of the imbedding by noting that the operators

\begin{equation}
S_0= \frac{1}{2} ( J_{11}+J_{22} -J_{33} -J_{44})
\end{equation}

\begin{equation}
S_+=\frac{1}{\sqrt2} (J_{13} + J_{24})
\end{equation}

\begin{equation}
S_-= \frac{1}{\sqrt2} (J_{31}+J_{42})
\end{equation}

\begin{equation}
I_0= \frac{1}{2} (J_{11} -J_{22} +J_{33}-J_{44})
\end{equation}

\begin{equation}
I_+=\frac{1}{\sqrt2} (J_{12} +J_{34})
\end{equation}

\begin{equation}
I_-=\frac{1}{\sqrt2} (J_{21}+J_{43})
\end{equation}
form an $SU(2)\times SU(2)$ subalgebra of the Skyrme model. 

So far we have only the even algebraic operators $J_{\alpha\beta}$ that are products of two fermionic operators $A$ and ${\cal{A}}$. To make a supersymmetric extension of the algebra, we need some odd algebraic operators that are made of products of odd number of fermionic operators. The even and odd operators are expected to form a superalgebra.

To obtain the superalgebra we now introduce the (odd) supercharge operators

\begin{equation}
Q_\gamma= \lambda {\cal{A}}_\gamma + \frac{1}{\lambda^*} {\cal{A}}_\mu^* {\cal{A}}_\gamma^* A_\mu^*
\end{equation}
and

\begin{equation}
\overline{Q}_\lambda= \lambda^* A_\lambda^* +\frac{1}{\lambda} {\cal{A}}_\nu A_\lambda A_\nu
\end{equation}
where $\lambda$ is an arbitrary complex parameter. Using the commutation relations we see that $J_{\alpha\beta}$, $Q_\gamma$ and $\overline{Q}_\gamma$ with ($\alpha, \beta, \gamma=1, \ldots, N$) form the superalgebra

\begin{equation}
\{ Q_\gamma, Q_\lambda \}= \{\overline{Q}_\gamma, \overline{Q}_\lambda \}=0
\end{equation}

\begin{equation}
\{ Q_\gamma, \overline{Q}_\lambda \}= J_{\gamma\lambda} -\delta_{\gamma\lambda} J_{\nu\nu}
\end{equation}

\begin{equation}
[J_{\alpha\beta},\overline{Q}_\gamma]=-\delta_{\alpha\gamma} \overline{Q}_\beta
\end{equation}

The quadratic invariant of the superalgebra is

\begin{equation}
C_2= J_{\mu\mu} J_{\nu\nu} -J_{\alpha\beta} J_{\beta\alpha} + [Q_\gamma, \overline{Q}_\gamma]
\end{equation}
which can be obtained by taking the supertrace of $\Omega^2$, where

\begin{equation}
\Omega= \left(
\begin{array}{cccc}
J_{11} & \cdots & J_{1N} & Q_1 \\
\vdots & \ddots & \vdots & \vdots \\
J_{N1} & \cdots & J_{NN} & Q_N  \\
-{\overline{Q}}_1 & \cdots & -{\overline{Q}}_N &  \sum_{\mu=1}^{N} J_{\mu\mu} 
\end{array} \right) 
\end{equation}
is a supertraceless matrix configuration with even and odd operator matrix elements. Other invariants of the superalgebra are $C_3=Str(\Omega^3), \ldots, C_N=Str(\Omega^N)$.

Among the irreducible representations of $SU(4)\subset U(4)$, there is a 15-dimensional boson multiplet ($\pi$, $\rho$, $\omega$) and a 20-dimensional fermion multiplet ($N$, $\Delta$). With the supersymmetric extension of $U(4)$ it is possible to extend Lagrangians which contain only the $\pi$, $\rho$ (and $\omega$) to the ones that contain mesons, nucleons and $\Delta$'s. The new Lagrangian must necessarily contain the Dirac and Rarita-Schwinger Lagrangians for $N$'s and $\Delta$'s as well as the usual meson Lagrangian. The Hamiltonian to be obtained then should resemble the casimir invariant $C_2$ shown above, and the supersymmetry breaking will bring the Hamiltonian to the form Eq.(\ref{eq:5s}), and the subsequent breaking of $U(4)$ into $SU(2) \times SU(2)$ will bring the Hamiltonian to the form of Eq.(\ref{eq:3s}).

In order to avoid the problem of predicting too high soliton masses when taking physical value of $F_\pi$ as an input parameter, many people subtract from the Hamiltonian a zero point energy without explaining its origin. We suggest that the subtraction of the zero point energy without explaining its origin. We suggest that the subtraction of the zero point energy may come from the remnant effect of supersymmetry after the supersymmetric theory has been broken down to a non-supersymmetric theory. It is known that in a field theory where supersymmetry is exact, the ground state energy is zero. The positive contribution from bosons is exactly canceled by the negative contributions from the fermions. In a theory with approximate hadronic supersymmetry, although we do not expect exact cancellation of bosonic and fermionic parts of the soliton energy, it is probable that there is a good balance of both sides so that the baryonic masses can be lowered significantly to give a reasonable value of the pion decay constant.

\section*{Conclusion and remarks}

Quark model with potentials derived from QCD, including the quark-diquark model for excited hadrons gave mass formulae in very good agreement with experiment and went a long way in explaining the approximate symmetries and supersymmetries of the hadronic spectrum, including the symmetry breaking mechanism. Near success of the minimal $SU(2)$ Skyrme model was modified to a generalized composite Skyrme model in which baryons are considered as composite objects made up of many $SU(2)$ configurations. In the special case of three $SU(2)$ configuration, the static properties of the soliton had excellent agreement with the experimental data. 

According to the large $N_c$ expansion, a more realistic model should include the vector mesons and the pions. Inclusion of low lying vector mesons $\omega$ and $\rho$ in a modified model, where the Lagrangian of the model was constructed by inserting $\gamma$- and $\sigma$- matrices that allowed us to get the usual meson physics, including the meson mass relations in perfect agreement with experiment. All the mesons were treated on equal footing so that they belong to a single meson multiplet.

The simplified model we built, in which a soliton ansatz to the Lagrangian ${\cal{L}}$ has not yet been found, gives us a good indication of the salient features of any soliton solutions to the full model. 

Magnetic properties are improved by adding vector mesons to the model. With the static approximation, the magnetic part of the configuration vanishes and the model is then becomes reducible to an $SU(4)$ Skyrme model that produces a variety of mass spectra for the solitons.  

Fermionic quantization for the collective coordinates of the skyrmion is introduced. It has the advantage that through the Pauli exclusion principle it it chops off the unphysical infinite tower of ($S=I\geq \frac{5}{2})$ particle states. With this approach, nucleons and $\Delta$'s can all be expressed as trilinear functions of the collective coordinates as we have shown in this paper, and the fermionic quantization approach allows a simple incorporation of supersymmetry into the model. Such a supersymmetric extension makes it possible to lower the soliton mass since fermions contribute negatively to the Hamiltonian. Imbedding of the $SU(2) \times SU(2)$ symmetry into $U(4)$ and $U(4/n)$ ($n=1,2$ in our case) is the first step towards a supersymmetric extension. $U(N)$ algebra is extended to a superalgebra for arbitrary $N$ by the introduction of supercharges, and the Casimir invariants are easily obtained by taking the supertrace of the powers of the configuration $\Omega$. 

The fermionic degrees of freedom must be introduced in such a way that an approximate symmetry under $SU(4/10)$ holds. It is to be seen if after the suggested modifications one will obtain a quantitatively viable skyrmionic description of hadronic physics at least as good as the 2-body potential model$^{\cite{85}, \cite{88}}$ inspired by QCD and reflecting its fundamental symmetries. It is highly desirable to find an explicit relationship between the relativistic quark model we built and the Skyrme model. These two models should have common hadronic symmetries. The approximate $SU(6)$ symmetry and $SU(6/21)$ supersymmetry have been established for the quark model, but not yet for the Skyrme model. We can build a supersymmetric Skyrme model based on the superalgebra we developed. Some preliminary work has been done in that direction, which will be the subject of another publication. It will be very interesting to compare the two different supersymmetry realizations of the quark model and the Skyrme model.

Strangeness should also be incorporated into the model. If we include strangeness in our model while treating all mesons equally, we would expect the mesons to form an $SU(6)$ multiplet. With such modifications we would expect that the predictions of the static properties of the solitons can be brought into much better agreement with experiment.

Also, it should in principle be possible to construct Fermi fields from Bose fields in $3+1$ dimensions in the light of these recent developments. The picture that one would like to see emerge is a duality between a field theory written in terms of quark and gluon fields that would accurately reflect the hilbert space structure at very small distances on one hand, and a dual theory expressed in terms of non-linear meson couplings whose soliton solutions would correspond to the structure of the Hilbert space as seen from a larger distance perspective on the other.  This would require two boson-fermion connections. First the quarks and gluons must give rise to the non-linear meson fields as composites and these latter must be reinterpreted in terms of baryons. It is not clear how fundamental this phenomenologically hoped for connection can be.

Other applications such as when one considers the pion fluctuations around a soliton one can study the meson-nucleon scatterings. In the many-skyrmion system, one can study dibaryon states, nucleon-nucleon interactions and properties of nuclei. Skyrme model is a viable model of baryons. Recent developments of the model give strong support to the belief that baryons can be regarded as topological solitons of a theory of interacting meson fields, and the low energy physics should be described non-perturbatively by an effective Lagrangian based on the underlying symmetries of QCD.

\end{document}